\documentclass[preprint]{elsarticle} 

\makeatletter
	\def\ps@pprintTitle{%
 	\let\@oddhead\@empty
	\let\@evenhead\@empty
	\def\@oddfoot{\centerline{\thepage}}%
	\let\@evenfoot\@oddfoot}
\makeatother

\usepackage{etoolbox}
\patchcmd{\MaketitleBox}{\footnotesize\itshape\elsaddress\par\vskip36pt}{\footnotesize\itshape\elsaddress\par\parbox[b][36pt]{\linewidth}{\vfill\hfill\textnormal{\today}\hfill\null\vfill}}{}{}%
\patchcmd{\pprintMaketitle}{\footnotesize\itshape\elsaddress\par\vskip36pt}{\footnotesize\itshape\elsaddress\par\parbox[b][36pt]{\linewidth}{\vfill\hfill\textnormal{\today}\hfill\null\vfill}}{}{}%

\usepackage[colorlinks=true,%
            breaklinks=true,%
            linkcolor=blue,
            urlcolor=blue,%
            citecolor=blue,%
            pdftitle={Title of paper}, 
            pdfkeywords={Keywords},	
            pdfauthor={Authors},		
            bookmarksopen=false,
            pdfpagemode=UseNone]{hyperref}

\usepackage[margin=0.9in]{geometry}
\usepackage{makecell}
\usepackage[T1]{fontenc}
\usepackage[english]{babel}
\usepackage{icomma}
\usepackage[latin1]{inputenc} 
\usepackage{subfiles}
\usepackage{xcolor}
\usepackage{tabularray}
\biboptions{numbers,sort&compress}
\usepackage[justification=centering]{subcaption}

\usepackage{amsmath}
\usepackage{pifont}

\usepackage{mathtools}
\usepackage{siunitx}
\usepackage{bbm}
\usepackage{amsmath}
\usepackage{amsfonts}
\usepackage{amsthm}
\usepackage{amssymb}
\usepackage{array}
\usepackage{algorithm} 
\usepackage{algorithmicx}
\usepackage{algpseudocode}
\usepackage{siunitx}
\usepackage{amsmath}
\usepackage{mathrsfs}
\usepackage{xpatch}

\usepackage{color}
\usepackage{url}
\usepackage{cleveref}
\usepackage{graphicx}
\usepackage{breakcites}
\usepackage{soul} 
\usepackage{enumitem}
\usepackage[title,titletoc,toc]{appendix}

\newtheoremstyle{mytheoremstyle}{5pt}{5pt}{\itshape}{}{\bfseries}{.}{.5em}{} 
\theoremstyle{mytheoremstyle}

\newtheoremstyle{myremarkstyle}{3pt}{3pt}{\itshape}{}{\bfseries}{.}{.5em}{} 
\theoremstyle{myremarkstyle}

\usepackage{pifont}

\newcommand{\Hquad}{\hspace{0.5em}}

\begin{document}
\begin{frontmatter}
    \title{Understanding cold electron impact on parallel-propagating whistler chorus waves via moment-based quasilinear theory}
    \author[1,2]{Opal Issan\corref{cor1}}\ead{oissan@ucsd.edu}
    \author[3]{Vadim Roytershteyn}
    \author[2]{Gian Luca Delzanno}
    \author[2]{Salomon Janhunen}
    
    \cortext[cor1]{Corresponding author}
    \address[1]{Department of Mechanical and Aerospace Engineering, University of California San Diego, La Jolla, CA, USA}
    \address[2]{T-5 Applied Mathematics and Plasma Physics Group, Los Alamos National Laboratory, Los Alamos, NM, USA}
    \address[3]{Space Science Institute, Boulder, CO 80301, USA}  
    \begin{abstract}
    Earth's magnetosphere hosts a wide range of collisionless particle populations that interact through various wave-particle processes.
    Among these, cold electrons, with energies below \SI{100}{\electronvolt}, often dominate the plasma density but remain poorly characterized due to measurement challenges such as spacecraft charging and photoelectron contamination. 
    Understanding the contribution of these cold populations to wave-particle interaction is of significant interest. 
    Recent kinetic simulations identified a secondary drift-driven instability in which parallel-propagating whistler-mode chorus waves excite oblique electrostatic whistler waves near the resonance cone and Bernstein-mode turbulence.
    These secondary modes enable a new channel of energy transfer from the parallel-propagating whistler wave to the cold electrons. 
    In this work, we develop a moment-based quasilinear theory of the secondary instabilities to quantify such energy exchange. 
    Our results show that these secondary instabilities persist for a wide range of parameters and, in many cases, lead to nearly complete damping of the primary wave.
    Such secondary instability might limit the amplitude of parallel-propagating whistler waves in Earth's magnetosphere and might explain why high-amplitude oblique whistler or electron Bernstein waves are rarely observed simultaneously with high-amplitude field-aligned whistler waves in the inner magnetosphere.
    \end{abstract}	
    \begin{keyword}
    drift-driven instabilities \sep quasilinear theory \sep whistler chorus waves \sep cold electrons 
    \end{keyword}
\end{frontmatter}

\section{Introduction}\label{sec:introduction}
Whistler-mode chorus waves are naturally occurring coherent emissions observed in Earth's magnetosphere, typically appearing in two distinct frequency bands: above and below half of the local electron cyclotron frequency~\cite{burtis_1969_chorus, meredith_2001_JGR}. 
Chorus waves play a crucial role in the dynamics of Earth's magnetosphere by locally accelerating and scattering energetic electrons through wave-particle interactions. These processes can lead to rapid precipitation and loss of radiation belt electrons into the atmosphere~\cite{thorne_2005_JGR, shprits_2008_review}. 
Beyond Earth's magnetosphere, whistler waves are also found in the solar wind~\cite{vasko_2020_soalr_wind_POP} and other planet magnetospheres, including Mars~\cite{teng_2023_chorus_mars}, Jupiter~\cite{menietti_2016_chorus_jupiter}, and Saturn~\cite{hospodarsky_2008_chorus_saturn}, demonstrating their relevance across diverse space plasma environments.

Chorus waves near the geomagnetic equator primarily propagate parallel to the background magnetic field (also referred to as \textit{field-aligned}). They are mainly driven by anisotropic keV-energy electrons, which have a higher perpendicular than parallel temperature~\cite{kennel_1966_trapped, sagdeev_1961_instability}. 
At mid to high magnetic latitudes, chorus waves are predominantly oblique due to propagation effects, including quasi-electrostatic modes near the resonance cone angle, and are typically observed during quiet to moderately disturbed geomagnetic conditions~\cite{agapitov_2013_oblique_cluster, li_2013_themis_whistler, teng_2019_van_allen_whistler}.
These highly oblique waves can scatter electrons more efficiently and resonate with a much broader spectrum of energies and pitch angles in comparison to parallel-propagating waves~\cite{artemyev_2016_oblique_SSR, li_2014_oblique_GRL, artemyev_2015_oblique_Nature}.
Upper band oblique waves are proposed to be generated by cyclotron resonance with anisotropic low parallel beta (warm) electrons~\cite{gary_2011_whistler_POP, hashimoto_1977_oblique_JPP, liu_2011_whistler_GRL, fu_2014_oblique_JGR}. 
Highly oblique lower-band chorus waves, near the resonance cone, are also observed near the geomagnetic equator~\cite{li_2016_GRL_van_allen}, and their dominant local generation mechanisms remain a topic of active debate.
Some competing hypotheses have been proposed, including Landau resonance with parallel electron beams in the keV~\cite{mourenas_2015_oblique_JGR, artemyev_2016_oblique_SSR} and \SI{100}{\electronvolt}~\cite{li_2016_GRL} energy ranges, cyclotron resonance with anisotropic keV electrons along with a plateau at approximately \SI{100}{\electronvolt} in the parallel velocity distribution~\cite{li_2016_GRL}, as well as nonlinear three-wave interactions, wherein a quasi-parallel lower-band chorus wave couples with a mildly oblique upper-band wave to generate a highly oblique lower-band wave~\cite{fu_2017_three_wave_GRL, teng_2018_oblique_GRL}.

Recently,~\citet{roytershteyn_2021_pop} proposed a new mechanism for generating highly oblique chorus waves via a secondary drift-driven instability of field-aligned chorus in the presence of cold electrons. 
Here, \textit{cold} refers to energies below \SI{100}{\electronvolt}, corresponding to the ``hidden'' particle populations, which remain poorly understood in the magnetosphere due to spacecraft charging and measurement contamination from photoelectrons with similar energies~\cite{delaznno_2021_cold_impact}.
\citet{roytershteyn_2021_pop} showed that the oscillating electric field of parallel-propagating chorus waves induces a polarization drift between cold electrons and ions, which can excite secondary electrostatic modes~\cite{muschietti_2017_stream_instabilities}. 
These include electrostatic oblique whistler modes near the resonance cone at frequencies similar to the primary field-aligned wave, along with short-wavelength perpendicular electron Bernstein electrostatic modes similar to electron cyclotron drift driven instability~(ECDI)~\cite{forslund_1970_ecdi}. 
The resulting secondary modes lead to the heating of the cold electrons and contribute to substantial damping of the primary field-aligned chorus wave.
\citet{li_2016_GRL} attribute plateau and beam-like electron features observed in conjunction with lower-band oblique chorus as their generation source.
In contrast, ~\citet{roytershteyn_2024_frontiers} reverse this causal link: secondary oblique waves lead to the formation of a plateau at (cold and warm electron) energies near the Landau resonance.
This mechanism complements other proposed origins of plateau formation~\cite{artemyev_2020_plateu_JGR}. 
The newly discovered channel of energy transfer from field-aligned chorus waves to cold electrons alters the bigger picture of chorus wave-induced electron precipitation in the presence of a cold electron population, which is the basis for some of the most promising radiation belt remediation strategies~\cite{carlsten_2019_rbr_IEEE}.

Although cold electrons often dominate the plasma density in regions where chorus waves are observed~\cite{delaznno_2021_cold_impact}, the drift-driven secondary instabilities described above were only recently identified in part because simulations resolving cold electron scales (e.g., Debye length) are challenging~\cite{roytershteyn_2021_pop}. For instance, the 2D3V simulation in~\citet{roytershteyn_2021_pop} required approximately 3 million CPU hours for a relatively small periodic domain, rendering tracking wave packets over global magnetospheric distances impractical. 
As a result, parametric studies clarifying the role of secondary instabilities are still lacking, and it remains an open question under which conditions these instabilities become important in the inner magnetosphere.

In this work, we develop a moment-based quasilinear theory~(QLT)~\cite{davidson_1972_qlt_book, yoon_2017_qlt, vedenov_1961_qlt, diamond_2010_qlt} description of the secondary electrostatic drift-driven instabilities to quantify the damping of the primary field-aligned whistler wave and the heating of cold electrons. 
The developed QLT framework offers insights into energy partitioning and the evolution and saturation characteristics of the secondary instabilities.
We verify the developed QLT approach with a fully non-linear particle-in-cell simulation. 
We then study the parametric dependence of the secondary instability on cold electrons and primary parallel-propagating whistler wave properties. 
The results show that the generation of oblique electrostatic waves can cause significant damping of the primary wave for a large range of parameters. 
The QLT results indicate that secondary electrostatic waves persist across a broad range of cold electron densities, resulting in a significant reduction of the primary wave's magnetic energy density (more than 75\%). 
Among these, the oblique electrostatic whistlers serve as the main cause of primary wave damping, producing at least five times stronger damping compared to the perpendicular short-wavelength modes. 
The oblique modes are excited in both upper- and lower-band chorus frequency with larger growth rates for lower frequencies, but the perpendicular modes are unstable only for a smaller range of frequencies in the lower-band range.
The modes have higher growth rates when the cold electron drift amplitude exceeds the cold electron thermal speed, thereby enabling the development of secondary instabilities at comparatively low whistler amplitudes, provided that the background electron population remains sufficiently cold (sub--eV to a few eV).
This mechanism may explain why quasi-parallel lower-band waves are scarce in regions dominated by highly oblique lower-band emissions~\cite{agapitov_2016_oblique_GRL} and why intense Bernstein-mode turbulence occurs in nearly the same region (from midnight to dawn, where the plasma sheet electrons are injected) as high amplitude chorus waves but rarely simultaneously~\cite{gao_2024_ech}. 

This paper is organized as follows. 
Section~\ref{sec:linear_and_QLT_secondary} describes the linear and moment-based QLT describing the drift-driven secondary instabilities. 
Section~\ref{sec:QLT_vs_PIC} compares the moment-based QLT results with PIC simulations. 
Section~\ref{sec:QLT_parameteric_scans} examines how the properties of the secondary modes change with the properties of the cold population and primary whistler wave using linear theory and moment-based QLT. 
Lastly, section~\ref{sec:conclusions} offers conclusions and an outlook toward future work.

\paragraph{Notation}{We adopt a Cartesian ($x,y,z$) reference frame. All quantities are in Gaussian units with a uniform magnetic field $\vec{B}_{0} = B_{0} \hat{z}$, $B_{0}>0$. The subscript `$s$' denotes the plasma species. In this work, we consider three species: cold electrons (`$c$'), hot electrons (`$h$'), and cold ions/protons (`$i$'). The subscript `$e$' refers to electron quantities, and the total electron density is $n_{e} = n_{c} + n_{h}$. The subscripts $\|$ and $\perp$ refer to the vector projections parallel and perpendicular to $\vec{B}_{0}$, respectively. The plasma is assumed to be quasi-neutral $n_{e} = n_{i}$. We denote the thermal velocity as $v_{ts} \coloneqq \sqrt{T_{s}/m_{s}}$, plasma frequency as $\omega_{ps} \coloneqq \sqrt{4\pi n_{s} q_{s}^2/m_{s}}$, cyclotron frequency as $\Omega_{cs} \coloneqq q_{s} B_{0}/m_{s}c$, skin depth as $d_{s} \coloneqq c/\omega_{ps}$, plasma beta $\beta_{s} \coloneqq 8\pi n_{s}T_{s}/B_{0}^2$, gyroradius $\rho_{s} \coloneqq v_{ts}/|\Omega_{cs}|$, and anisotropy level as $A_{s} \coloneqq T_{\perp s}/T_{\| s} - 1$, where $c$ is the speed of light in vacuum, $e$ is the positive elementary charge, and $q_{s}, m_{s}, T_{s}, n_{s}$ are the charge, rest mass, temperature, and density of species $s$, respectively. We refer to the field-aligned whistler wave as the ``primary'' or ``driver'' wave. The wave normal angle, i.e., the angle between the wavevector $\vec{k}$ and the background magnetic field $\vec{B}_{0}$, is denoted by $\theta_{k}$. The velocity vector cylindrical coordinate system centered about $\vec{B}_{0}$ is denoted by $\vec{v} \coloneqq [v_{\perp}\cos(\theta_{v}) \Hquad v_{\perp} \sin(\theta_{v}) \Hquad v_{\|}]^{\top}$, where $\theta_{v}$ is the gyrophase angle. }

\section{Quasilinear theory of secondary electrostatic drift-driven instability}\label{sec:linear_and_QLT_secondary}
We start by briefly reviewing the linear theory of the secondary drift-driven instabilities in section~\ref{sec:linear_secondary}, as detailed in \citet{roytershteyn_2021_pop} and~\citet{roytershteyn_2024_frontiers}. 
We then derive the moment-based QLT equations in section~\ref{sec:QLT_secondary}, which describe the temporal evolution of cold electron heating and the simultaneous damping of the primary whistler wave by the excitation of electrostatic modes. 
In both linear theory and QLT, we neglect the hot electron response and assume that the primary field-aligned whistler wave is given.

\subsection{Linear theory}\label{sec:linear_secondary}
A plasma subject to an oscillating electric field perpendicular to the background magnetic field generates a transverse relative drift between electrons and ions~\cite[\S 6.2.3]{akhiezer_1975_book}. 
Here, the relative drift is driven by a monochromatic parallel-propagating whistler wave, which is right-hand polarized and characterized by the following electric field:
\begin{align*}
    \vec{E}_{W}(z, t) &\coloneqq  |E_{W}| \left[\sin(\omega_{0} t - k_{\|0}z) \Hquad -\cos(\omega_{0} t- k_{\|0}z) \Hquad 0\right]^{\top}, 
\end{align*}
where $\omega_{0} \lesssim \Omega_{ce}$~($k_{\|0}$) is the driver frequency (wavenumber) and $|E_{W}|$ is the amplitude of the electric field.
From this point forward, we do not include spatial variations of the driving field since the secondary instabilities correspond typically to short-wavelength fluctuations, i.e. $k_{\|} \gg k_{\|0} \sim 1/d_{e}$, where $k_{\|0}$ and $k_{\|}$ are the parallel wavenumbers associated with the primary whistler and the secondary instabilities, respectively~\cite{roytershteyn_2021_pop}. 
The equations of motion for the drift $\vec{V}_{Ds}(t)$ driven by the electric field of the primary whistler wave are 
\begin{equation}\label{eom}
    \frac{\mathrm{d} \vec{V}_{Ds}(t)}{\mathrm{d}t}  = \frac{q_{s}}{m_{s}} \left[\vec{E}_{W}(t) + \frac{\vec{V}_{Ds}(t) \times \vec{B}_{0}}{c}\right].
\end{equation}
Let $V_{\perp, Ds}(t) \coloneqq V_{x, Ds}(t) + i V_{y, Ds}(t)$, such that Eq.~\eqref{eom} yields
\begin{align*}
    \frac{\mathrm{d}V_{\perp, Ds}(t)}{\mathrm{d} t} &= -\frac{q_{s}}{m_{s}}\left[|E_{W}|i\exp(i\omega_{0}t) + \frac{i B_{0}V_{\perp, Ds}(t)}{c}\right].
\end{align*}
We assume the solution is of the form $V_{\perp, Ds}(t) \propto \exp(i\omega_{0} t)$, which results in
\begin{equation}\label{Vds}
   \boxed{V_{\perp, Ds}(t) = -\frac{q_{s}}{m_{s}}\frac{|E_{W}|}{\omega_{0} + \Omega_{cs}} \exp(i\omega_{0} t)} = -\frac{\omega_{0}}{k_{\|0}} \frac{|B_{W}|}{B_{0}} \frac{\Omega_{cs}}{\omega_{0}  + \Omega_{cs}} \exp(i\omega_{0} t),
\end{equation}
such that $|B_{W}|$ is the magnetic field amplitude of the primary wave magnetic field, i.e. $\vec{B}_{W}(z, t)$. 
In Eq.~\eqref{Vds}, we use that $|B_{W}| = \frac{c k_{\|0}}{\omega_{0}}|E_{W}|$ from Faraday's law, i.e. $\nabla_{\vec{x}} \times \vec{E}_{W}(z, t) = -\partial_{t} \vec{B}_{W}(z, t)/c$.
Because the polarization drift is mass-dependent, a relative drift between the electrons and ions arises.
For a primary whistler wave with frequency $\omega_{0}= \alpha |\Omega_{ce}|$ and $\sqrt{m_{e}/m_{i}} < \alpha < 1$, the electron drift magnitude is much larger than the ion drift magnitude, i.e. $|\vec{V}_{De}|/|\vec{V}_{Di}| \sim \frac{m_{i}}{m_{e}} \frac{\alpha}{1 - \alpha } \gg 1$. Unlike hot electrons, the cold electron drift can exceed their thermal speeds $|\vec{V}_{Dc}| \gtrsim v_{tc}$, leading to secondary drift-driven instabilities between the cold electrons and ions. 
The secondary instabilities can develop at relatively low whistler amplitude if the background electrons are sufficiently cold because the energy available to the instability is related to~$|\vec{V}_{Dc}|/v_{tc} \propto |B_{W}|/\sqrt{T_{c}} $~\cite{roytershteyn_2021_pop}. 
Since we linearize Eq.~\eqref{eom}, from superposition, Eq.~\eqref{Vds} is a good approximation also for a quasi-monochromatic primary wave with $[\Delta \omega]_{0} \ll \omega_{0}$ and $[\Delta k_{\|}]_{0} \ll k_{\|0} \ll k_{\|}$, where $[\Delta \omega]_{0}$ and $[\Delta k_{\|}]_{0}$ are the the primary wave frequency bandwidth and wavenumber bandwidth, respectively.

We begin deriving the dispersion relation with the electrostatic Vlasov-Poisson equations in a magnetized plasma, viz., 
\begin{align}
    \left[ \partial_{t}  + \vec{v} \cdot \nabla_{\vec{x}} + \frac{q_{s}}{m_{s}} \left(\delta \vec{E}(\vec{x}, t) + \vec{E}_{W}(t) + \frac{\vec{v} \times \vec{B}_{0}}{c}\right)\cdot \nabla_{\vec{v}} \right] f_{s}(\vec{x}, \vec{v}, t) = 0, \label{vlasov-equation} \\
     \nabla_{\vec{x}} \cdot \delta \vec{E}(\vec{x}, t) \coloneqq -\nabla_{\vec{x}}^{2} \delta \phi(\vec{x}, t) = 4\pi \sum_{s} q_{s} \int \mathrm{d}^3 v f_{s}(\vec{x}, \vec{v}, t), \label{poisson-equation}
\end{align} 
where $\delta \vec{E}(\vec{x}, t)$ is the secondary modes' electric field, $\delta \phi(\vec{x}, t)$ is the electrostatic potential, and $f_{s}(\vec{x}, \vec{v}, t)$ is the particle distribution function of species $s$.
A transformation into the frame of reference co-drifting (or co-oscillating) with cold electrons can be obtained by a change of coordinates $(\vec{x}, \vec{v}) \to (\vec{x}', \vec{v}')$, where 
\begin{equation}\label{co-drifting_frame}
    \vec{x}' \coloneqq \vec{x} - \int_{0}^{t} \vec{V}_{Dc}(\tau) \mathrm{d} \tau \qquad \mathrm{and} \qquad \vec{v}' \coloneqq \vec{v} - \vec{V}_{Dc}(t),
\end{equation}
such that the cold electron electrostatic Vlasov equation~\eqref{vlasov-equation} in the co-drifting frame is
\begin{equation}\label{vlasov-cold-co-drifting}
    \left[\partial_{t}  + \vec{v}'  \cdot \nabla_{\vec{x}'}  - \frac{e}{m_{e}} \left[\delta \vec{E}(\vec{x}', t) + \frac{\vec{v}' \times \vec{B}_{0}}{c} \right] \cdot \nabla_{\vec{v}'} \right]f_{c}(\vec{x}', \vec{v}', t)=0,
\end{equation}
such that $f_{c}(\vec{x}', \vec{v}', t) = F_{0c}(\vec{v}') + \delta f_{c}(\vec{x}', \vec{v}', t)$. 
We assume the cold electron equilibrium distribution function $F_{0c}(\vec{v}')$ is bi-Maxwellian and gyrotropic, i.e. $[\vec{v}' \times \vec{B}_{0}] \cdot \nabla_{\vec{v}'} F_{0c} \propto \partial_{\theta_{v'}} F_{0c}= 0$, such that 
\begin{equation}\label{bi-maxwellian}
    F_{0c}(\vec{v}') \coloneqq \frac{n_{c}}{\pi^{3/2} \alpha_{\perp c}^2 \alpha_{\| c}} \exp \left( - \frac{{v_{\|}'}^2}{\alpha_{\| c}^2} - \frac{{v_{\perp}'}^2}{\alpha_{\perp c}^2}  \right) ,\qquad
    \alpha_{\perp c}^{2} \coloneqq \frac{2 T_{\perp c}}{m_{e}}\qquad \mathrm{and} \qquad 
    \alpha_{\| c}^{2} \coloneqq \frac{2 T_{\| c}}{m_{e}}.
\end{equation}
After linearizing Eq.~\eqref{vlasov-cold-co-drifting}, the resulting cold electron linear response with a bi-Maxwellian equilibrium, see Eq.~\eqref{bi-maxwellian}, in the co-drifting frame is
\begin{align}
    \delta \tilde{n}_{c}(\vec{k}, \omega) &\coloneqq \int  \mathrm{d}^3 v' \int  \mathrm{d}^{3} x' \int \mathrm{d} t  \delta f_{c}(\vec{x}', \vec{v}', t) \exp(i\omega t -i \vec{k}\cdot \vec{x}') \label{electron-response-co-drifting}\\
    &= \frac{e n_{c}}{T_{\| c}}  \delta \tilde{\phi}(\vec{k}, \omega) \left[ 1 +   \sum_{n=-\infty}^{\infty} \Gamma_{n}(\lambda_{c}) Z(\xi_{c}^{n}) \left[\xi_{c}^{0} +  \frac{n |\Omega_{ce}|}{k_{\|} \alpha_{\| c}} \left[\frac{\alpha_{\| c}^2}{\alpha_{\perp c}^2} -1 \right]\right]\right], \nonumber
\end{align}
where $\omega \coloneqq \omega_{r} + i \gamma$ is the wave frequency, $\vec{k} = [k_{\perp} \Hquad 0 \Hquad k_{\|}]^{\top}$ is the wavevector, $Z$ is the plasma dispersion function~\cite{fried_conte_1961}, $\Gamma_{n}(\lambda_{c}) \coloneqq \exp(-\lambda_{c}) I_{n}(\lambda_{c})$, $I_{n}$ is the modified Bessel function of the first kind, $\lambda_{c} \coloneqq k_{\perp}^2 \alpha_{\perp c}^2/2\Omega_{ce}^2$, $\xi_{c}^{n} \coloneqq [\omega - n|\Omega_{ce}|]/|k_{\|}| \alpha_{\| c}$, $\delta \tilde{\phi}(\vec{k}, \omega)$ is the electrostatic potential in the co-drifting frame.
We adopt the convention that $\omega_{r} > 0$.
The algebraic steps to derive the electrostatic response for magnetized plasmas with bi-Maxwellian distributions, including integration over unperturbed orbits and Bessel function identities, are described in~\citet[\S 11]{stix_1992_theory} and~\citet[\S 2]{gary_1993_theory}. 

We assume that the ions are unmagnetized because their drift due to the primary whistler wave is negligible. 
Therefore, the ion response with an isotropic Maxwellian equilibrium in the laboratory frame is
\begin{equation}\label{ion_response_lab}
    \delta \hat{n}_{i}(\vec{k}, \omega) \coloneqq \int \mathrm{d}^3 v f_{i}(\vec{x}, \vec{v}, t) =  \frac{e n_{i}}{2T_{i}} \delta \hat{\phi}(\vec{k}, \omega) Z'\left(\frac{\omega}{\sqrt{2} v_{ti} |\vec{k}|}\right).
\end{equation}
Following~\citet{kaw_and_lee_1973}, we can relate the Fourier components of any quantity in the co-drifting (with cold electrons) frame $\tilde{A}$ to those in the stationary (ion) frame $\hat{A}$, i.e. 
\begin{align*}
    \tilde{A}(\vec{k}, \omega) &= \int \mathrm{d} t \int \mathrm{d}^{3}x \exp\left(i \omega t - i \vec{k}\cdot\vec{x}' \right) \tilde{A}(\vec{x}', t) =  \int \mathrm{d} t \int \mathrm{d}^{3}x \exp\left(i \omega t - i \vec{k}\cdot\vec{x}  + \frac{ik_{\perp}  |\vec{V}_{Dc}|}{\omega_{0}} \sin(\omega_{0} t)\right) \hat{A}(\vec{x}, t) \\
    &= \sum_{n=-\infty}^{\infty}  J_{n}\left(\frac{k_{\perp} |\vec{V}_{Dc}|}{\omega_{0}} \right) \hat{A}( \vec{k}, \omega+n\omega_{0}),
\end{align*}
where $J_{n}$ is the Bessel function of the first kind. In the above expression, we use the Jacobi-Anger identity~\cite[\S 9.1.42--43]{abramowitz_1964_math}:
\begin{equation*}
    \exp(ia\sin(x))= \sum_{n=-\infty}^{\infty} J_{n}(a) \exp(inx).
\end{equation*}
Similarly, the inverse transform is 
\begin{equation*}
    \hat{A}(\vec{k}, \omega) = \sum_{n=-\infty}^{\infty} J_{n}\left(\frac{k_{\perp} |\vec{V}_{Dc}|}{\omega_{0}} \right) \tilde{A}(\vec{k}, \omega - n \omega_{0}).
\end{equation*}
Therefore, the ion response~\eqref{ion_response_lab} in the co-drifting frame is 
\begin{equation}\label{ion-response-co-drifting}
    \delta \tilde{n}_{i}(\vec{k}, \omega) = \frac{e n_{i}}{2 T_{i}} \sum_{m=-\infty}^{\infty} \sum_{n=-\infty}^{\infty} J_{m}\left(\frac{k_{\perp} |\vec{V}_{Dc}|}{\omega_{0}}\right) J_{n}\left(\frac{k_{\perp} |\vec{V}_{Dc}|}{\omega_{0}}\right) Z' \left( \frac{\omega + m \omega_{0}}{\sqrt{2} v_{ti} |\vec{k}|}\right) \delta \tilde{\phi}\left(\vec{k}, \omega + [m - n] \omega_{0}\right).
\end{equation}
Inserting the ion response~\eqref{ion-response-co-drifting} and cold electron reponse~\eqref{electron-response-co-drifting} in the co-drifting frame in the Poisson equation~\eqref{poisson-equation} results in 
\begin{align}\label{full-dispersion-relation}
    |\vec{k}|^2 \delta \tilde{\phi}(\vec{k}, \omega) + 2 \frac{n_{c}}{n_{e}} \frac{\omega_{pe}^2}{\alpha_{\| c}^2}  \delta \tilde{\phi}(\vec{k}, \omega) \left[ 1 +   \sum_{n=-\infty}^{\infty} \Gamma_{n}(\lambda_{c}) Z(\xi_{c}^{n}) \left[\xi_{c}^{0} +  \frac{n |\Omega_{ce}|}{k_{\|} \alpha_{\| c}} \left[ \frac{\alpha_{\| c}^2}{\alpha_{\perp c}^2} - 1 \right]\right]\right] \\
    = \frac{\omega_{pi}^2}{2 v_{ti}^2} \sum_{m=-\infty}^{\infty} \sum_{n=-\infty}^{\infty} J_{m}\left(\frac{k_{\perp} |\vec{V}_{Dc}|}{\omega_{0}}\right) J_{n}\left(\frac{k_{\perp} |\vec{V}_{Dc}|}{\omega_{0}}\right) Z' \left( \frac{\omega + m \omega_{0}}{\sqrt{2} v_{ti} |\vec{k}|}\right) \delta \tilde{\phi}\left(\vec{k}, \omega + [m -n] \omega_{0}\right). \nonumber
\end{align}
We write Eq.~\eqref{full-dispersion-relation} in a more compact form as
\begin{align}
    F(\vec{k}, \omega) \delta \tilde{\phi}(\vec{k}, \omega) &= \sum_{m'=-\infty}^{\infty} \Theta_{0, m'}(\vec{k}, \omega) \delta \tilde{\phi}(\vec{k}, \omega + m'\omega_{0}), \label{f-theta-equation}\\
    F(\vec{k}, \omega) &\coloneqq |\vec{k}|^2  + 2 \frac{n_{c}}{n_{e}} \frac{\omega_{pe}^2}{\alpha_{\| c}^2}  \left[ 1 +   \sum_{n=-\infty}^{\infty} \Gamma_{n}(\lambda_{c}) Z(\xi_{c}^{n}) \left[\xi_{c}^{0} +  \frac{n |\Omega_{ce}|}{k_{\|} \alpha_{\| c}} \left[ \frac{\alpha_{\| c}^2}{\alpha_{\perp c}^2} -1\right]\right]\right], \nonumber \\
    \Theta_{n, m'}(\vec{k}, \omega) &\coloneqq \frac{\omega_{pi}^2}{2 v_{ti}^2} \sum_{m=-\infty}^{\infty}  J_{m-n}\left(\frac{k_{\perp} |\vec{V}_{Dc}|}{\omega_{0}}\right)  J_{m -m'}\left(\frac{k_{\perp} |\vec{V}_{Dc}|}{\omega_{0}}\right) Z' \left( \frac{\omega + m \omega_{0}}{\sqrt{2} v_{ti} |\vec{k}|}\right). \nonumber
\end{align}
Equation~\eqref{f-theta-equation} can be evaluated at any $\omega$ including sidebands $\omega + n \omega_{0}$, such that 
\begin{equation*}
    F(\vec{k}, \omega + n \omega_{0}) \delta \tilde{\phi}(\vec{k}, \omega + n\omega_{0}) = \sum_{m'=-\infty}^{\infty} \Theta_{0, m'}(\vec{k}, \omega + n \omega_{0}) \delta \tilde{\phi}(\vec{k}, \omega + (m' + n)\omega_{0}) =  \sum_{m'=-\infty}^{\infty} \Theta_{n, m'}(\vec{k}, \omega) \delta \tilde{\phi}(\vec{k}, \omega + m'\omega_{0}).
\end{equation*}
Therefore, solving for the full dispersion relation leads to solving for $\mathrm{det}(\mathcal{D}(\vec{k}, \omega)) = 0$, where we truncate the infinite matrix to $\mathcal{D}(\vec{k}, \omega) \in \mathbb{C}^{2N+1 \times 2N+1}$, $N \in \mathbb{N}$ is the number of sidebands included in each direction, and $\mathcal{D}_{n, m}(\vec{k}, \omega) \coloneqq F(\vec{k}, \omega + n\omega_{0}) \delta_{n, m} -  \Theta_{n, m}(\vec{k}, \omega)$, where $\delta_{n, m}$ is the Kronecker delta function. 
The dispersion relation can be simplified further by only including the frequencies satisfying $|\omega + m^{*} \omega_{0}| \sim \sqrt{2} v_{ti} |\vec{k}|$ with $m^{*}\in \mathbb{Z}$:
\begin{align}\label{approx-dispersion-relation}
    |\vec{k}|^2  + 2 \frac{n_{c}}{n_{e}} \frac{\omega_{pe}^2}{\alpha_{\| c}^2}  \left[ 1 +   \sum_{n=-\infty}^{\infty}  \Gamma_{n}(\lambda_{c}) Z(\xi_{c}^{n}) \left[\xi_{c}^{0} +  \frac{n |\Omega_{ce}|}{k_{\|} \alpha_{\| c}} \left[ \frac{\alpha_{\| c}^2}{\alpha_{\perp c}^2} -1\right]\right]\right] \\
    = \frac{\omega_{pi}^2}{2 v_{ti}^2}  \left[J_{m*}\left(\frac{k_{\perp} |\vec{V}_{Dc}|}{\omega_{0}}\right)\right]^2  Z' \left( \frac{\omega + m^{*} \omega_{0}}{\sqrt{2} v_{ti} |\vec{k}|}\right). \nonumber
\end{align}
This approximation assumes the electron response intersects with only one driver harmonic and ignores couplings to other modes. 
This approximation is justified in the parameter regime considered in this study by numerically comparing the dispersion relation results of Eq.~\eqref{approx-dispersion-relation} with those of Eq.~\eqref{f-theta-equation} in section~\ref{sec:parametric_linear}.
For the range of parameters discussed in this work, the unstable modes correspond to short-wavelength ECDI-like modes, leading to perpendicular heating of the cold electrons and oblique electrostatic whistler, where the refractive index approaches infinity, which heats the cold electrons in both parallel and perpendicular directions. 

As shown by~\citet{roytershteyn_2024_frontiers}, we can gain insight into the nature of the instability by deriving a simple expression for the oblique modes growth rate $\gamma$ via simplifying Eq.~\eqref{approx-dispersion-relation} by assuming $|\xi_{c}|\gg1$, expanding the ion response in the small argument limit $Z'(\xi) \approx - 2 - 2i \sqrt{\pi} \xi$, and $\lambda_{c} \ll 1$ with $\Gamma_{0}(\lambda_{c}) \approx 1 - \lambda_{c}$ and $\Gamma_{\pm1}(\lambda_{c}) \approx \lambda_{c}/2$, i.e.
\begin{equation*}
    k_{\|}^2\left[1 - \frac{n_{c}}{n_{e}}\frac{\omega_{pe}^2}{\omega^2} \right] + k_{\perp}^2 \left[ 1 - \frac{n_{c}}{n_{e} }\frac{\alpha_{\perp c}^2}{\alpha_{\|c}^2}\frac{\omega_{pe}^2}{\omega^2 - \Omega_{ce}^2} \right] +\frac{\omega_{pi}^2}{ v_{ti}^2}  \left[J_{m^{*}}\left(\frac{k_{\perp} |\vec{V}_{Dc}|}{\omega_{0}}\right)\right]^2 \left[ 1 + \frac{i \sqrt{\pi}  [\omega + m^{*} \omega_{0}]}{\sqrt{2} v_{ti} |\vec{k}|}\right] = 0,
\end{equation*}
and by assuming $\gamma /\omega_{r} \ll 1$, we get 
\begin{equation}\label{simplified-growth-rate-oblique-whistler}
    [\text{oblique whistler}] \qquad \frac{\gamma}{\omega_{r}} = -\frac{\sqrt{\pi}}{2 v_{ti}^3}\frac{n_{e}}{n_{c}} \frac{m_{e}}{m_{i}}  \frac{\omega_{r} + m^{*} \omega_{0}}{\sqrt{2} |\vec{k}| }\left[J_{m^{*}}\left(\frac{k_{\perp} |\vec{V}_{Dc}|}{\omega_{0}}\right)\right]^2  \left[ \frac{k_{\|}^2}{\omega_{r}^2} +  \frac{\alpha_{\perp c}^2}{\alpha_{\| c}^2} \frac{\omega_{r}^2 k_{\perp}^2}{(\omega_{r}^2 - \Omega_{ce}^2)^{2}} \right]^{-1}.
\end{equation}
Therefore, we observe from Eq.~\eqref{simplified-growth-rate-oblique-whistler} that the secondary electrostatic instability occurs when $\omega_{r} < -m^{*}\omega_{0}$. 
In the limit of perpendicular propagation ($k_{\|} \ll k_{\perp}$) and by the first-order asymptotic expansion of the plasma dispersion function $Z(\xi) \approx -1/\xi$ for $|\xi| \gg 1$, the cold electron component of the dispersion relation in Eq.~\eqref{approx-dispersion-relation} can be further simplified to
\begin{equation}\label{dispersion_relation_ecdi}
    k_{\perp}^2 - 4 \frac{n_{c}}{n_{e}} \frac{\omega_{pe}^2}{\alpha_{\perp c}^2} \sum_{n=1}^{\infty} \frac{n^2 \Omega_{ce}^2 \Gamma_{n}(\lambda_{c})}{\omega^2 - n^2 \Omega_{ce}^2} = \frac{\omega_{pi}^2}{2v_{ti}^2} \left[J_{m^{*}}\left(\frac{k_{\perp} |\vec{V}_{Dc}|}{\omega_{0}}\right)\right]^2 Z'\left(\frac{\omega + m^{*}\omega_{0}}{\sqrt{2} v_{ti}k_{\perp}}\right),
\end{equation}
see~\cite[\S 11]{stix_1992_theory} for detailed derivation. We can also gain insight to the ECDI-like modes instability by simplifying Eq.~\eqref{dispersion_relation_ecdi} with $\lambda_{c} \gg 1$, approximating the ion response in the small argument limit $Z'(\xi) \approx - 2 - 2i \sqrt{\pi} \xi$, and by assuming $\gamma /\omega_{r} \ll 1$, which results in 
\begin{equation}\label{simplified-growth-rate-ECDI}
    [\text{ECDI-like}] \qquad \frac{\gamma}{\omega_{r}} = - \frac{\sqrt{\pi}}{8 v_{ti}^3} \frac{n_{e}}{n_{c}} \frac{m_{e}}{m_{i}} \frac{\omega_{r} + m^{*} \omega_{0}}{\sqrt{2} |\vec{k}|} \left[J_{m^{*}}\left(\frac{k_{\perp} |\vec{V}_{Dc}|}{\omega_{0}}\right)\right]^2 \left[\frac{\Gamma_{1}(\lambda_{c})\Omega_{ce}^2 \omega_{r}^2}{\alpha_{\perp c}^2(\omega_{r}^2 - \Omega_{ce}^2)^2} \right]^{-1}.
\end{equation}
Similar to Eq.~\eqref{simplified-growth-rate-oblique-whistler}, Eq.~\eqref{simplified-growth-rate-ECDI} shows that the instability occurs when $\omega_{r} < -m^{*}\omega_{0}$. We further discuss the parametric dependence of the simplified dispersion relations in Eqns.~\eqref{simplified-growth-rate-oblique-whistler} and~\eqref{simplified-growth-rate-ECDI} in section~\ref{sec:parametric_linear}.

\subsection{Moment-based quasilinear theory}\label{sec:QLT_secondary}
Quasilinear theory~(QLT), developed in the 1960s~\cite{vedenov_1961_qlt, drummond_1962_qlt}, describes the slow evolution of the particle distribution function and its relaxation back to a marginally stable state. 
It is applicable in regimes with small-amplitude, broadband, randomly phased waves, enabling a statistical description of velocity-space diffusion~\cite{diamond_2010_qlt}.
We employ the moment-based QLT approach, also known as macroscopic QLT~\cite{davidson_1972_qlt_book}, which involves taking velocity moments of the QLT equations and assuming a bi-Maxwellian equilibrium (see~\cite{yoon_2017_qlt} for a review in the context of anisotropy-driven instabilities).

We begin by allowing the cold electron equilibrium distribution function $F_{0c}(\vec{v}, t)$ to evolve in time.
The cold electron QLT evolution equations in the co-drifting frame are
\begin{align}
    \partial_{t} F_{0c}(\vec{v}', t) &= \frac{1}{v_{\perp}'} \partial_{v_{\perp}'} \left[v_{\perp}' \left[ \mathcal{D}_{\perp \perp} \partial_{v_{\perp}'} F_{0c} + \mathcal{D}_{\perp \|} \partial_{v_{\|}'} F_{0c}\right] \right] + \partial_{v_{\|}'} \left[ \mathcal{D}_{\perp \|} \partial_{v_{\perp}'} F_{0c} + \mathcal{D}_{\| \|} \partial_{v_{\|}'} F_{0c}\right], \label{qlt_evolution_pde}\\
    \partial_{t} |\delta \vec{\tilde{E}}(\vec{k}, t)|^2 &= 2 \gamma  |\delta \vec{\tilde{E}}(\vec{k}, t)|^2, \label{dEdt}
\end{align}
where the diffusion tensors are
\begin{equation}\label{diffusion_tensors}
\begin{bmatrix}
\mathcal{D}_{\perp \perp}  \\
\mathcal{D}_{\perp \|}  \\
\mathcal{D}_{\| \|}  
\end{bmatrix}\coloneqq i \frac{e^2}{m_{e}^2} \sum_{n=-\infty}^{\infty} \int \mathrm{d}^3 k \frac{|\delta \vec{\tilde{E}}(\vec{k}, t)|^2}{|\vec{k}|^2}  \frac{J_{n}^2(k_{\perp}v_{\perp}'/\Omega_{ce})}{\omega - k_{\|} v_{\|}' - n\Omega_{ce}}  \begin{bmatrix}
n^2 \Omega_{ce}^2/{v_{\perp}'}^2 \\
n \Omega_{ce} k_{\|}/v_{\perp}' \\
k_{\|}^2  
\end{bmatrix},
\end{equation}
see~\cite[\S 8.5]{davidson_1972_qlt_book}. It is convenient to derive the QLT equations in the co-drifting frame, as it is equivalent to the electrostatic QLT equations in a magnetized plasma, excluding the presence of the primary wave.
The diffusion tensors in Eq.~\eqref{diffusion_tensors} can be decomposed into non-resonant and resonant components $\mathcal{D} = \mathcal{D}^{nr} + \mathcal{D}^{r}$, where the resonant components are
\begin{equation}\label{resonant_diffusion_tensors}
\begin{bmatrix}
\mathcal{D}^{r}_{\perp \perp}  \\
\mathcal{D}^{r}_{\perp \|}  \\
\mathcal{D}^{r}_{\| \|}  
\end{bmatrix}\coloneqq \pi \frac{e^2}{m_{e}^2} \sum_{n=-\infty}^{\infty} \int \mathrm{d}^{3} k \frac{|\delta \vec{\tilde{E}}(\vec{k}, t)|^2}{|\vec{k}|^2}  J_{n}^2\left(\frac{k_{\perp}v_{\perp}'}{\Omega_{ce}}\right) \delta(\mathrm{Re}\{\omega\} - k_{\|} v_{\|}' - n\Omega_{ce}) \begin{bmatrix}
n^2 \Omega_{ce}^2/{v_{\perp}'}^2 \\
n \Omega_{ce} k_{\|}/v_{\perp}' \\
k_{\|}^2  
\end{bmatrix}.
\end{equation}
In the above expression, we utilize the Plemelj formula~\cite[\S 8.4.2]{davidson_1972_qlt_book}.
The resonant diffusion operator $\mathcal{D}^{r}$ describes irreversible exchange of energy between the resonant particles and the waves, and the non-resonant diffusion operator $\mathcal{D}^{nr}$ describes reversible exchange of energy supported by particles oscillating or quivering in the presence of a wave field~\cite[\S 3.3.1]{diamond_2010_qlt}.
The parallel and perpendicular second-order moment components can be manipulated via the product rule and Eq.~\eqref{qlt_evolution_pde} as
\begin{align*}
    \frac{\mathrm{d} K_{\perp c}(t)}{\mathrm{d} t} &\coloneqq \frac{m_{e}}{2} \int \mathrm{d}^{3} v' {v_{\perp}'}^2 \partial_{t} F_{0c}(\vec{v}', t) = -2\pi 
    m_{e} \int_{-\infty}^{\infty} \mathrm{d} v_{\|}' \int_{0}^{\infty} \mathrm{d} v_{\perp}' v_{\perp}'^2 \left[\mathcal{D}_{\perp \perp } \partial_{v_{\perp}'}  + \mathcal{D}_{\| \perp} \partial_{v_{\|}'} \right] F_{0c}(\vec{v}', t),  \\
    \frac{\mathrm{d} K_{\| c}(t)}{\mathrm{d} t} &\coloneqq m_{e} \int \mathrm{d}^{3}v' v_{\|}'^2 \partial_{t} F_{0c}(\vec{v}', t)  = -4\pi m_{e}  \int_{-\infty}^{\infty} \mathrm{d} v_{\|}' \int_{0}^{\infty} \mathrm{d} v_{\perp}' v_{\|}' v_{\perp}' \left[ \mathcal{D}_{\| \perp} \partial_{v_{\perp}'} + \mathcal{D}_{\| \|} \partial_{v_{\|}'}\right] F_{0c}(\vec{v}', t).
\end{align*}
In the above, we used $\mathrm{d}^{3}v \coloneqq v_{\perp} \mathrm{d} v_{\|} \mathrm{d} v_{\perp} \mathrm{d} \theta_{v}$ and integration by parts.
Similarly, temperature due to resonant particle heating is computed by the resonant component of the diffusion tensors~\eqref{resonant_diffusion_tensors}:
\begin{align*}
    \frac{\mathrm{d} T_{\perp c}(t)}{\mathrm{d} t} &= -\frac{2\pi 
    m_{e}}{n_{c}} \int_{-\infty}^{\infty} \mathrm{d} v_{\|}' \int_{0}^{\infty} \mathrm{d} v_{\perp}' v_{\perp}'^2 \left[\mathcal{D}_{\perp \perp }^{r} \partial_{v_{\perp}'}  + \mathcal{D}_{\| \perp}^{r} \partial_{v_{\|}'} \right] F_{0c},  \\
    \frac{\mathrm{d} T_{\| c}(t)}{\mathrm{d} t} &= -\frac{4\pi m_{e}}{n_{c}}  \int_{-\infty}^{\infty} \mathrm{d} v_{\|}' \int_{0}^{\infty} \mathrm{d} v_{\perp}' v_{\|}' v_{\perp}' \left[ \mathcal{D}_{\| \perp}^{r} \partial_{v_{\perp}'} + \mathcal{D}_{\| \|}^{r} \partial_{v_{\|}'}\right] F_{0c}.
\end{align*}
Next, we assume that the cold electron equilibrium $F_{0c}(\vec{v}, t)$ is bi-Maxwellian for all time, see Eq.~\eqref{bi-maxwellian} except that we allow $T_{\perp c}$ and $T_{\| c}$ (and correspondingly $\alpha_{\perp c}$ and $\alpha_{\| c}$) to evolve in time. This could be regarded as the lowest-order approximation~\cite{yoon_2017_qlt}. Then, by taking moments of Eq.~\eqref{qlt_evolution_pde}, the second-order moment and temperature evolutions are given by 
\begin{align}
    \frac{\mathrm{d} K_{\perp c}(t)}{\mathrm{d} t} &= \frac{1}{2\pi} \frac{n_{c}}{n_{e}} \frac{\omega_{pe}^2}{\alpha_{\| c}^2} \int \mathrm{d}^3 k  \frac{|\delta \vec{\tilde{E}}(\vec{k}, t)|^2}{|\vec{k}|^2} \mathrm{Im}\left\{\sum_{n=-\infty}^{\infty}n|\Omega_{ce}|   \Gamma_{n}(\lambda_{c}) \left[\xi_{c}^{0} + \frac{n|\Omega_{ce}|}{k_{\|} \alpha_{\| c} } \left[\frac{\alpha_{\| c}^2}{\alpha_{\perp c}^2} - 1\right] \right] Z(\xi_{c}^{n})\right\},\label{dKperpdt} \\
    \frac{\mathrm{d} K_{\| c}(t)}{\mathrm{d} t} &= \frac{1}{\pi} \frac{n_{c}}{n_{e}} \frac{\omega_{pe}^2}{\alpha_{\| c}^2} \int \mathrm{d}^3 k \frac{|\delta \vec{\tilde{E}}(\vec{k}, t)|^2}{|\vec{k}|^2}\mathrm{Im} \left\{\omega +  \sum_{n=-\infty}^{\infty}  \Gamma_{n}(\lambda_{c}) \left[ \omega  + n |\Omega_{ce}|\left[\frac{\alpha_{\| c}^2}{\alpha_{\perp c}^2} -1 \right]   \right] \xi_{c}^{n}Z(\xi_{c}^{n}) \right\}, \label{dKpardt} \\
    \frac{\mathrm{d} T_{\perp c}(t)}{\mathrm{d} t} &= \frac{\omega_{pe}^2}{2\sqrt{\pi}n_{e}\alpha_{\| c}^2} \ \int \mathrm{d}^3 k \frac{|\delta \vec{\tilde{E}}(\vec{k}, t)|^2}{|\vec{k}|^2}   \sum_{n=-\infty}^{\infty} n |\Omega_{ce}| \Gamma_{n}(\lambda_{c})\left[\mathrm{Re}\{\xi_{c}^{0}\} + \frac{n|\Omega_{ce}|}{k_{\|} \alpha_{\| c}} \left[\frac{\alpha_{\| c}^2}{\alpha_{\perp c}^2} -1 \right] \right] \exp\left(-\mathrm{Re}\{\xi_{c}^{n}\}^2 \right), \label{dTperpdt} \\
    \frac{\mathrm{d} T_{\| c}(t)}{\mathrm{d} t} &= \frac{\omega_{pe}^2}{  \sqrt{\pi}n_{e}\alpha_{\| c}^2} \int \mathrm{d}^3 k \frac{|\delta \vec{\tilde{E}}(\vec{k}, t)|^2}{|\vec{k}|^2} \sum_{n=-\infty}^{\infty}  \Gamma_{n}(\lambda_{c}) \left[ \omega_{r} + n|\Omega_{ce} |\left[\frac{\alpha_{\| c}^2 }{\alpha_{\perp c}^2 } -1\right] \right] \mathrm{Re}\{\xi_{c}^{n}\} \exp\left(-\mathrm{Re}\{\xi_{c}^{n}\}^2 \right).\label{dTpardt}
\end{align}
For perpendicular electrostatic waves $k_{\|} \ll k_{\perp}$, the perpendicular second-order moment and temperature evolution simplify to 
\begin{align}
    \frac{\mathrm{d} K_{\perp c}(t)}{\mathrm{d} t} &=  -\frac{1}{4\pi} \int_{0}^{\infty} \mathrm{d} k_{\perp} k_{\perp} |\delta \vec{\tilde{E}}(k_{\perp}, t)|^2  \mathrm{Im} \left\{\omega \left[ 1- \frac{\omega_{pi}^2 }{2 k_{\perp}^2 v_{ti}^2} \left[J_{m*}\left(\frac{k_{\perp} |V_{Dc}|}{ \omega_{0}}\right) \right]^2 Z{'}\left(\frac{\omega + m_{*} \omega_{0}}{\sqrt{2} k_{\perp} v_{ti}}\right) \right]\right\}, \label{dKperpdt_ecdi}\\
    \frac{\mathrm{d} T_{\perp c}(t)}{\mathrm{d} t} &=   \frac{\omega_{pi}^2}{8 \pi n_{c} v_{ti}^2}  \int_{0}^{\infty} \mathrm{d} k_{\perp} k_{\perp} \omega_{r} \frac{ |\delta \vec{\tilde{E}}(k_{\perp}, t)|^2 }{k_{\perp}^2}\left[J_{m*}\left(\frac{k_{\perp} |V_{Dc}|}{ \omega_{0}}\right)\right]^2  \mathrm{Im} \left\{Z{'}\left(\frac{\omega_{r} + m_{*} \omega_{0}}{\sqrt{2} k_{\perp} v_{ti}}\right)\right\}, \label{Tperpdt_ecdi}
\end{align}
where we used the first-order asymptotic expansion of the plasma dispersion function $Z(\xi) \approx -1/\xi$ for $|\xi| \gg 1$ and inserted the dispersion relation in Eq.~\eqref{dispersion_relation_ecdi}.
We can compute the damping of the primary whistler wave and the drift of the cold electrons via conservation of energy in the laboratory frame:
\begin{equation}\label{dBdt}
    \frac{\mathrm{d} |B_{W}(t)|^2}{\mathrm{d}t} = - \frac{8\pi \left[\frac{\mathrm{d} K_{\perp c}(t)}{\mathrm{d} t} + \frac{1}{2} \frac{\mathrm{d} K_{\| c}(t)}{\mathrm{d} t} + \frac{1}{4\pi} \int \mathrm{d}^3 k \gamma |\delta \vec{\tilde{E}}(\vec{k}, t)|^2 \right]}{1 + 4\pi\left|\frac{\omega_{0}}{k_{\|0}} \frac{1}{B_{0}} \frac{|\Omega_{ce}|}{\omega_{0}  -|\Omega_{ce}|}\right|^2} ,
\end{equation}
see~\ref{sec:energy_conservation} for a detailed derivation transforming from quantities in the laboratory and co-drifting frame (since the co-drifting frame is non-inertial and does not conserve total energy). 
The main assumptions in Eq.~\eqref{dBdt} are that only cold electrons, primary whistler waves, and the secondary electrostatic waves contribute to the energy conservation. We treat the ions as stationary because the secondary instability develops on electron timescales.
From Eq.~\eqref{Vds}, the drift velocity amplitude is proportional to the primary wave magnetic field amplitude, i.e.
\begin{equation}\label{dVdt}
    \frac{\mathrm{d}|V_{Dc}(t)|^2}{\mathrm{d}t} =  \left[\frac{\omega_{0}}{k_{\|0}} \frac{|\Omega_{ce}|}{\omega_{0} - |\Omega_{ce}|} \right]^2 \frac{\mathrm{d} }{\mathrm{d}t}\frac{|B_{W}(t)|^2}{B_{0}^2}.
\end{equation}
To quantify the damping of the primary whistler and the heating of the cold electrons due to secondary instabilities, we solve Eqns.~\eqref{dEdt}, \eqref{dKperpdt}--\eqref{dTpardt} (or \eqref{dKperpdt_ecdi}--\eqref{Tperpdt_ecdi}), \eqref{dBdt}, and~\eqref{dVdt}, in conjunction with the simplified dispersion relation (Eq.~\eqref{approx-dispersion-relation} for the oblique electrostatic whistler and Eq.~\eqref{dispersion_relation_ecdi} for the ECDI-like perpendicular modes) that is evaluated at each time step.

\section{Comparison of quasilinear theory with particle-in-cell simulation}\label{sec:QLT_vs_PIC}
We compare the developed moment-based QLT with a PIC simulation performed using the Vector Particle-In-Cell code~\cite{bowers_2008_VPIC}, which solves numerically the system of relativistic Vlasov-Maxwell equations.
It is important to highlight that the primary whistler wave drives the secondary instabilities, regardless of the specific source exciting the primary wave (e.g., heat-flux~\cite{gary_1994_whistler_heat_flux} and antennas~\cite{stenzel_2019_whistler_antenna}). 
In this setup, the initial anisotropy of the hot electron population excites the parallel propagating whistler wave. 
We consider a low-anisotropy 2D3V simulation presented in~\citet{roytershteyn_2021_pop}.
We describe the PIC and moment-based QLT simulation setup in section~\ref{sec:simulation-setup-PIC-QLT} and compare their results in section~\ref{sec:PIC-numerical}.

\subsection{Particle-in-cell and quasilinear theory simulation setup}\label{sec:simulation-setup-PIC-QLT}
We initialize the cold electron and ion populations as isotropic Maxwellian distributions ($A_{c}=A_{i} = 0$) with equal temperature ($T_{i} = T_{c}$), and the hot electron population as an anisotropic bi-Maxwellian distribution with $A_{h} = 2$.
The initialization parameters for the PIC simulation are $n_{c} = 0.8n_{e}$, $\omega_{pe} = 4 |\Omega_{ce}|$, $T_{c} = \SI{1}{\electronvolt}$, $\alpha_{c} = 0.0079 d_{e} |\Omega_{ce}|$, and $T_{\| h}/T_{c} = 2000$. 
The maximum growth rate of the primary whistler wave has a frequency $\omega_{0} \approx 0.5 |\Omega_{ce}|$ and wavenumber $k_{\|0} d_{e} \approx 1$, see~\ref{sec:QLT_primary} for the dispersion relation results of the primary whistler anisotropy instability.
The PIC discretization parameters are: a periodic domain of size $L_{x} \times L_{z} = 0.2 \pi d_{e} \times 2 \pi d_{e}$, a grid resolution of $n_{x} \times n_{z} = 540 \times 5000$ cells, $N_{ppc} = 10^{4}$ particles per cell per species, and a timestep of $\Delta t \omega_{pe} \approx 8.4 \times 10^{-4}$. 
The spatial domain is relatively short in the $\hat{z}$-direction, resolving only a single mode of the primary whistler wave.
The most unstable mode is a field-aligned whistler wave, as the hot electron parallel beta ($\beta_{\|h} \approx 0.128$) is above the critical beta value ($\beta^{*}_{\|h} \approx 0.004$) for the given cold electron density $n_{c}=0.8n_{e}$, see~\cite[Eq.~(2)]{gary_2012_linear} for the critical beta empirical relation.
Because the QLT derived in section~\ref{sec:QLT_secondary} assumes a given primary whistler wave, we initialize the moment-based QLT simulations with PIC estimates at $t|\Omega_{ce}| \approx 700$, where $|\vec{V}_{Dc}(t=0)|/v_{tc} \approx 0.92$, $|B_{W}(t=0)|^2/d_{e}^3 m_{e}^2 \Omega_{ce}^4 e^{-2} \approx 4 \times 10^{-4}$, and $|\delta \vec{\tilde{E}}(\vec{k}, t=0)|^2/ m_{e}^2 \Omega_{ce}^4 e^{-2}\approx 10^{-11}$. 
For the moment-based QLT simulations, we use the implicit temporal integrator described in~\citet{petzold_1983_temporal_integrator}, which automatically switches in time from the nonstiff implicit Adams method to the stiff backward differentiation formulas, with absolute and relative error tolerances set to $10^{-6}$. 
Additionally, we solve the dispersion relations numerically using the Newton-Raphson method. 
We use $N_{k}=50$ for the QLT resolution for estimating the wavenumber integrals at a fixed wave normal angle with $\theta_{k} = 90^{\circ}$ for perpendicular modes and $\theta_{k} = \arccos(\omega_{0}/|\Omega_{ce}|)$ for the oblique modes.
We assume the two secondary instabilities (oblique and perpendicular modes) are independent of one another, such that their cumulative effect is given by their superposition. 
This is motivated by the PIC simulations, which show that the perpendicular and oblique modes saturate at different times (see Figures 4 and 8 in~\citet{roytershteyn_2021_pop}). 

\subsection{Particle-in-cell vs. quasilinear theory numerical results}\label{sec:PIC-numerical}

\begin{figure}
    \centering
    \includegraphics[width=0.65\linewidth]{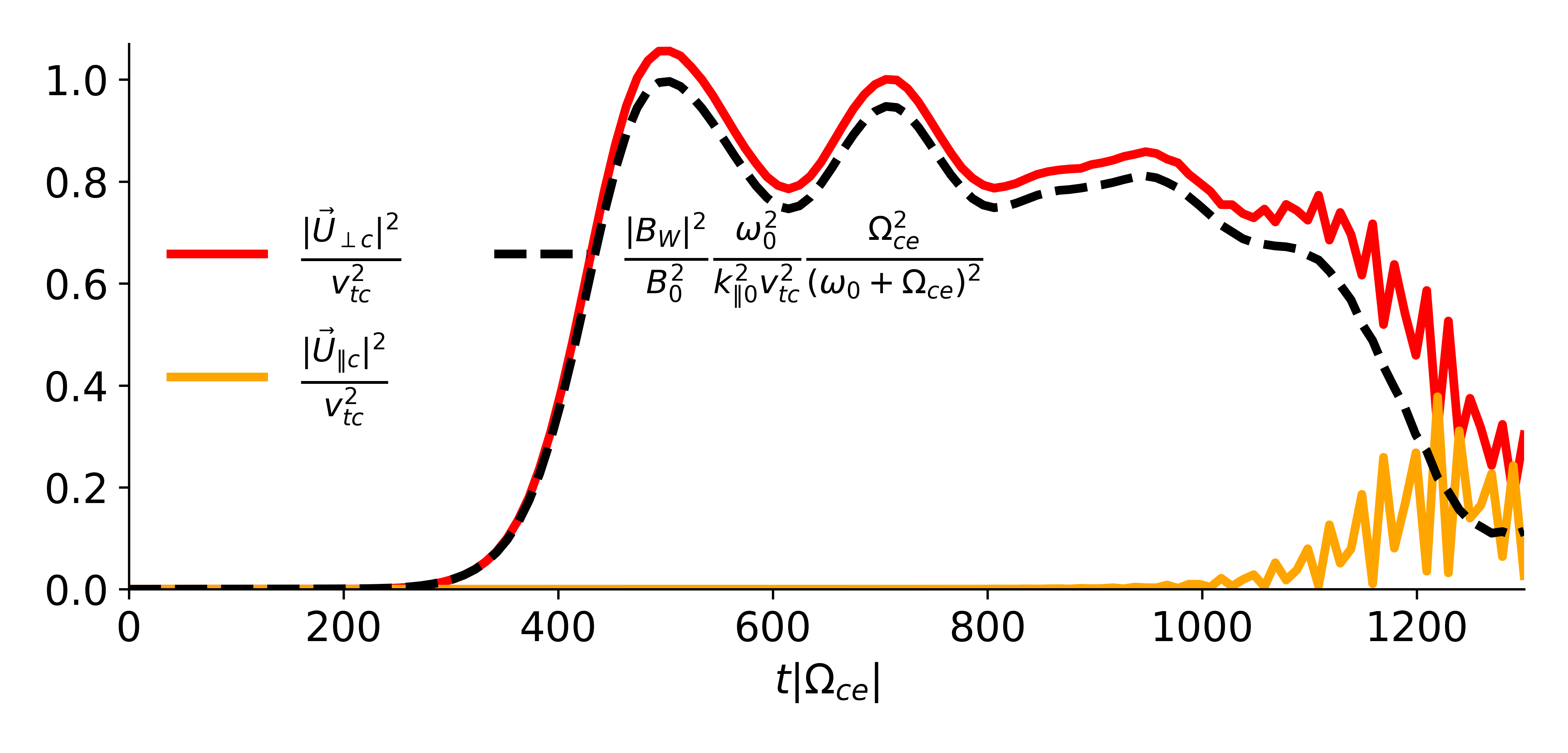}
    \caption{Primary wave normalized magnetic energy density $|B_{W}|^2$ and the normalized cold electron average bulk velocity $\vec{U}_{c}(t) \coloneqq \int \mathrm{d}^3 v \int \mathrm{d}^3 x \vec{v} f_{c}(\vec{x}, \vec{v}, t)$ from the 2D3V PIC simulation. The primary wave magnetic energy density is proportional to the perpendicular cold electron drift, which validates Eq.~\eqref{Vds}, until the onset of electrostatic turbulence, which transfers energy to cold electrons.}
    \label{fig:PIC-drift-magnitude}
\end{figure}

\begin{figure}
    \centering
        \begin{subfigure}{0.495\textwidth}
        \caption{Oblique frequency and growth rate}
        \includegraphics[width=\textwidth]{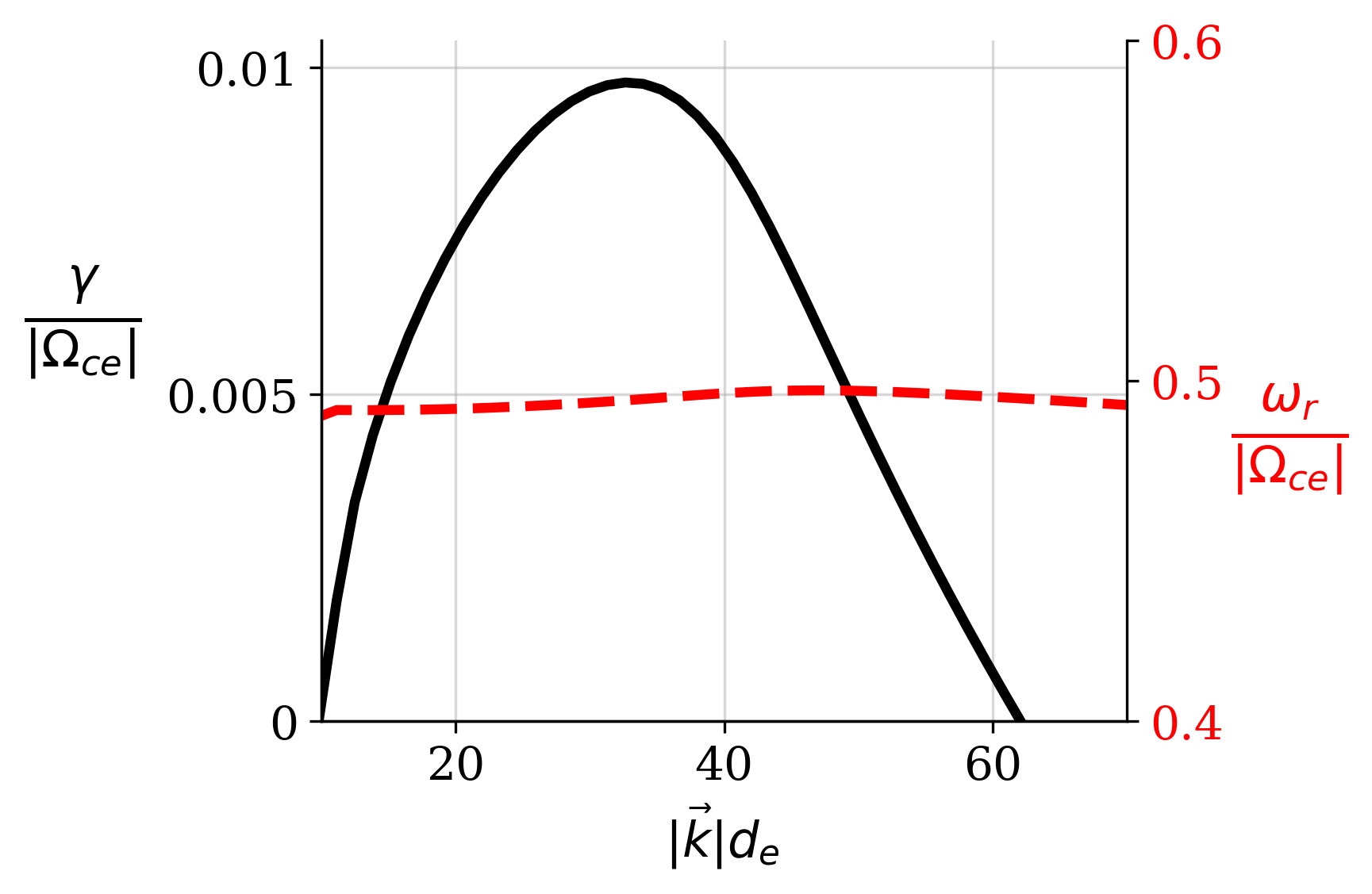}
        \label{fig:dispersion-oblique-low-anisotropy}
    \end{subfigure}
    \begin{subfigure}{0.495\textwidth}
        \caption{Perpendicular frequency and growth rate}
        \includegraphics[width=\textwidth]{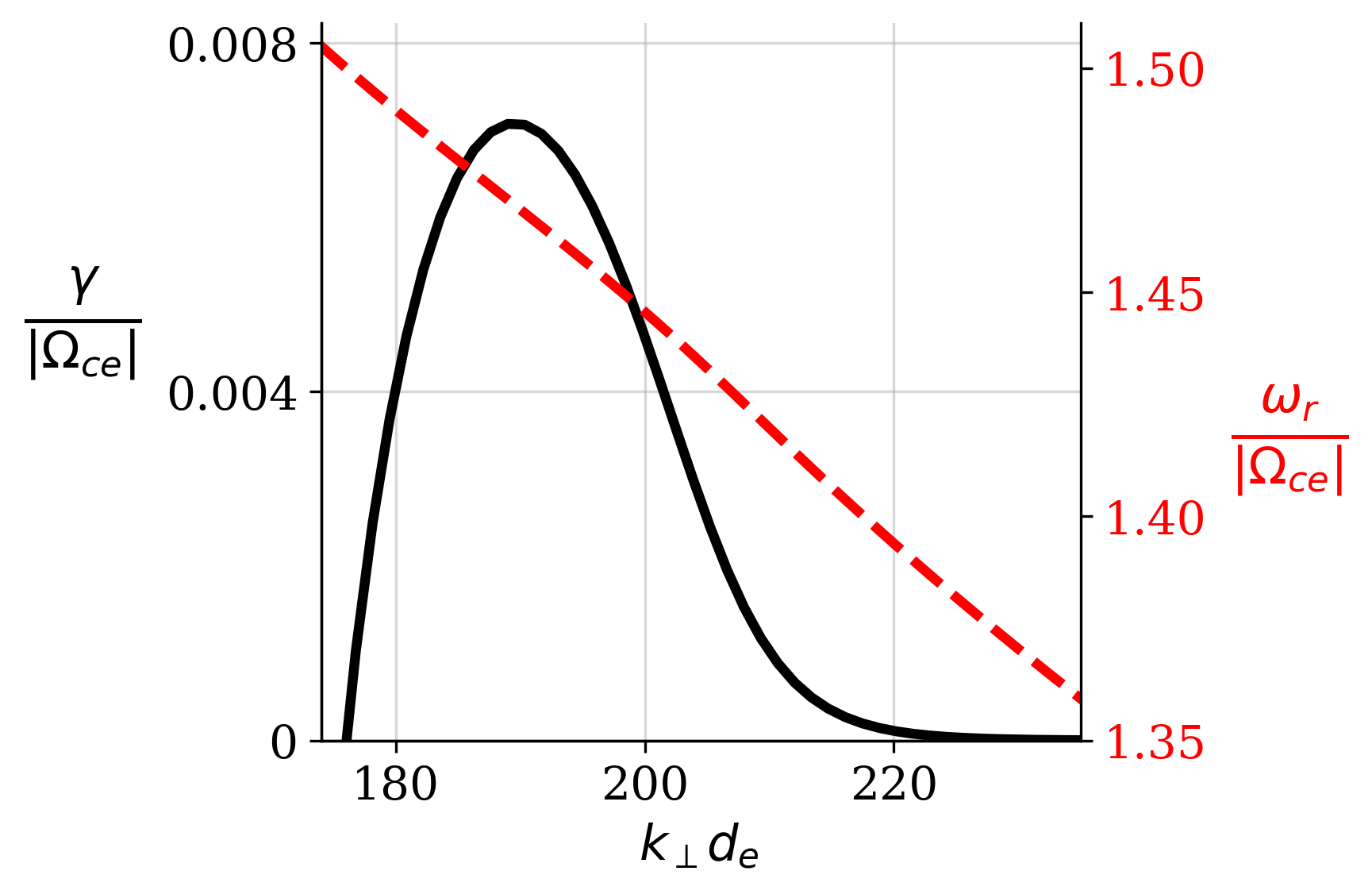}
        \label{fig:dispersion-perpendicular-low-anisotropy}
    \end{subfigure}
    \begin{subfigure}{0.495\textwidth}
        \caption{PIC spectrum at $t|\Omega_{ce}|=850$}
        \includegraphics[width=\textwidth]{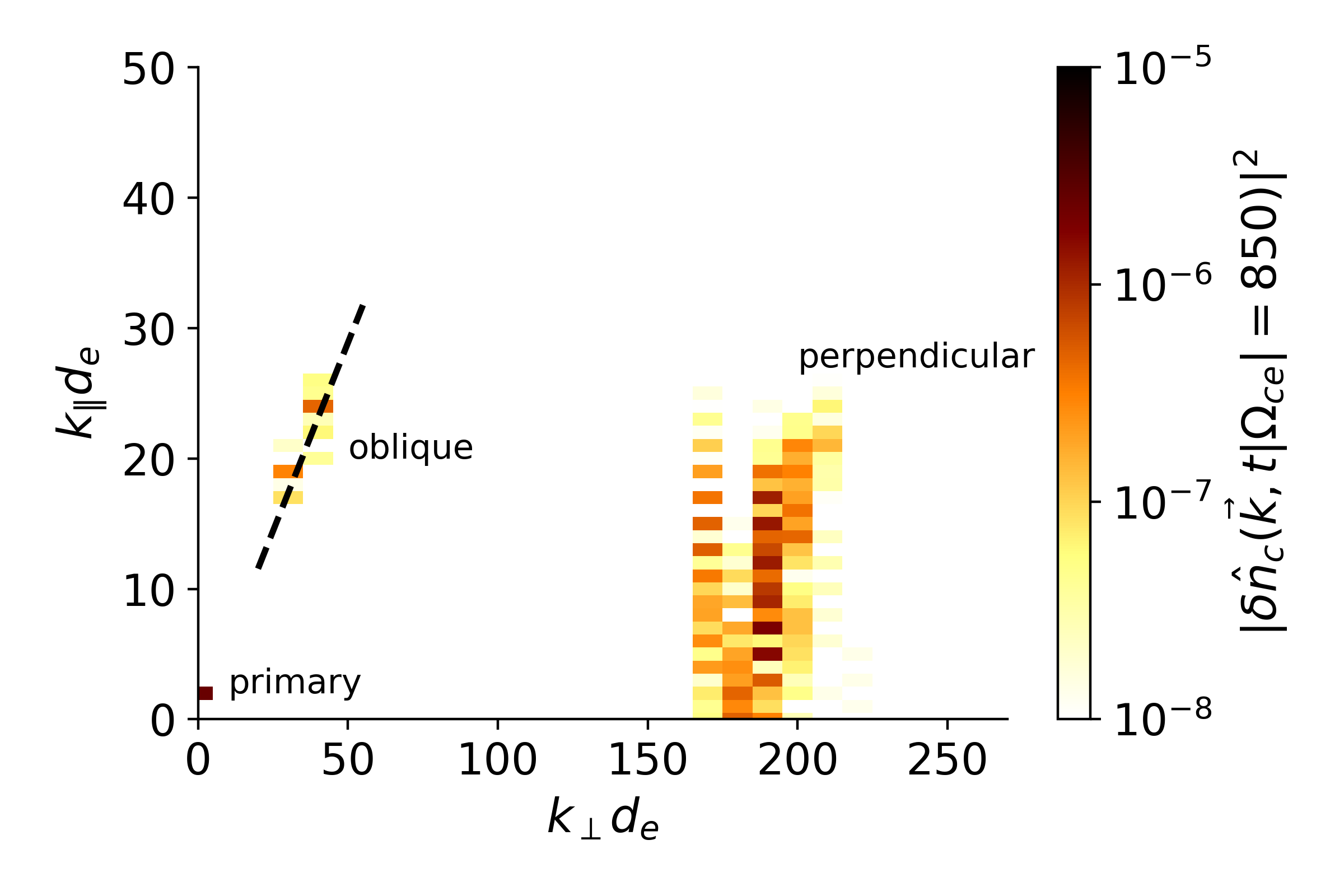}
        \label{fig:FFT-nc-low-anisotropy-850}
    \end{subfigure}
    \hspace{-12pt}
    \begin{subfigure}{0.495\textwidth}
        \caption{PIC spectrum at $t|\Omega_{ce}| = 1000$}
        \includegraphics[width=\textwidth]{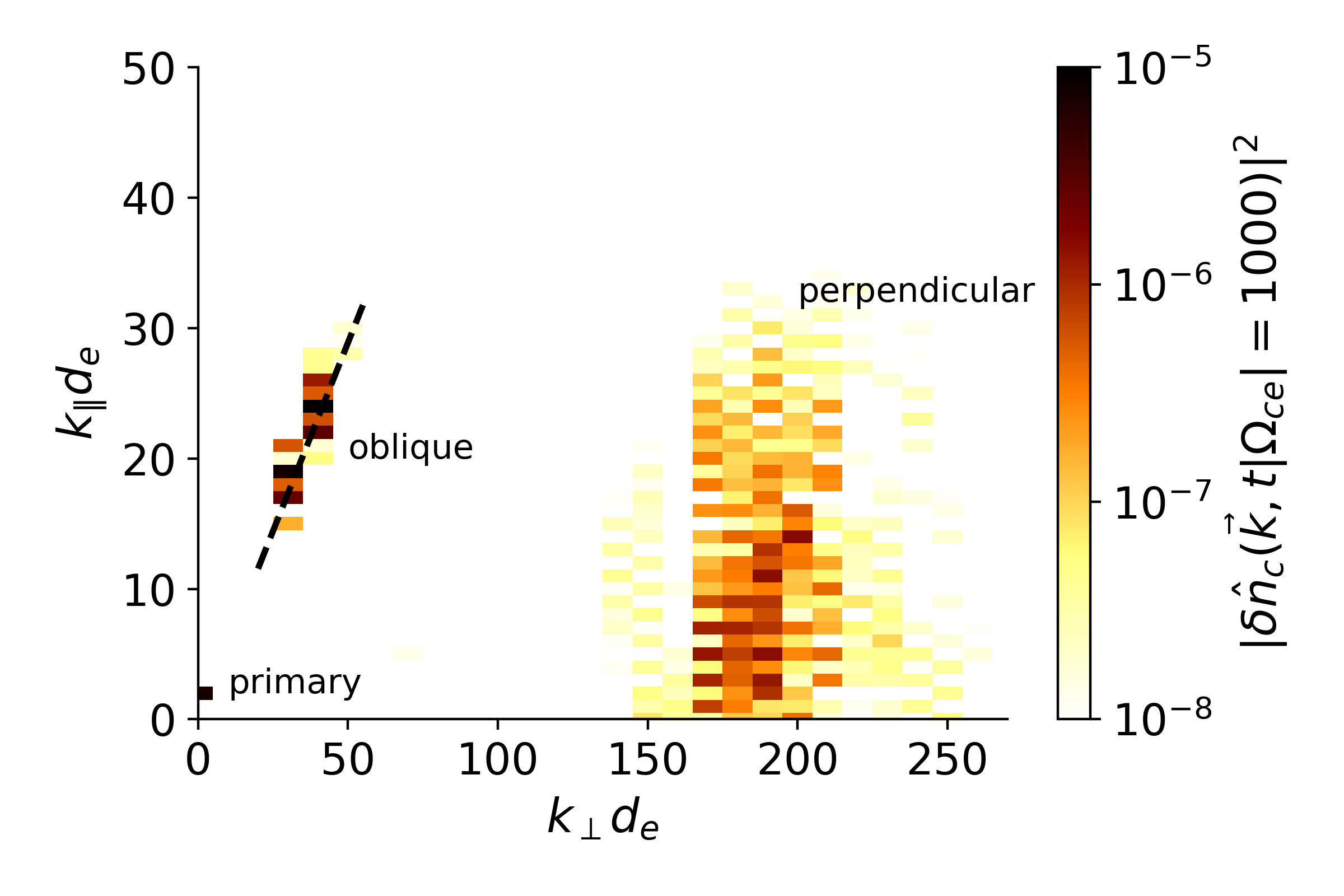}
        \label{fig:FFT-nc-low-anisotropy-1000}
    \end{subfigure}
    \caption{Frequency and growth rate for the PIC simulation parameters, shown for (a)~oblique and (b)~perpendicular modes. The oblique whistler has a frequency close to the primary field-aligned whistler, $\omega_{0} \approx 0.5 |\Omega_{ce}|$, near the resonance cone, while the perpendicular modes correspond to short-wavelength perturbations. Subfigures~(c) and~(d) show the cold electron density spectra in $(k_{\perp}, k_{\|})$ from the PIC simulation at $t|\Omega_{ce}| = 850$ and $t|\Omega_{ce}| = 1000$.}
    \label{fig:fft_and_dispersion_low_anisotropy}
\end{figure}

The magnetic energy density of the primary wave and cold electron bulk velocity from the PIC simulation are shown in Figure~\ref{fig:PIC-drift-magnitude}. 
The primary wave magnetic energy density grows until it reaches saturation and starts to damp at $t|\Omega_{ce}| \approx 700$ due to the onset of the secondary instabilities. 
The cold electron perpendicular bulk velocity squared is proportional to the magnetic energy density of the primary wave until $t|\Omega_{ce}| \approx 1000$ (as estimated in Eq.~\eqref{Vds}), where electrostatic turbulence transfers momentum to the cold electrons. 
The bulk velocity also grows in the parallel direction at $t|\Omega_{ce}| \approx 1000$ from the oblique electrostatic wave. 
%
The frequency and growth rate estimated at $t|\Omega_{ce}| \approx 700$ from Eqns.~\eqref{approx-dispersion-relation}--\eqref{dispersion_relation_ecdi} are shown in Figures~\ref{fig:dispersion-oblique-low-anisotropy} and \ref{fig:dispersion-perpendicular-low-anisotropy}. 
The oblique whistler wave has a frequency similar to the primary field-aligned whistler wave $\omega_{0} \approx 0.5 |\Omega_{ce}|$ with a wave normal angle near the resonance cone $\theta_{k} = 60^{\circ}$. The perpendicular modes are short wavelength $k_{\perp} d_{e} \approx 200$ and have a comparable growth rate in comparison to the oblique modes. 
Figures~\ref{fig:FFT-nc-low-anisotropy-850} and \ref{fig:FFT-nc-low-anisotropy-1000} show the wavenumber spectrum of the cold electron density fluctuations from the PIC simulation after the whistler wave saturates at times $t|\Omega_{ce}| = 850$ and $t|\Omega_{ce}|= 1000$, respectively. 
The black dashed line marks the resonance cone. 
The wavenumbers excited in the PIC simulation resemble the linear dispersion relation range of unstable wavenumbers shown in Figures~\ref{fig:dispersion-oblique-low-anisotropy} and~\ref{fig:dispersion-perpendicular-low-anisotropy}. 

\begin{figure}
    \centering
    \begin{subfigure}{0.49\textwidth}
        \caption{Cold electron heating}
        \includegraphics[width=\textwidth]{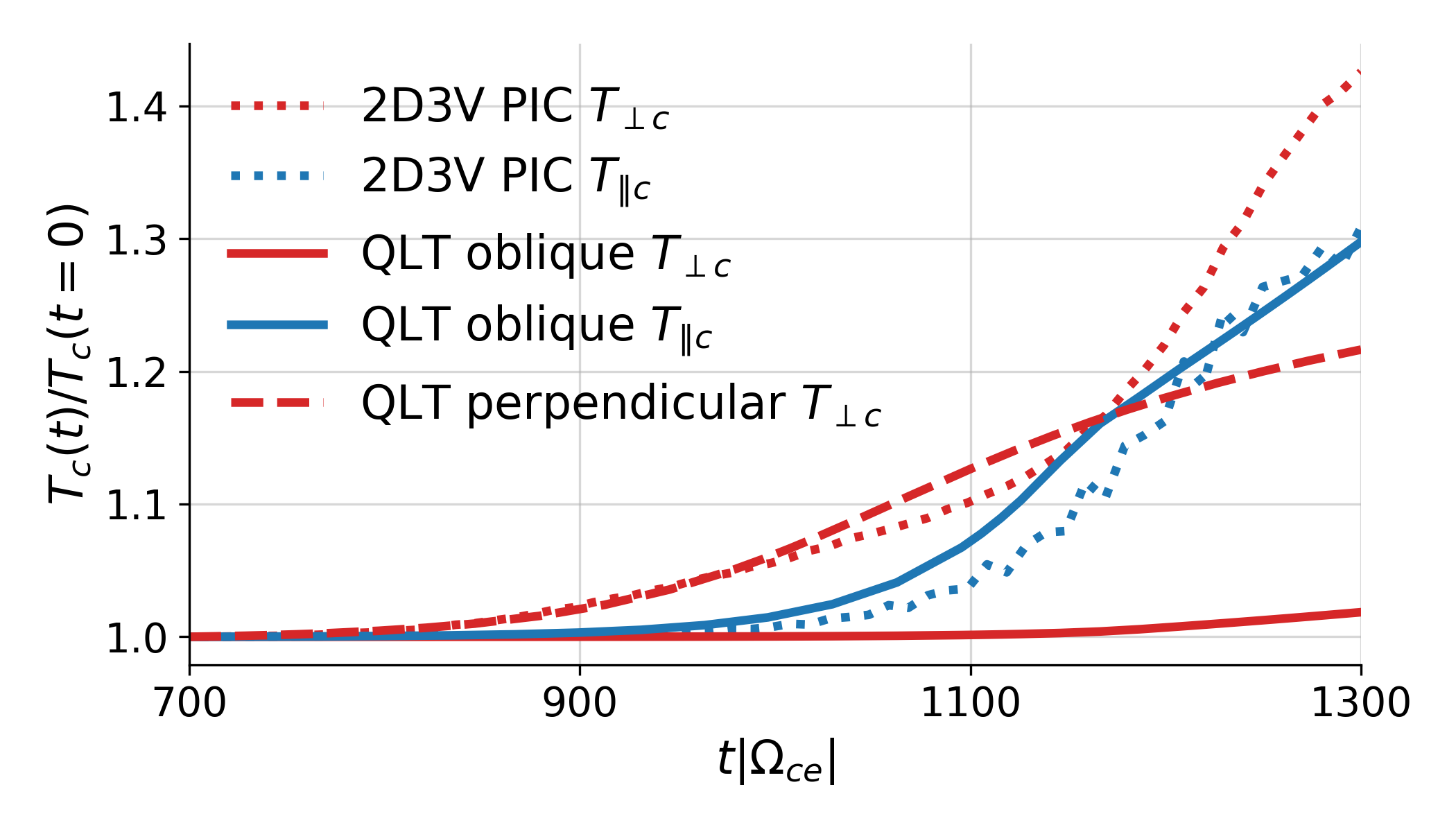}
    \end{subfigure}
    \hspace{-12pt}
    \begin{subfigure}{0.49\textwidth}
        \caption{Primary whistler wave damping}
        \includegraphics[width=\textwidth]{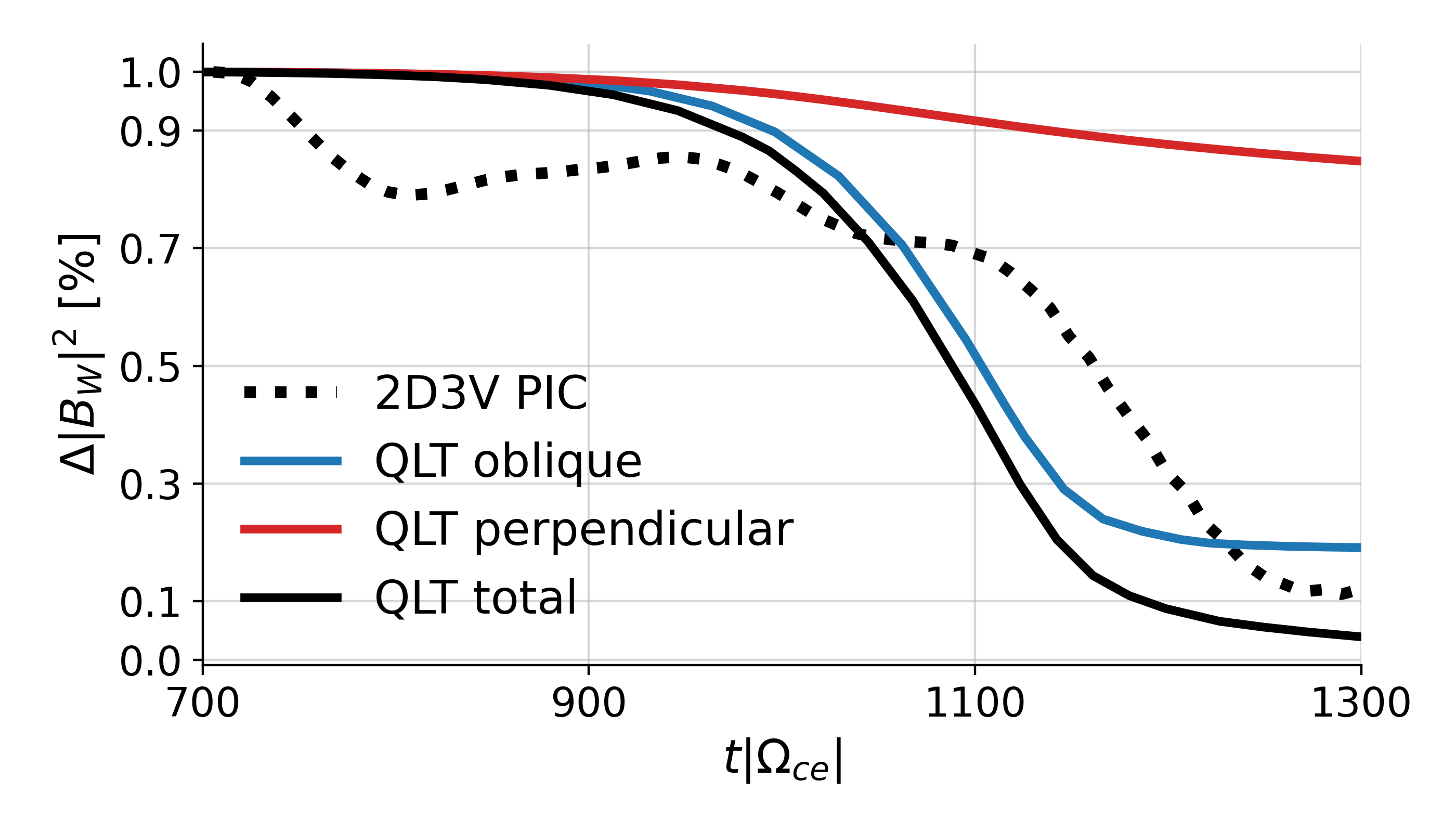}
    \end{subfigure}
        \caption{Comparison of PIC and QLT results: (a) cold electron heating and (b) primary whistler wave damping, i.e. $\Delta |B_{W}|^2 = |B_{W}(t)|^2/|B_{W}(t|\Omega_{ce}|=700)|^2$. The moment-based QLT results qualitatively resemble the PIC results, capturing the overall trends accurately; however, QLT slightly underestimates the cold electron heating (resonant interactions) in the perpendicular direction and slightly overestimates the whistler wave damping due to overestimation of non-resonant interactions. }
    \label{fig:dB_dT_PIC_vs_QLT_low_anisotropy}
\end{figure}

\begin{figure}
    \centering
    \begin{subfigure}{0.45\textwidth}
        \caption{$t|\Omega_{ce}| = 800$}
        \includegraphics[width=\textwidth]{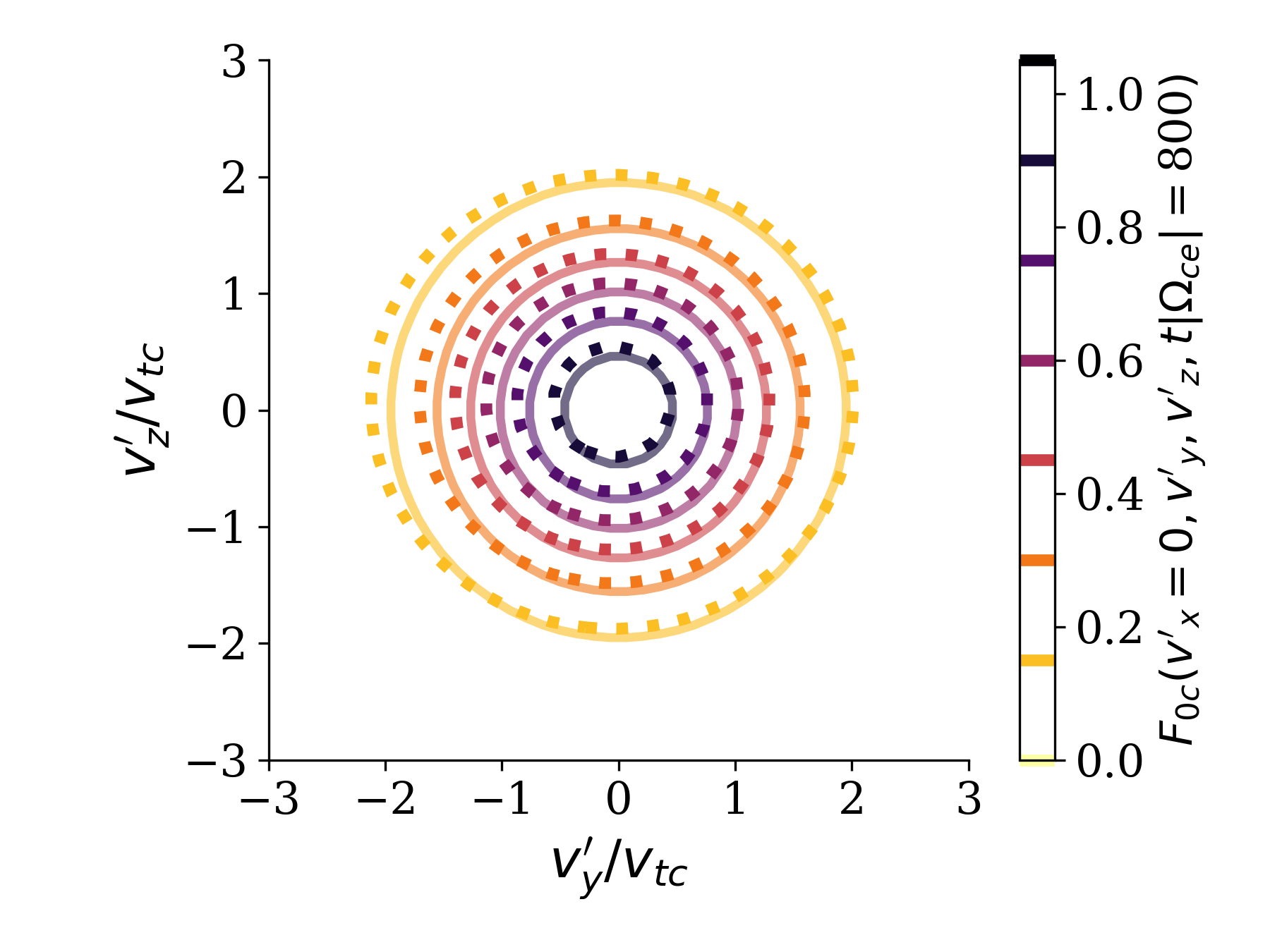}
    \end{subfigure}
    \begin{subfigure}{0.45\textwidth}
        \caption{$t|\Omega_{ce}| = 1300$}
        \includegraphics[width=\textwidth]{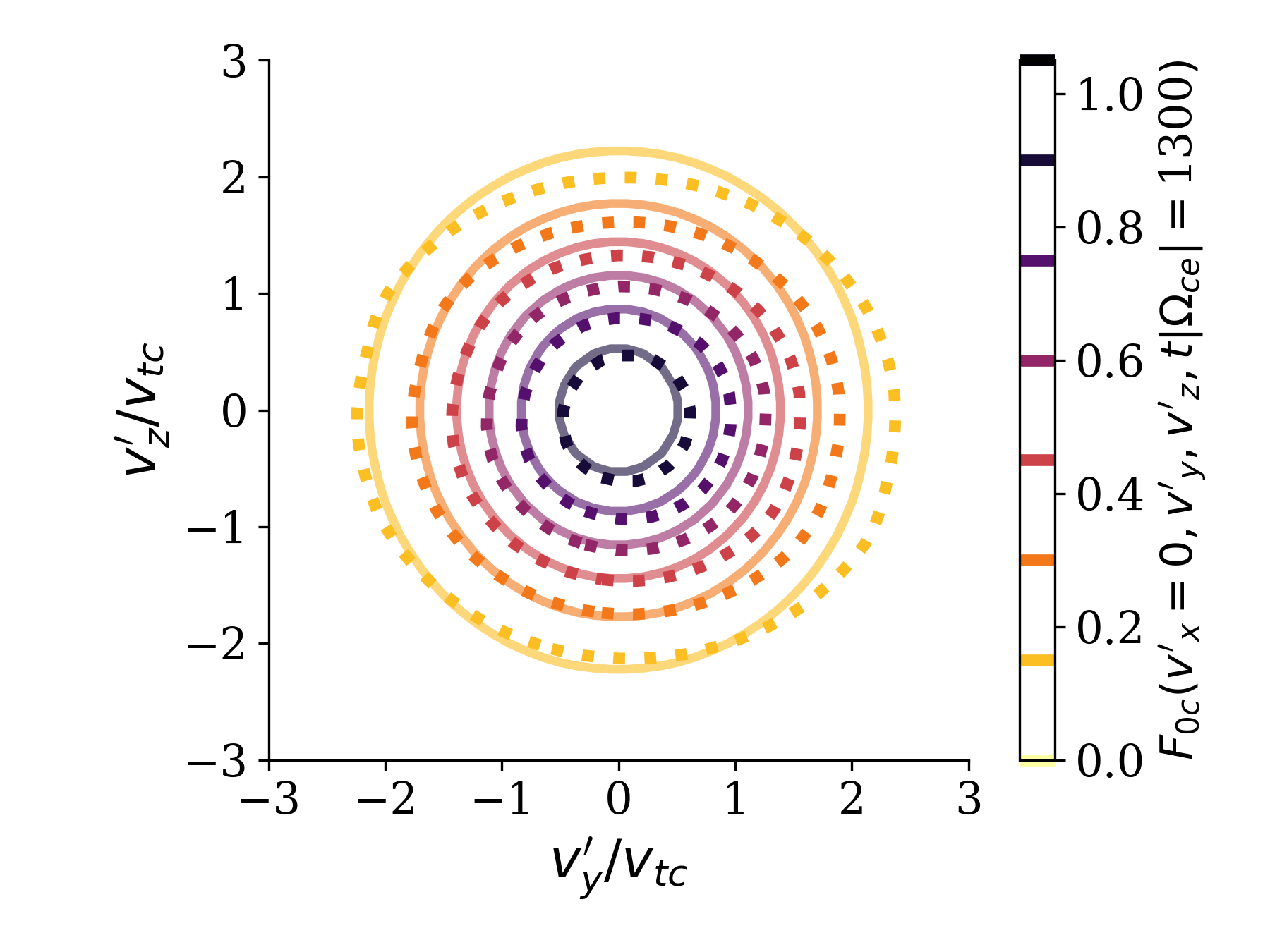}
    \end{subfigure}
    \caption{The (normalized) cold electron equilibrium distribution function in the co-drifting frame $F_{0c}(v'_{x} = 0,  v_{y}', v_{z}', t)$  from the PIC simulation (dotted) and QLT simulations (solid) at times: (a) $t|\Omega_{ce}| = 800$  and (b) $t|\Omega_{ce}| = 1300$. Overall, the QLT and PIC results show good agreement, and the distribution function remains approximately bi-Maxwellian throughout the simulation, validating a fundamental assumption of the moment-based QLT approach.}
    \label{fig:pdf_PIC_vs_QLT_low_anisotropy}
\end{figure}

The moment-based QLT and PIC cold electron heating and damping of the primary whistler wave are shown in Figure~\ref{fig:dB_dT_PIC_vs_QLT_low_anisotropy}. 
The PIC simulation demonstrates that the primary wave magnetic energy density $|B_{W}|^2$ decreases by $90\%$ between the times of $t|\Omega_{ce}| = 700$ and $t|\Omega_{ce}| = 1300$.
The moment-based QLT simulations result in $80\%$ damping due to the excitation of oblique electrostatic whistlers and $15\%$ damping due to the excitation of ECDI-like perpendicular modes. If we sum the two damping percentages, then QLT predicts a total of $95\%$ damping, which reproduces quite well the PIC results. 
In this case, the oblique instability contributed five times more than the perpendicular instability in damping the primary whistler wave. 
The secondary modes heat the cold electron population, where PIC predicts $\times 1.4$ heating in the perpendicular direction and $\times 1.3$ in the parallel direction by $t|\Omega_{ce}| \approx 1300$. The QLT heating is slightly lower for perpendicular heating at approximately $\times 1.2$ and in agreement in the parallel direction with $\times 1.3$. 
The moment-based QLT predicts a slight overestimation of primary wave damping and a slight underestimation of cold electron heating, primarily due to an overestimation of non-resonant interactions. 
This is because the primary whistler wave dampens due to the increase in cold electron energy and electrostatic energy, see Eq.~\eqref{dBdt}.
The QLT results (not shown here) and the PIC results (see Figure~\ref{fig:full_energy} in~\ref{sec:energy_conservation}) show that the fluctuations in the electrostatic energy do not significantly impact the total energy. 
Therefore, if QLT overestimates the primary wave damping and underestimates the cold electron heating, then it must be overestimating the cold electron kinetic energy gain (non-resonant interactions). 
Figure~\ref{fig:pdf_PIC_vs_QLT_low_anisotropy} shows the cold electron equilibrium distribution function at $t|\Omega_{ce}| \in \{800, 1300\}$ in the co-drifting reference frame, as obtained from QLT and PIC simulations. 
The PIC simulation shows that the cold electron equilibrium remains approximately bi-Maxwellian throughout this simulation, confirming a key assumption in the moment-based QLT approach. 
We note that the recent study by~\citet{roytershteyn_2024_frontiers} shows that there exists a plateau in the cold electron distribution for the high-anisotropy cases (corresponding to higher amplitude primary waves), yet such non-Maxwellian features are moderate for low-anisotropy cases (such as the one presented here). 
In general, the good agreement between the moment-based QLT developed here and the PIC simulations gives us confidence that the theory captures the most important features of the saturation of the secondary instabilities.

\section{Secondary instabilities sensitivity to cold electron and primary whistler wave properties}\label{sec:QLT_parameteric_scans}
Earth's inner magnetosphere contains electron populations in the sub-eV and eV energy range originating from the ionosphere~\cite{delaznno_2021_cold_impact}. 
However, measuring cold electrons in the magnetosphere is challenging due to contamination from spacecraft-emitted photoelectrons and secondary electrons, which fall within the same energy range; therefore, the properties of these particle populations are not well characterized.
We therefore study the impact of the cold electron properties along with the primary whistler wave properties on the onset of the secondary instability via linear theory in section~\ref{sec:parametric_linear} and on the 
saturation properties via moment-based QLT in section~\ref{sec:parametric_QLT}. 
In the following parametric studies, we vary one parameter at a time while fixing all other parameters to match the conditions of the PIC simulation presented in section~\ref{sec:QLT_vs_PIC}, i.e. $n_{c} = 0.8 n_{e}$, $\omega_{0} = 0.5 |\Omega_{ce}|$, 
$\omega_{pe} = 4 |\Omega_{ce}|$, 
$\alpha_{\perp c} = \alpha_{\parallel c} = 0.0079 d_{e}|\Omega_{ce}|$, 
$v_{ti} = v_{tc}/\sqrt{1836}$, and
$|\vec{V}_{Dc}| = 0.92 v_{tc}$. By varying the primary wave frequency $\omega_{0}$, the drift amplitude is adjusted according to $|\vec{V}_{Dc}|/v_{tc} = 0.46|\Omega_{ce}|/|\omega_{0}-\Omega_{ce}|$ as given in Eq.~\eqref{Vds}.

\subsection{Parametric study via linear theory}\label{sec:parametric_linear}
\begin{figure}
    \centering
    \begin{subfigure}{0.45\textwidth}
        \centering
        \caption{Perpendicular growth rate at $\theta_{k} = 90^{\circ}$ \\ Eq.~\eqref{f-theta-equation} $\Rightarrow$ full dispersion relation}
        \includegraphics[width=\textwidth]{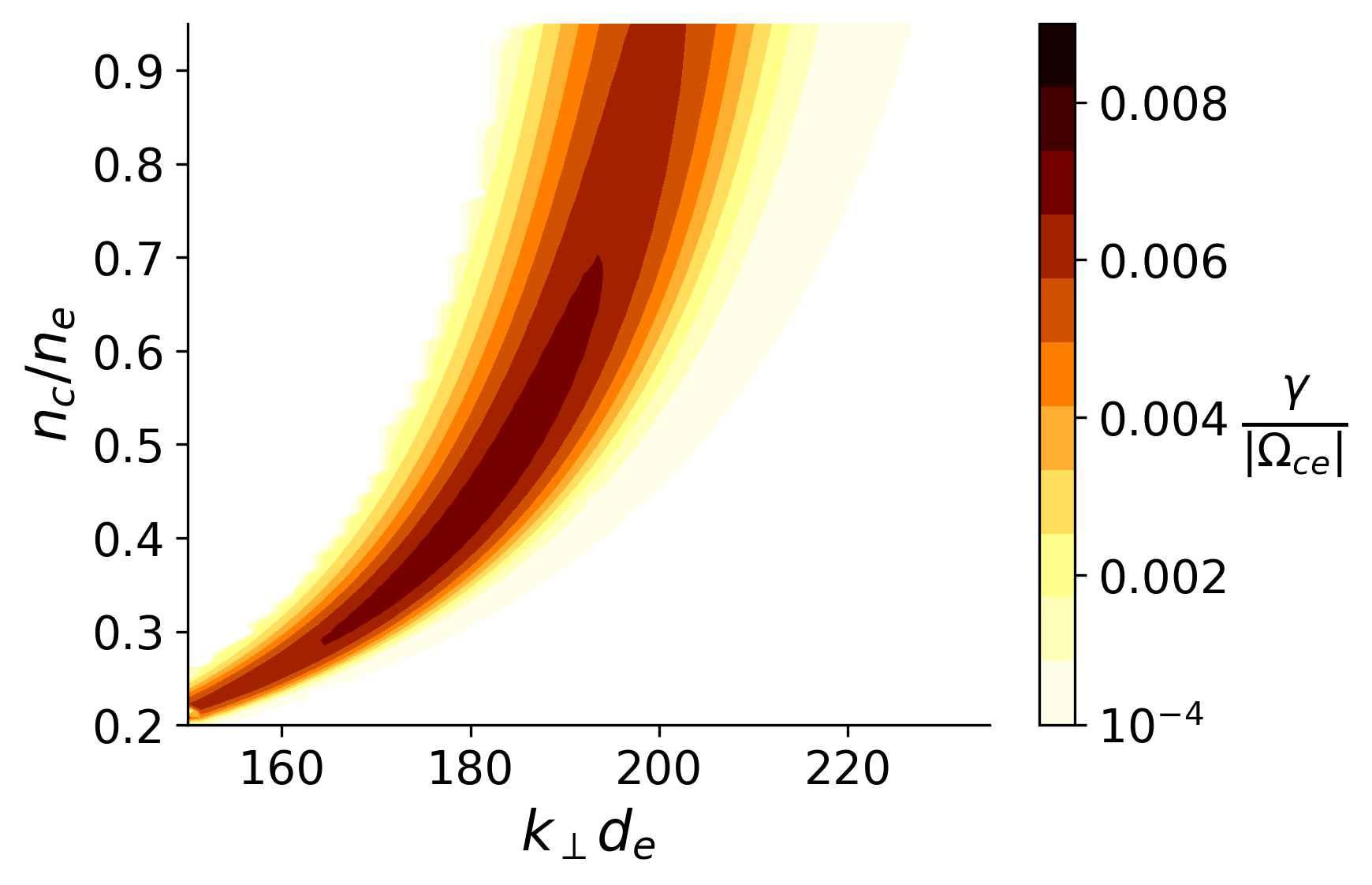}
        \label{fig:fig5-a}
    \end{subfigure}
        \begin{subfigure}{0.45\textwidth}
        \centering
        \caption{Oblique growth rate at $\theta_{k}=60^{\circ}$\\ Eq.~\eqref{f-theta-equation} $\Rightarrow$ full dispersion relation}
        \includegraphics[width=\textwidth]{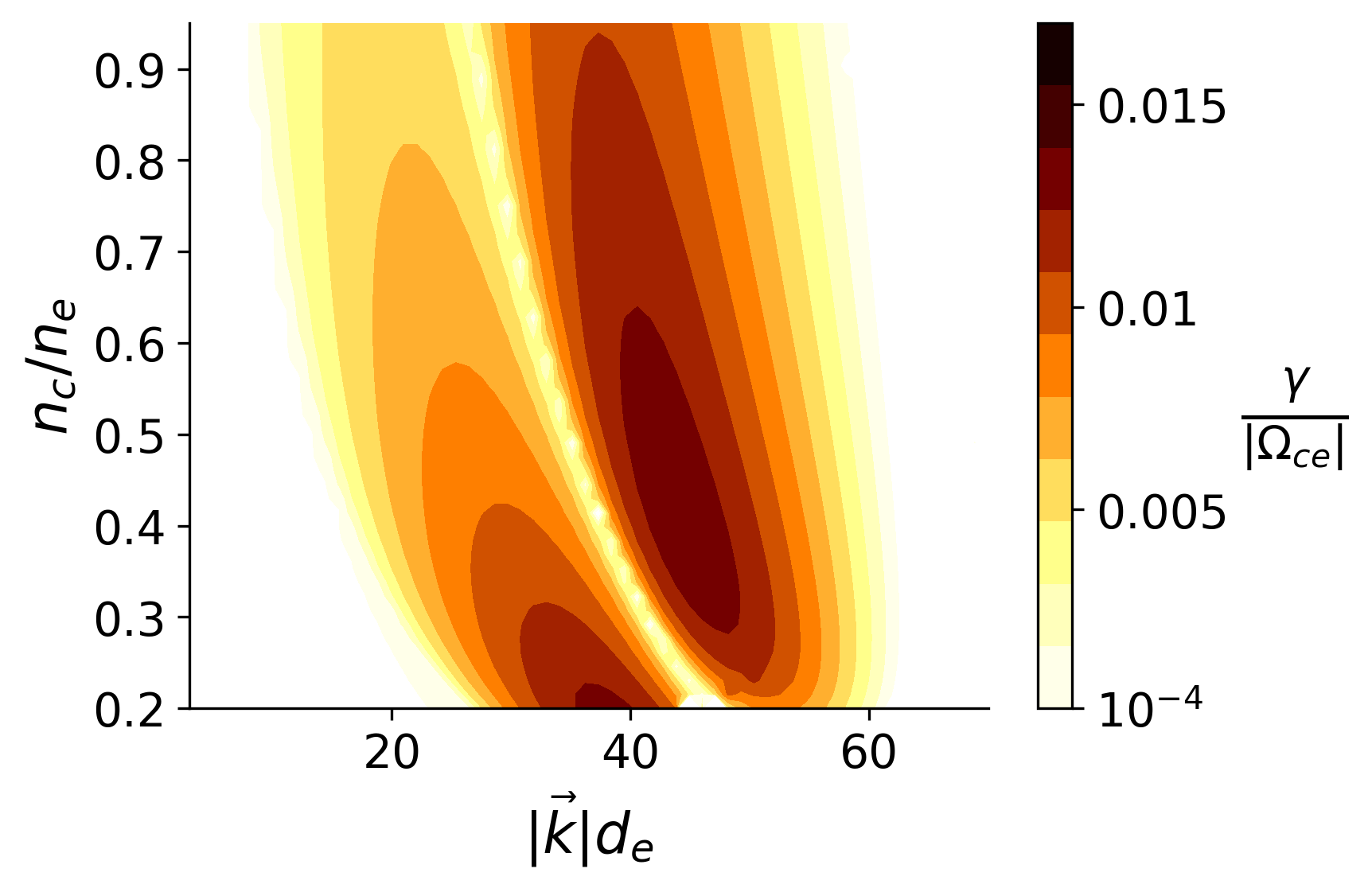}
        \label{fig:fig5-b}
    \end{subfigure}
    \begin{subfigure}{0.45\textwidth}
        \centering
        \caption{Perpendicular growth rate at $\theta_{k} = 90^{\circ}$  \\ Eq.~\eqref{dispersion_relation_ecdi} $\Rightarrow$ approximation}
        \includegraphics[width=\textwidth]{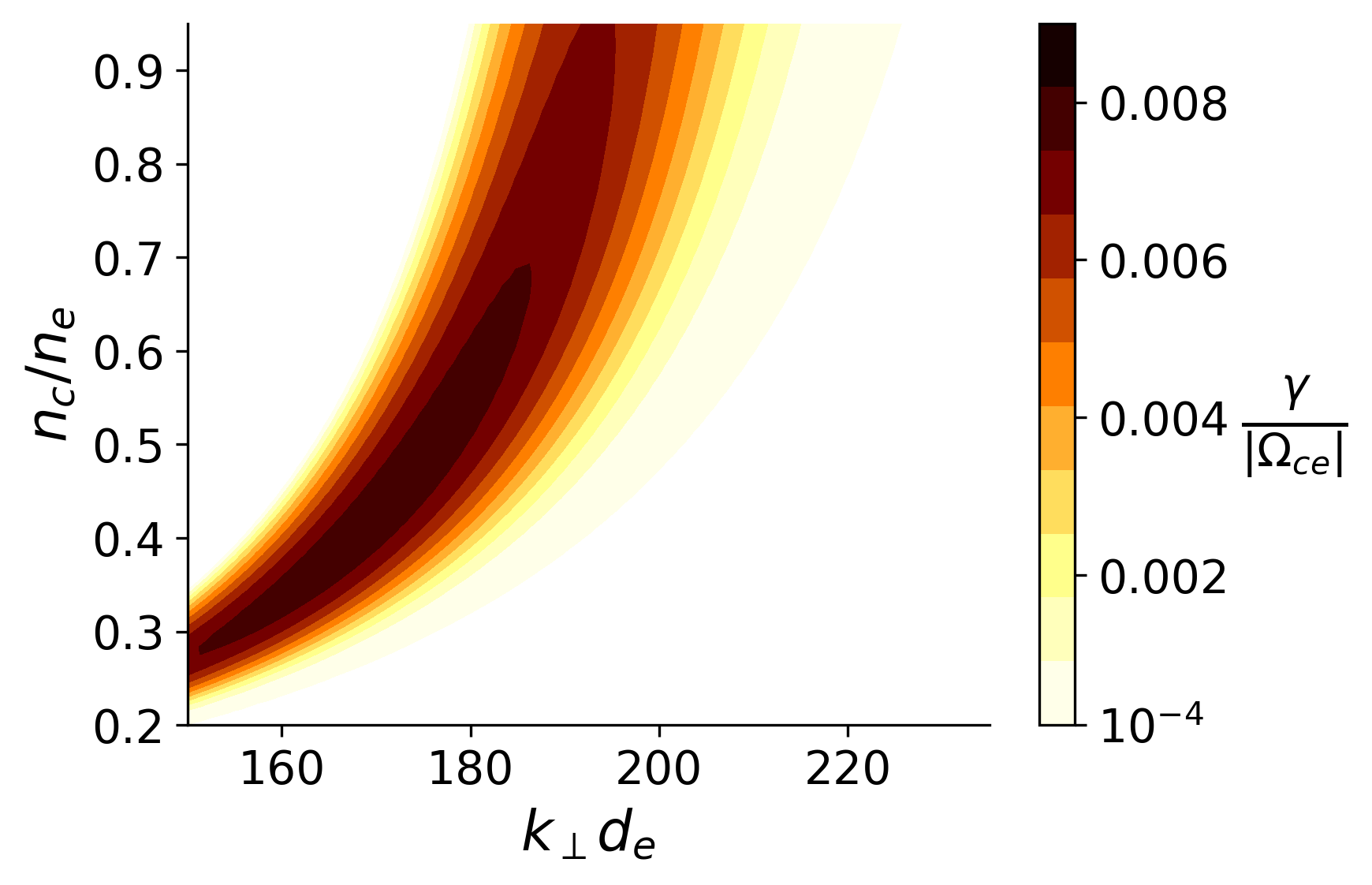}
        \label{fig:fig5-c}
    \end{subfigure}
        \begin{subfigure}{0.45\textwidth}
        \centering
        \caption{Oblique growth rate at $\theta_{k}=60^{\circ}$ \\ Eq.~\eqref{approx-dispersion-relation} $\Rightarrow$ approximation}
        \includegraphics[width=\textwidth]{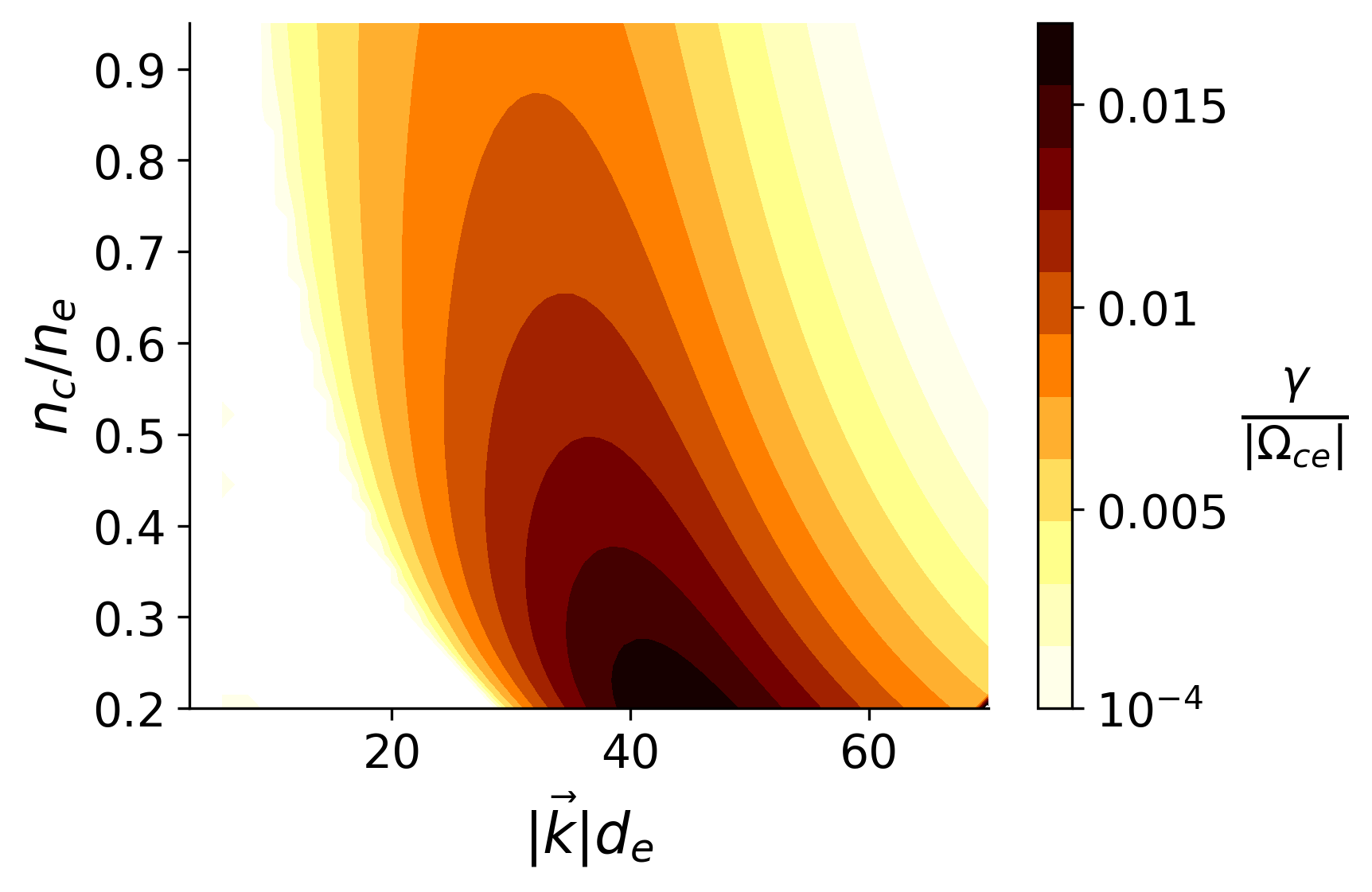}
        \label{fig:fig5-d}
    \end{subfigure}
    \caption{Growth rate of the secondary instabilities as a function of cold electron density $n_{c}/n_{e}$ and wavenumber amplitude $|\vec{k}| d_e$. 
    Subfigures~(a/b) show results from the full dispersion relation in Eq.~\eqref{f-theta-equation}, while subfigures~(c/d) show results from the corresponding approximations in Eq.~\eqref{dispersion_relation_ecdi} and Eq.~\eqref{approx-dispersion-relation}. 
    The approximate solutions show good qualitative agreement with the full dispersion relation.
    The results in subfigures~(a/b) indicate that the growth rate is the largest when $0.3 \lesssim n_{c}/n_{e} \lesssim 0.6$.}
    \label{fig:growth-rate-denisty-scan}
\end{figure}

We begin by studying the impact of cold electron density on the secondary instabilities using linear theory. 
Figure~\ref{fig:growth-rate-denisty-scan} shows the growth rates of the instabilities as a function of the ratio of cold electron density to total electron density $n_{c}/n_{e}$ and wavenumber amplitude $|\vec{k}| d_e$ by solving Eq.~\eqref{approx-dispersion-relation} for oblique modes and Eq.~\eqref{dispersion_relation_ecdi} for perpendicular modes along with the full dispersion relation in Eq.~\eqref{f-theta-equation}. 
From Figures~\ref{fig:fig5-a}--\ref{fig:fig5-b}, the instabilities have largest growth rates when $0.3 \lesssim n_{c}/n_{e} \lesssim 0.6$.
The simplified dispersion relation estimates in Eqns.~\eqref{simplified-growth-rate-oblique-whistler} and~\eqref{simplified-growth-rate-ECDI} indicate that the growth rate increases with decreasing $n_{c}/n_{e}$, i.e. $\gamma \propto \left[n_{c}/n_{e}\right]^{-1}$, yet the numerical results of the full dispersion relation~\eqref{f-theta-equation} show a more complex dependence. 
For the perpendicular instability, the range of unstable modes also becomes broader as $n_{c}/n_{e}$ increases.
Moreover, as $n_{c}/n_{e}$ increases, the perpendicular instability moves towards larger wavenumbers~$k_{\perp}$ and the oblique whistler modes move towards smaller wavenumbers~$|\vec{k}|$. 
%
%
Not shown here, the maximum growth rate for the oblique instability is along the resonance cone for the considered range of $n_{c}/n_{e}$, and the excited modes span a small range of wave normal angles with $\theta_{k} \in [57^{\circ}, 60^{\circ}]$ for the considered range of densities $n_{c}/n_{e} \in [0.2, 0.95]$. 
Similarly, for the same density range, the quasi-perpendicular modes span a small range of wave normal angles with $\theta_{k} \in [84^{\circ}, 90^{\circ}]$.
Figure~\ref{fig:growth-rate-denisty-scan} also compares the full dispersion relation in Eq.~\eqref{f-theta-equation} with $N=5$ sidebands on each side of a given frequency and its approximations in Eq.~\eqref{approx-dispersion-relation} and~\eqref{dispersion_relation_ecdi}, which assumes the electron Bernstein mode intersects only with one driver harmonic and ignores couplings to other modes.
We set the oblique mode approximation~\eqref{approx-dispersion-relation} with $m^{*} = -1$ and the perpendicular mode approximation~\eqref{dispersion_relation_ecdi} with $m^{*} = -3$.
Neglecting sidebands can result in a modest overestimation of the growth rate, particularly at low cold plasma densities. 
Moreover, the full dispersion relation produces oblique modes with two peaks in wavenumber space, as shown in Figure~\ref{fig:fig5-b}, whereas the approximation yields only a single peak, as shown in Figure~\ref{fig:fig5-d}.
Nevertheless, the approximate solutions in Eqns.~\eqref{approx-dispersion-relation} and~\eqref{dispersion_relation_ecdi} show good qualitative agreement with the full dispersion relation.

\begin{figure}
    \centering
    \includegraphics[width=0.5\linewidth]{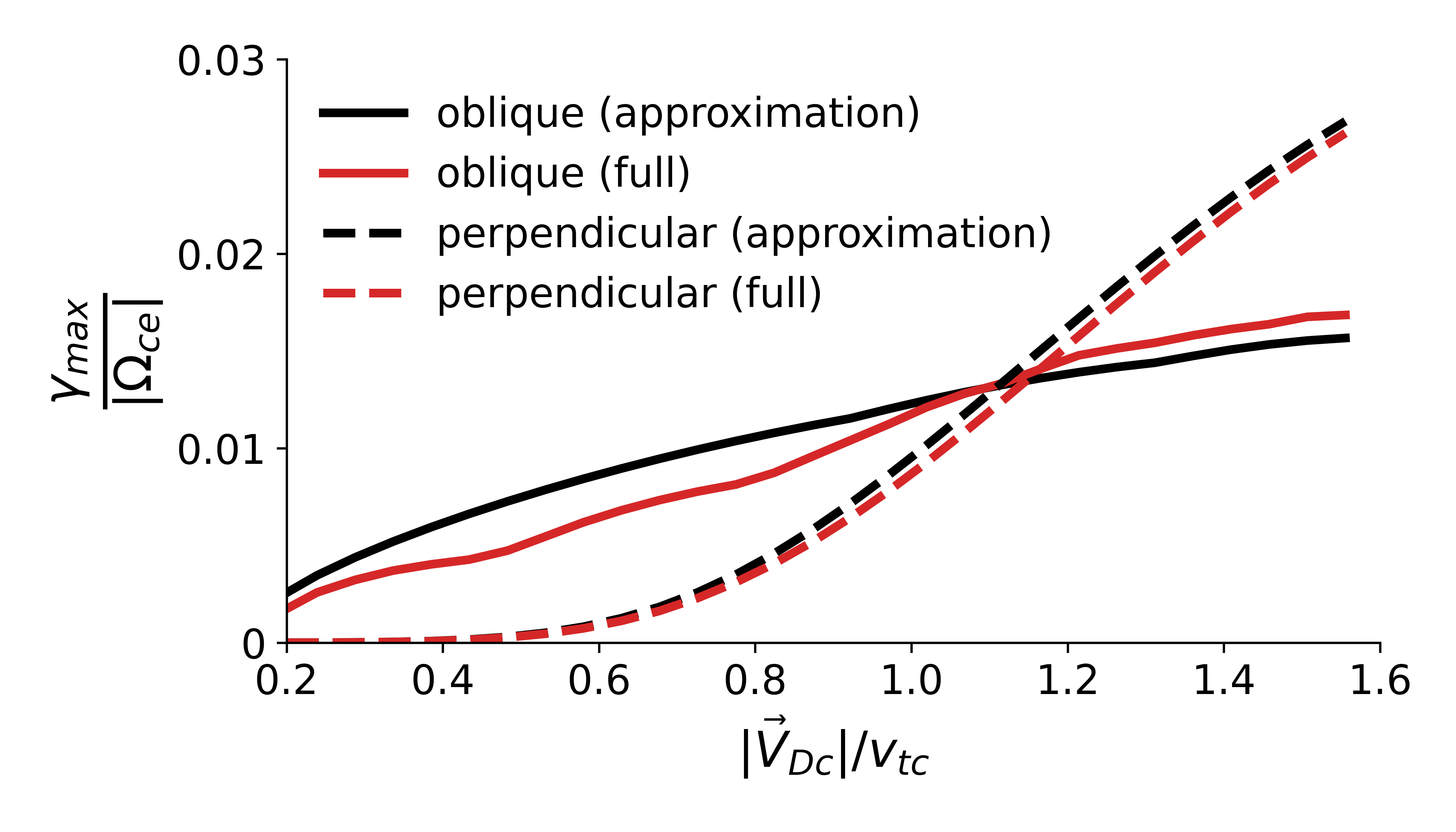}
    \caption{Maximum growth rate over wave normal angles as a function of the ratio of induced cold-electron drift to their thermal velocity $|\vec{V}_{Dc}|/v_{tc}$. Two distinct instability branches are compared: oblique whistler (dashed lines) and ECDI-like perpendicular (solid lines) modes. The results suggest a relatively low threshold for the onset of secondary instabilities, with oblique modes dominating at moderate drifts $0.2 \leq |\vec{V}_{Dc}|/v_{tc}\leq 1$ and perpendicular modes prevailing at higher drifts $|\vec{V}_{Dc}|/v_{tc} \geq 1$. }
    \label{fig:drift_dependence_growth_rate}
\end{figure}

The key parameter controlling the secondary instabilities is the ratio of induced cold-electron drift to their thermal velocity $|\vec{V}_{Dc}|/v_{tc}$. 
Figure~\ref{fig:drift_dependence_growth_rate} demonstrates the perpendicular modes and oblique electrostatic whistler maximum growth rate over wave normal angle ($10^{\circ}$ range from purely perpendicular and the corresponding resonance cone angle) as a function of $|\vec{V}_{Dc}|/v_{tc}$. 
The results are obtained by solving both the approximate dispersion relations in Eqns.~\eqref{approx-dispersion-relation} and~\eqref{dispersion_relation_ecdi} and the full dispersion relation in Eq.~\eqref{f-theta-equation}, which result in comparable growth rates and similar dependence on $|\vec{V}_{Dc}|/v_{tc}$. 
The oblique whistler waves have larger maximum growth rates in comparison to perpendicular modes when $0.2 \leq |\vec{V}_{Dc}|/v_{tc} \leq 1$. However, for larger drift values $|\vec{V}_{Dc}|/v_{tc} \geq 1$, the perpendicular modes eventually surpass the oblique modes in growth rate. 

\begin{figure}
    \centering
    \begin{subfigure}{0.45\textwidth}
        \centering
        \caption{Perpendicular maximum growth rate \\ Eq.~\eqref{f-theta-equation} $\Rightarrow$ full dispersion relation}
        \includegraphics[width=\textwidth]{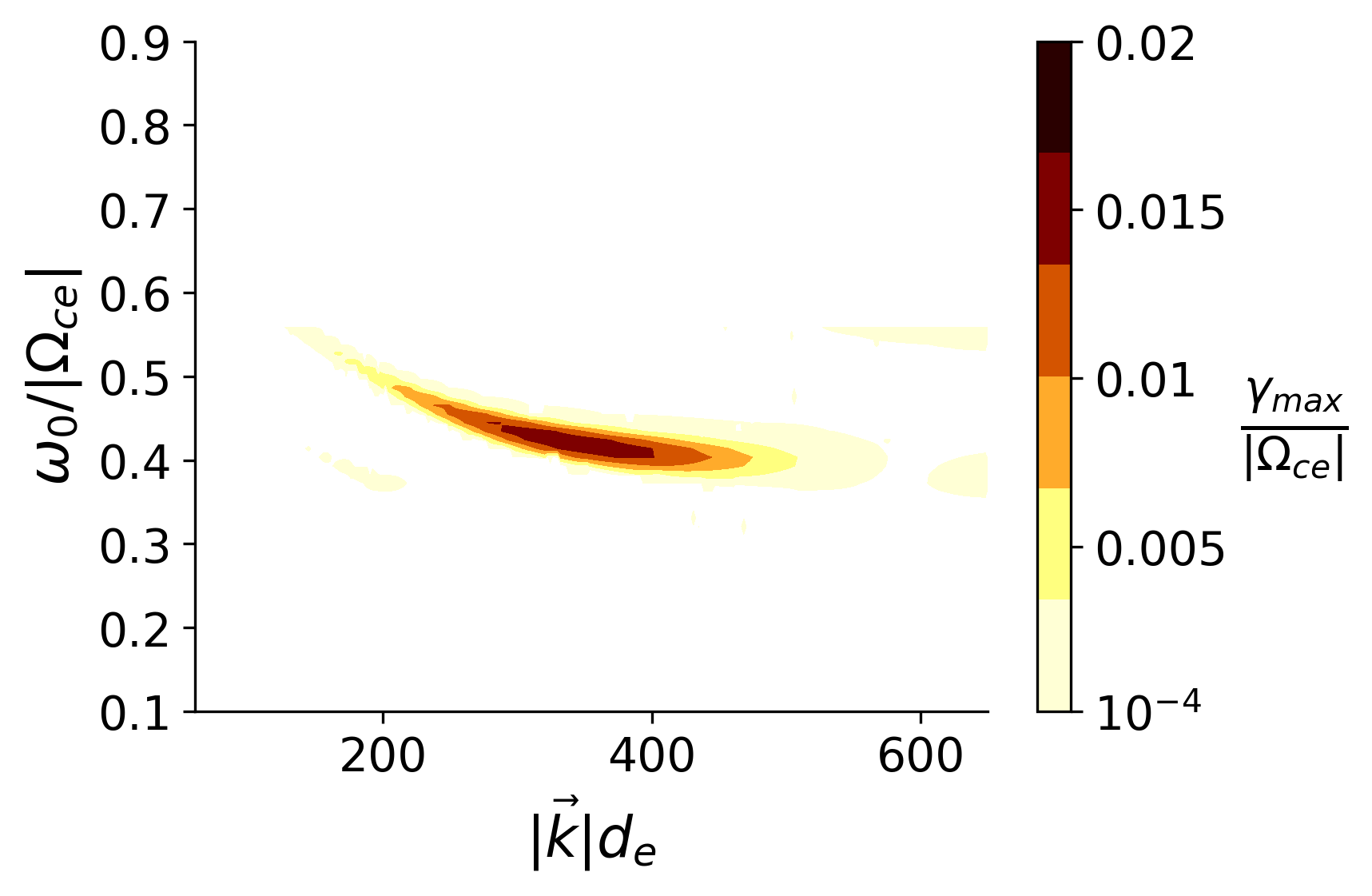}
        \label{fig:fig7-a}
    \end{subfigure}
    \begin{subfigure}{0.45\textwidth}
        \centering
        \caption{Oblique maximum growth rate \\ Eq.~\eqref{f-theta-equation} $\Rightarrow$ full dispersion relation}
        \includegraphics[width=\textwidth]{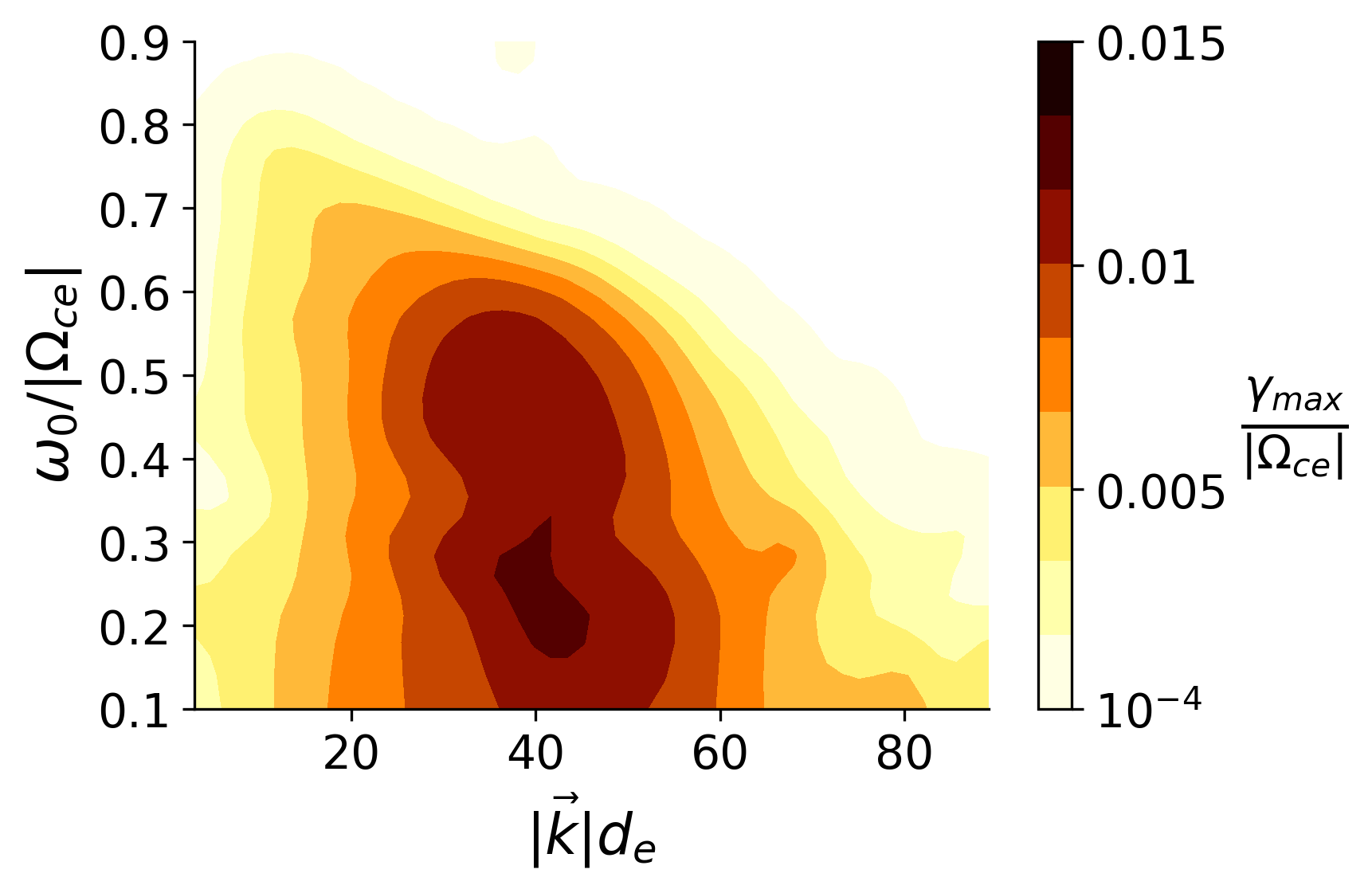}
        \label{fig:fig7-b}
    \end{subfigure}
    \begin{subfigure}{0.45\textwidth}
        \centering
        \caption{Perpendicular maximum growth rate \\ Eq.~\eqref{dispersion_relation_ecdi} $\Rightarrow$ approximation}
        \includegraphics[width=\textwidth]{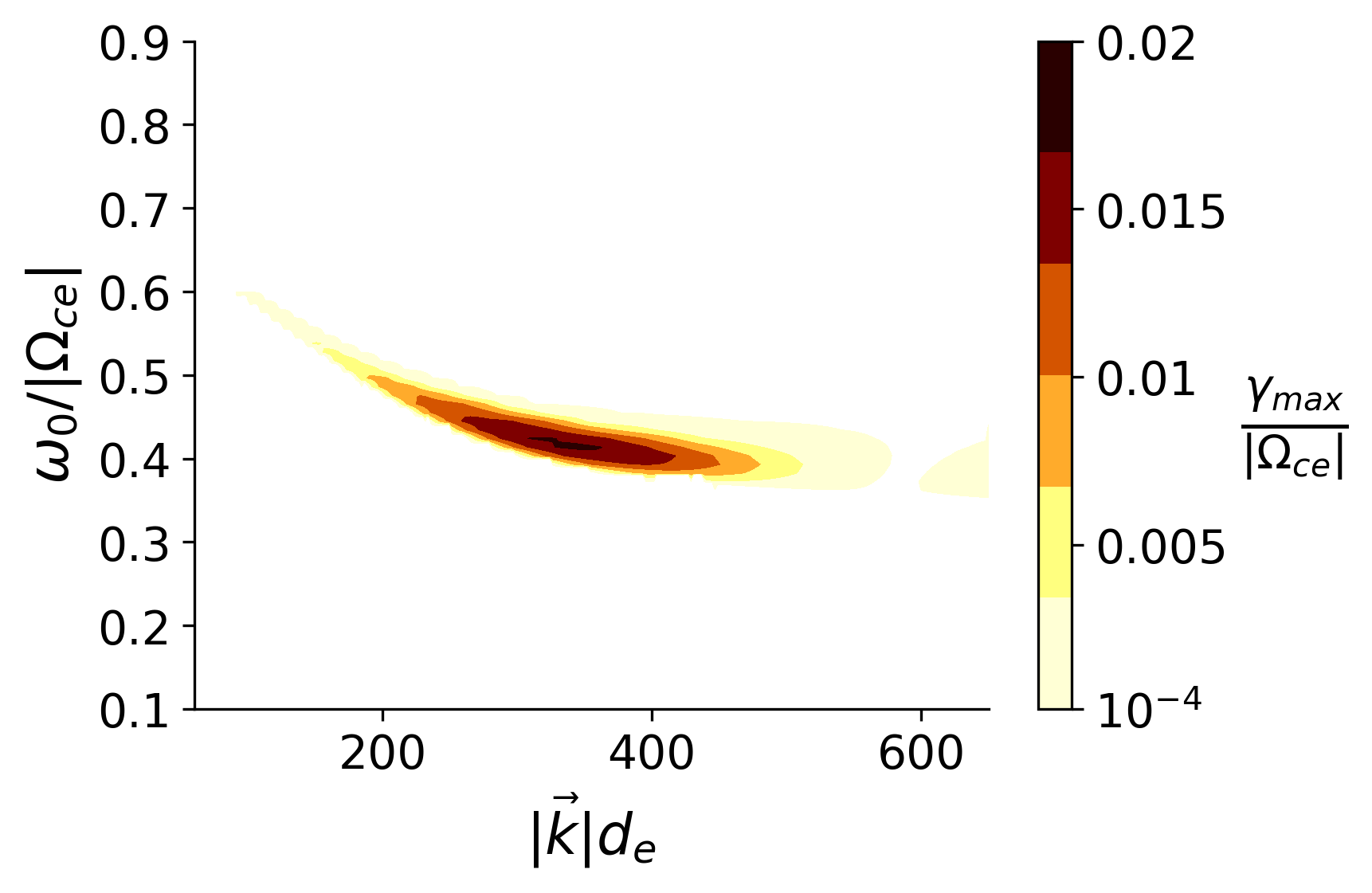}
        \label{fig:fig7-c}
    \end{subfigure}
    \begin{subfigure}{0.45\textwidth}
        \centering
        \caption{Oblique maximum growth rate \\ Eq.~\eqref{approx-dispersion-relation} $\Rightarrow$ approximation}
        \includegraphics[width=\textwidth]{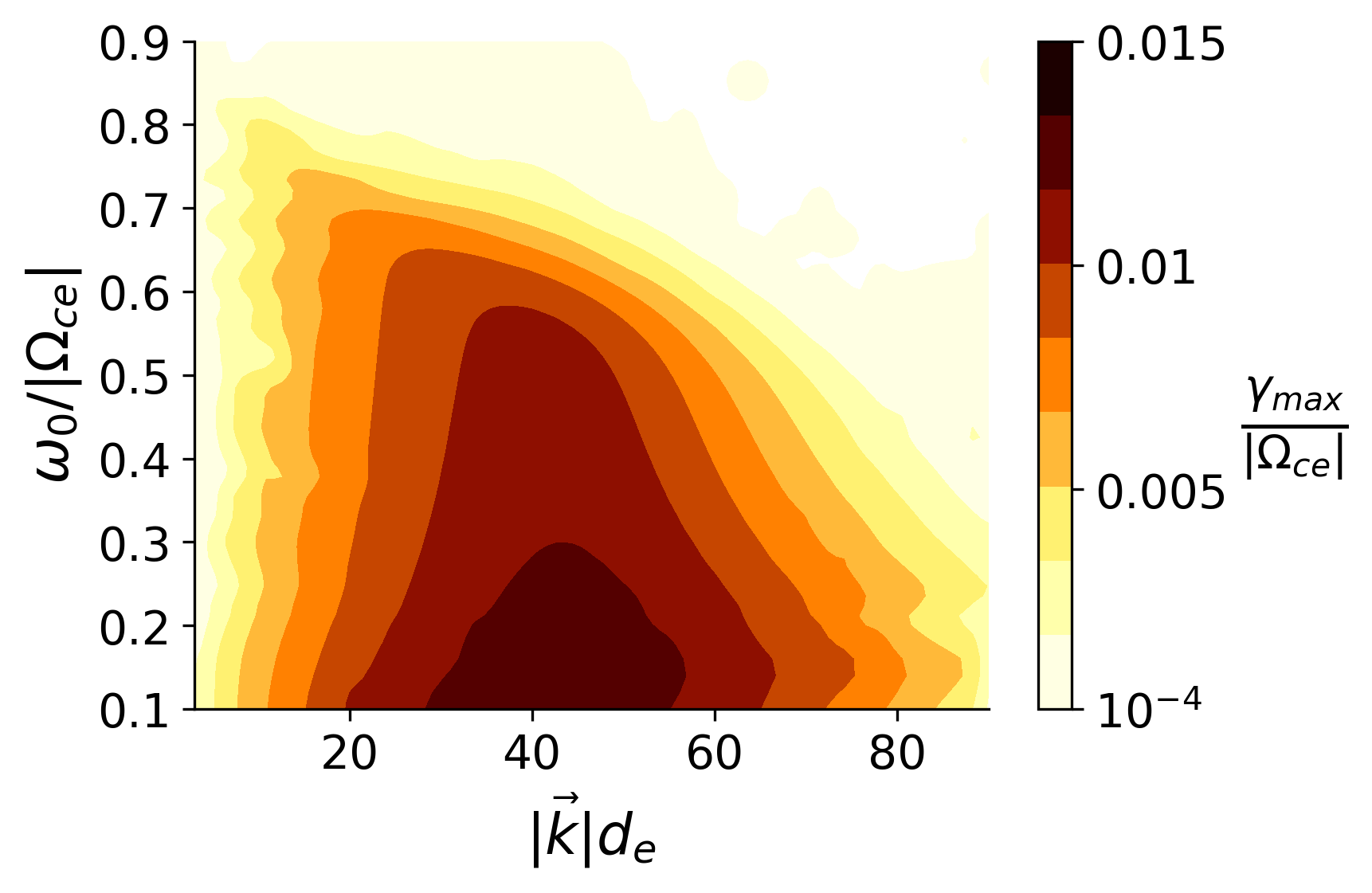}
        \label{fig:fig7-d}
    \end{subfigure}
    \caption{Maximum growth rate over wave normal angle as a function of the primary whistler wave frequency $\omega_{0}$ and wavevector amplitude $|\vec{k}| d_{e}$. The cold electron drift amplitude depends on the primary frequency and is set according to $|\vec{V}_{Dc}|/v_{tc} = 0.46|\Omega_{ce}|/|\omega_{0}-\Omega_{ce}|$ (see Eq.~\eqref{Vds}). Perpendicular electrostatic modes are unstable for lower-band primary waves with $0.35 \lesssim \omega_{0} \lesssim 0.55$, whereas oblique electrostatic modes remain unstable over the entire $0.1 \lesssim \omega_{0} \lesssim 0.9$, with larger growth rates at smaller $\omega_{0}$.  }
    \label{fig:growth-rate-omega0-scan}
\end{figure}

Figure~\ref{fig:growth-rate-omega0-scan} demonstrates the secondary modes' dependence on the primary whistler wave frequency $\omega_{0}$. 
As we vary $\omega_{0}$, the drift amplitude also varies as $|\vec{V}_{Dc}|/v_{tc} = 0.46|\Omega_{ce}|/|\omega_{0}-\Omega_{ce}|$, see Eq.~\eqref{Vds}.
The perpendicular and oblique results are maximized over a range of wave normal angles ($10^{\circ}$ range from purely perpendicular and the corresponding resonance cone angle). 
The perpendicular modes are only unstable for $0.35 \lesssim \omega_{0} \lesssim 0.55$, whereas the oblique modes are unstable over the entire frequency range and are most unstable for lower-band primary waves $\omega_{0} \lesssim 0.5$. 
The maximum growth rate for the oblique modes has an inverse relationship to the primary frequency $\omega_{0}$, which can be explained algebraically from Eq.~\eqref{simplified-growth-rate-oblique-whistler}.
The amplitude of the cold electron drift is $|V_{Dc}| \propto |\Omega_{ce}|/[|\omega_{0} - |\Omega_{ce}||]$, see Eq.~\eqref{Vds}, such that higher primary frequencies result in larger drifts.
However, from Figure~\ref{fig:growth-rate-omega0-scan}, the maximum perpendicular wavenumber decreases fast with increasing $\omega_{0}$ for both perpendicular and oblique modes, such that the Bessel function argument $k_{\perp} |V_{Dc}|/\omega_{0}$ decreases with increasing $\omega_{0}$. 
Therefore, since $J_{1}(x) \approx x/2$ for small argument, the growth rate is larger for smaller~$\omega_{0}$, as shown in Figures~\ref{fig:fig7-b} and~\ref{fig:fig7-d}.
Here again, the results obtained by solving the approximate dispersion relations in Eqs.~\eqref{approx-dispersion-relation} and~\eqref{dispersion_relation_ecdi} are in good agreement with those from the full dispersion relation in Eq.~\eqref{f-theta-equation}, thereby motivating their adoption in the next section within the QLT framework.

\subsection{Parametric study via moment-based quasilinear theory}\label{sec:parametric_QLT}
\begin{figure}
    \centering
    \begin{subfigure}{0.4\textwidth}
        \centering
        \caption{Heating and initial maximum growth rate vs. $\frac{n_{c}}{n_{e}}$}
        \includegraphics[width=\textwidth]{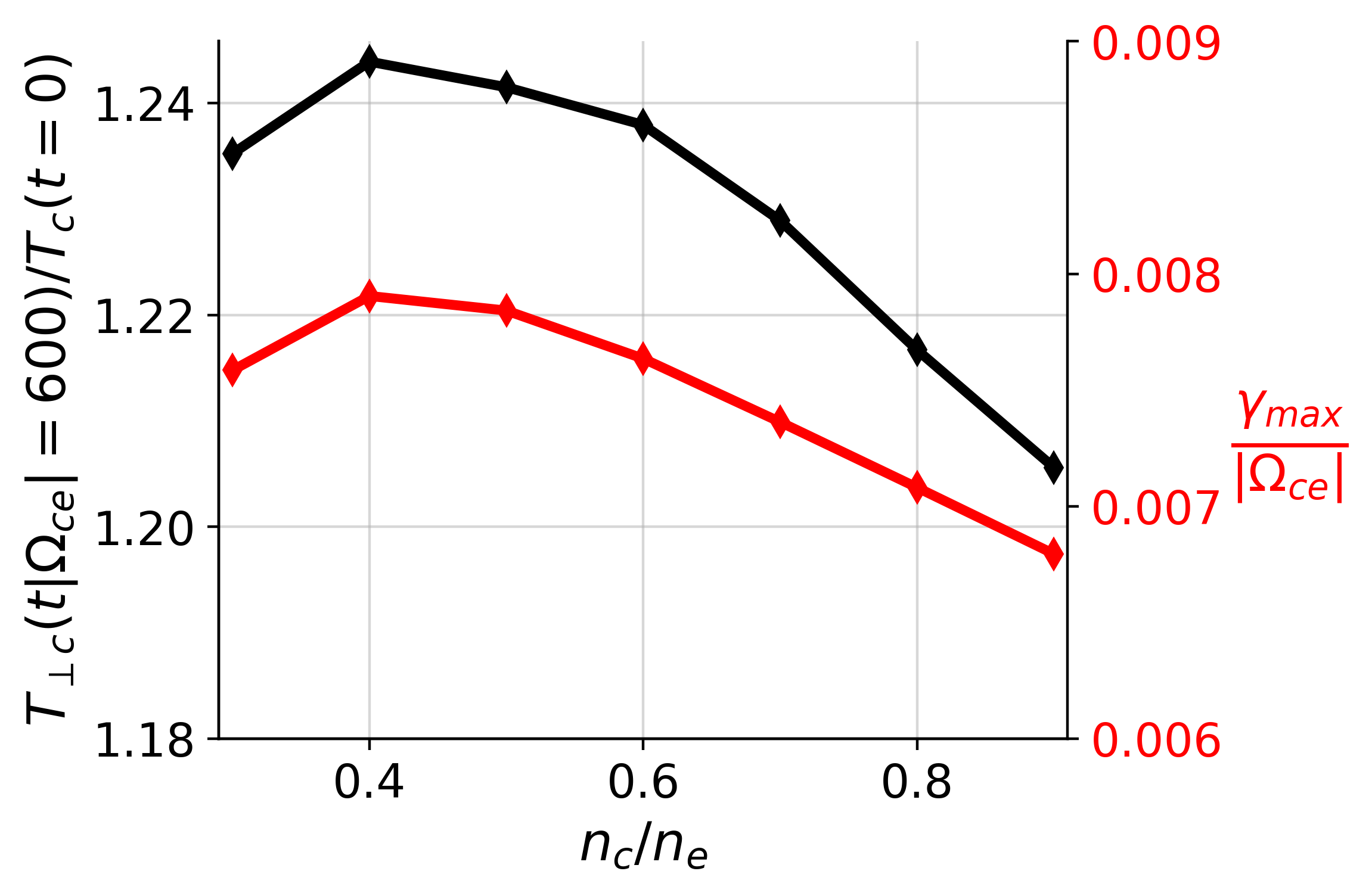}
        \label{fig:reduction-in-heating}
    \end{subfigure}
    \begin{subfigure}{0.4\textwidth}
        \centering
        \caption{Primary whistler wave damping vs. $\frac{n_{c}}{n_{e}}$}
        \includegraphics[width=\textwidth]{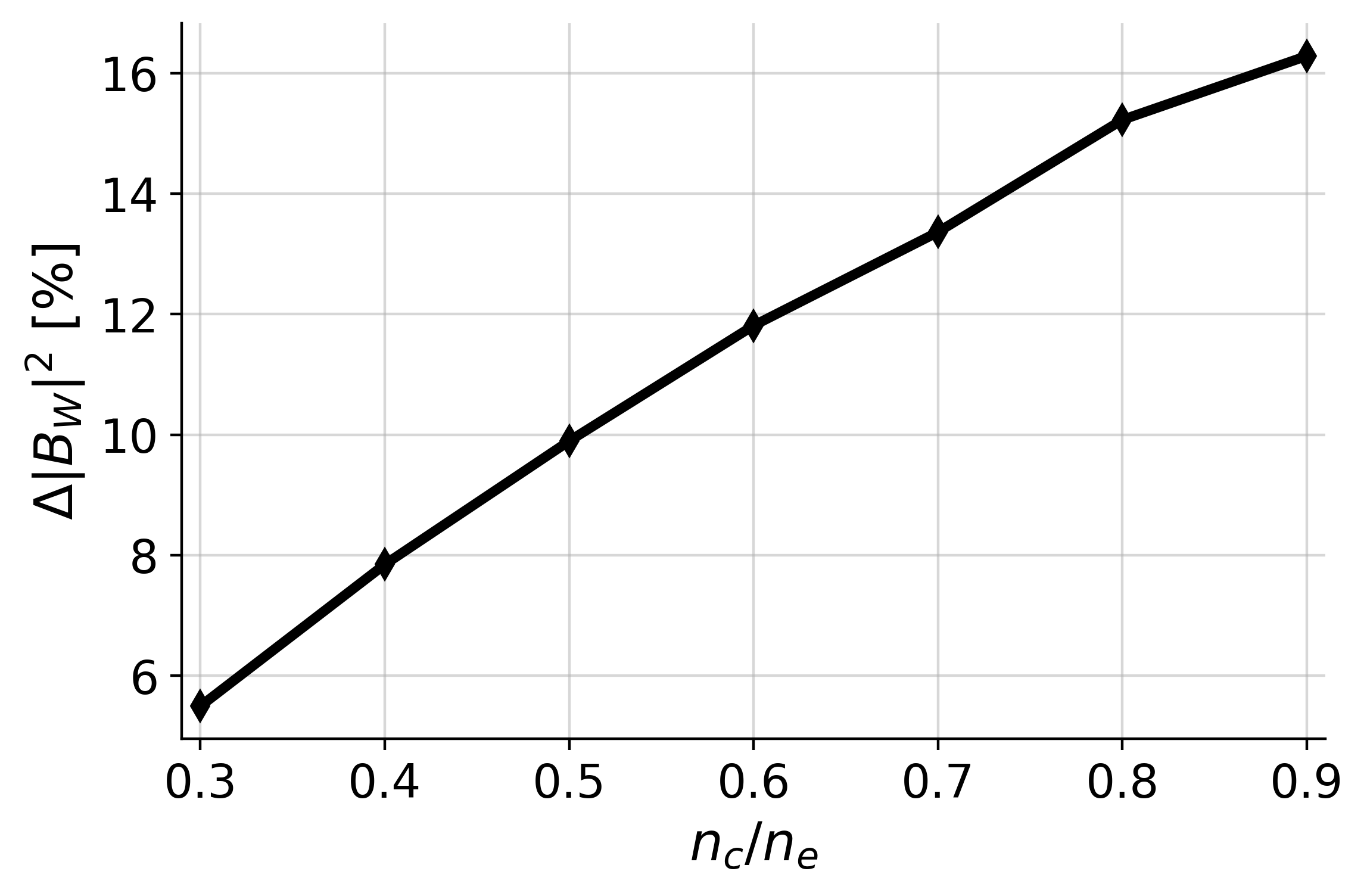}
    \end{subfigure}
    \begin{subfigure}{0.4\textwidth}
        \centering
        \caption{Heating and initial maximum growth rate vs. $T_{c}$}
        \includegraphics[width=\textwidth]{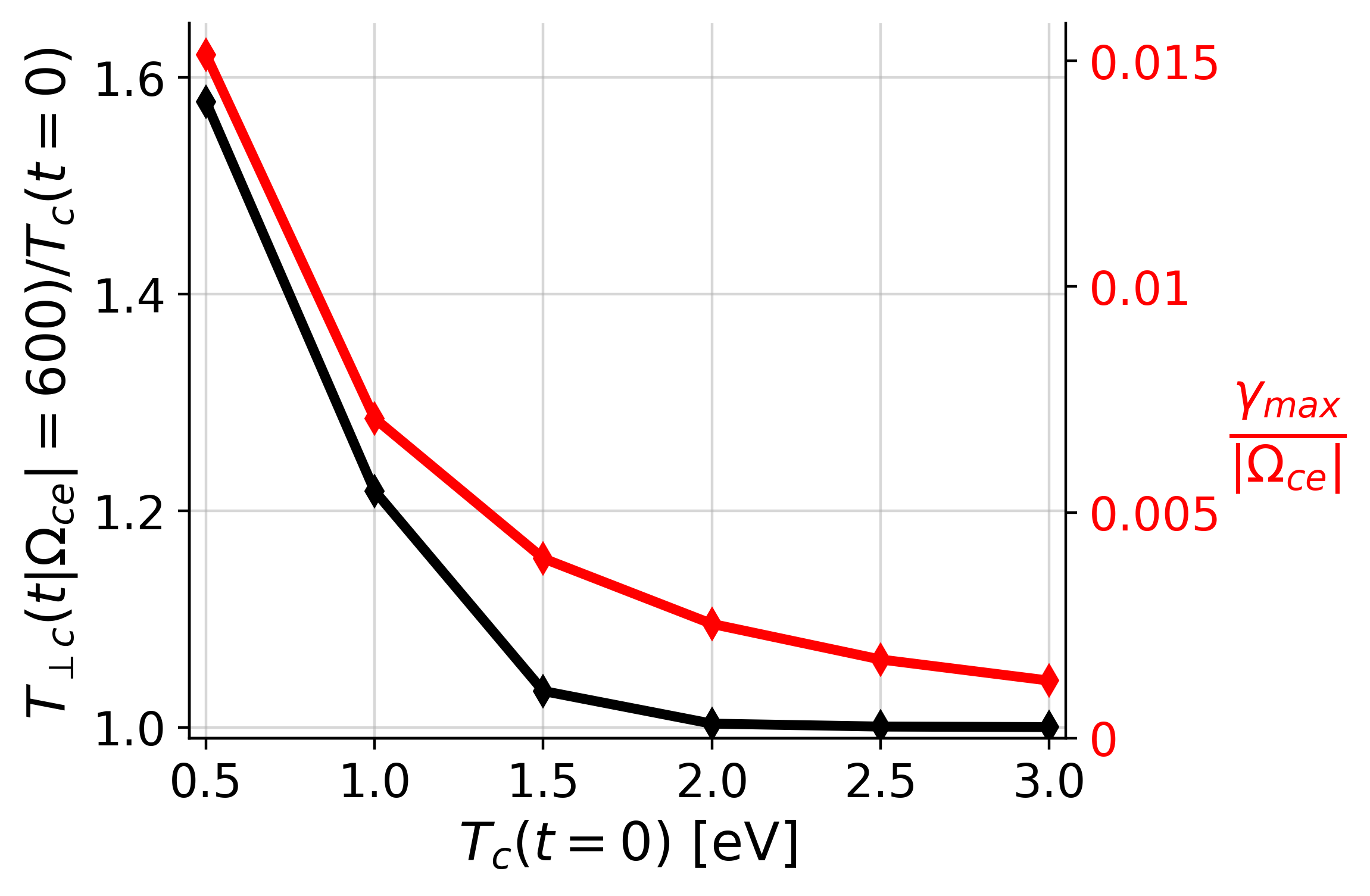}
    \end{subfigure}
    \begin{subfigure}{0.4\textwidth}
        \centering
        \caption{Primary whistler wave damping vs. $T_{c}$}
        \includegraphics[width=\textwidth]{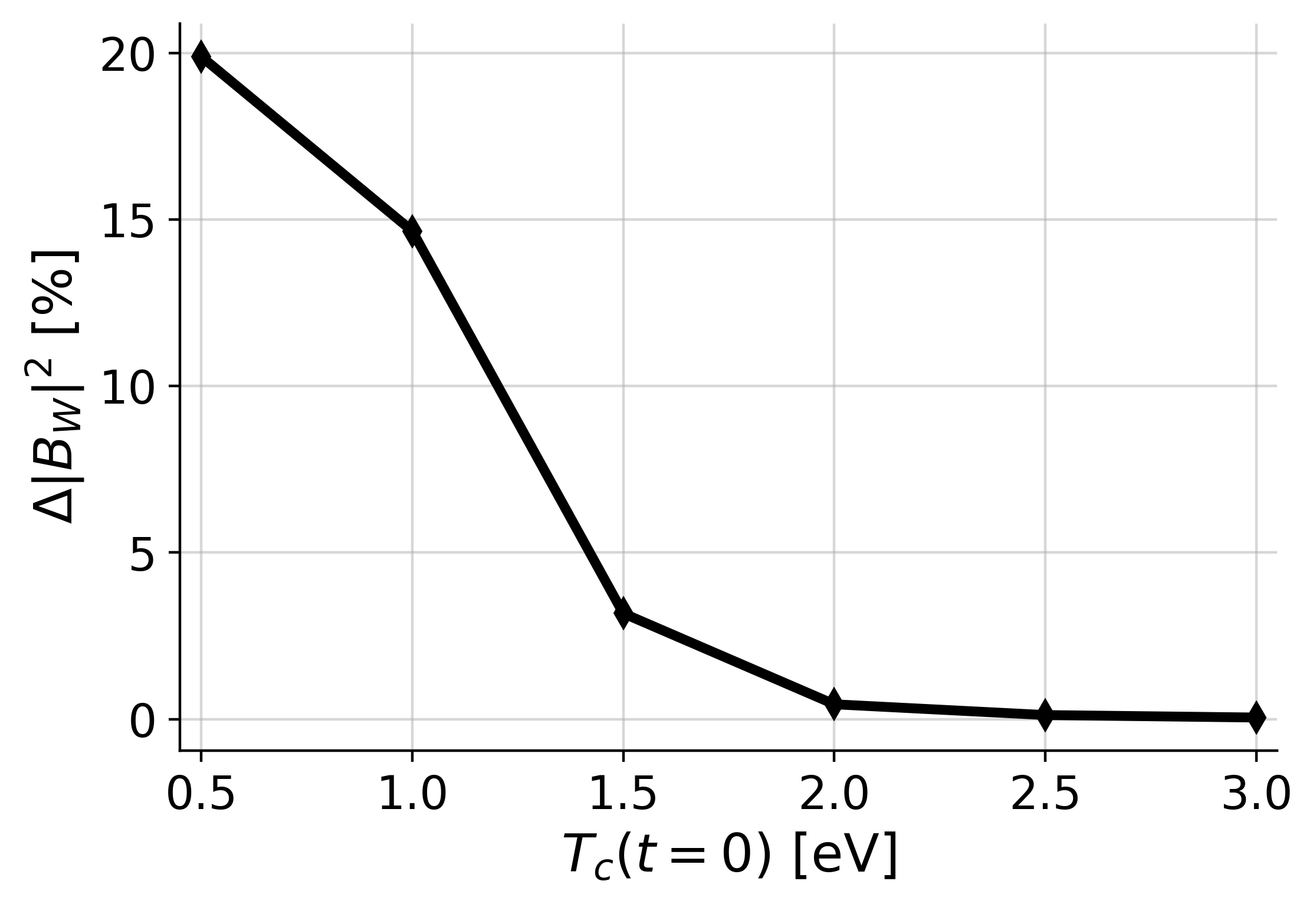}
    \end{subfigure}
    \begin{subfigure}{0.4\textwidth}
        \centering
        \caption{Heating and initial maximum growth rate vs. $\omega_{0}$}
        \includegraphics[width=\textwidth]{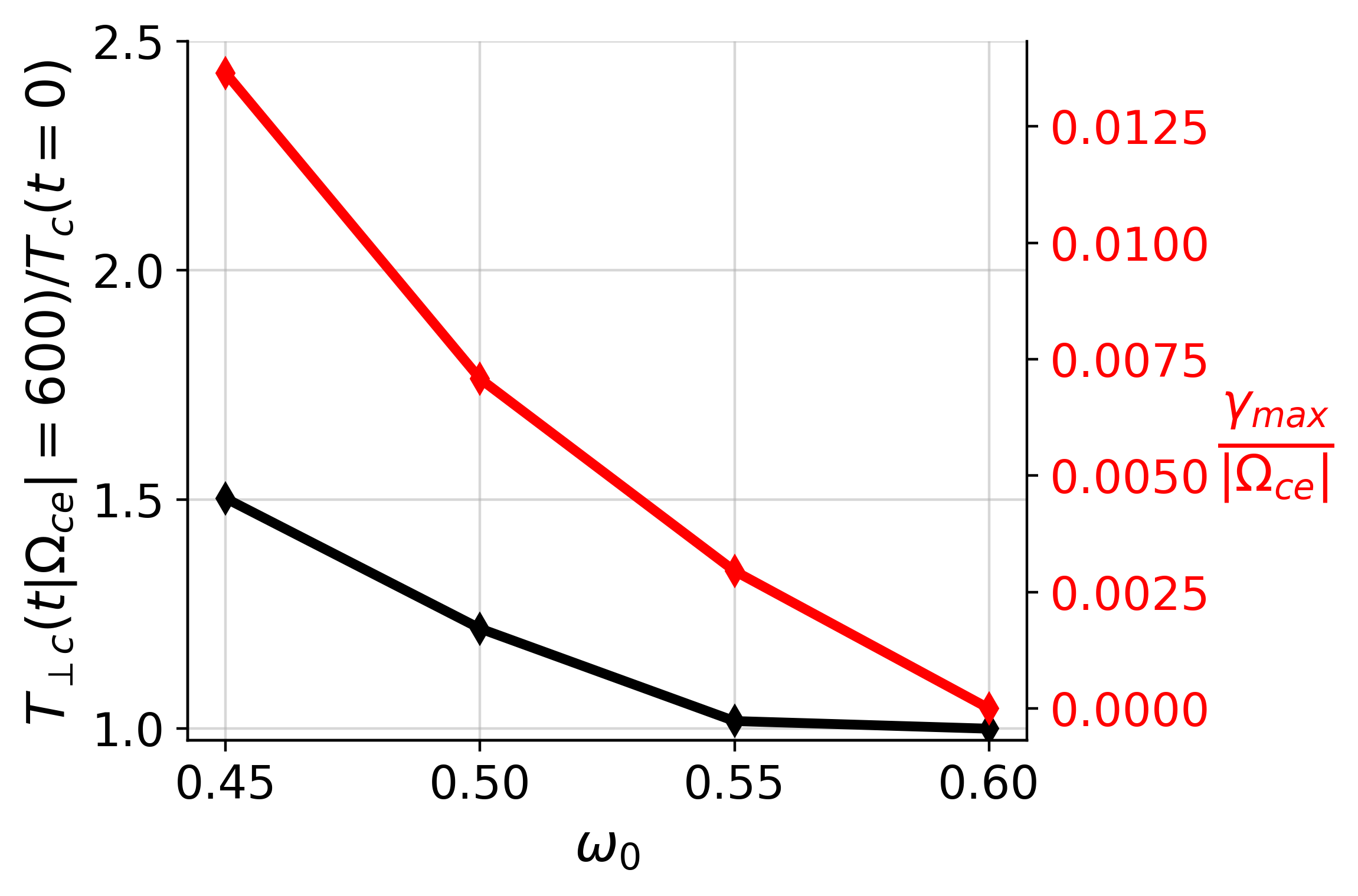}
    \end{subfigure}
    \begin{subfigure}{0.4\textwidth}
        \centering
        \caption{Primary whistler wave damping vs. $\omega_{0}$}
        \includegraphics[width=\textwidth]{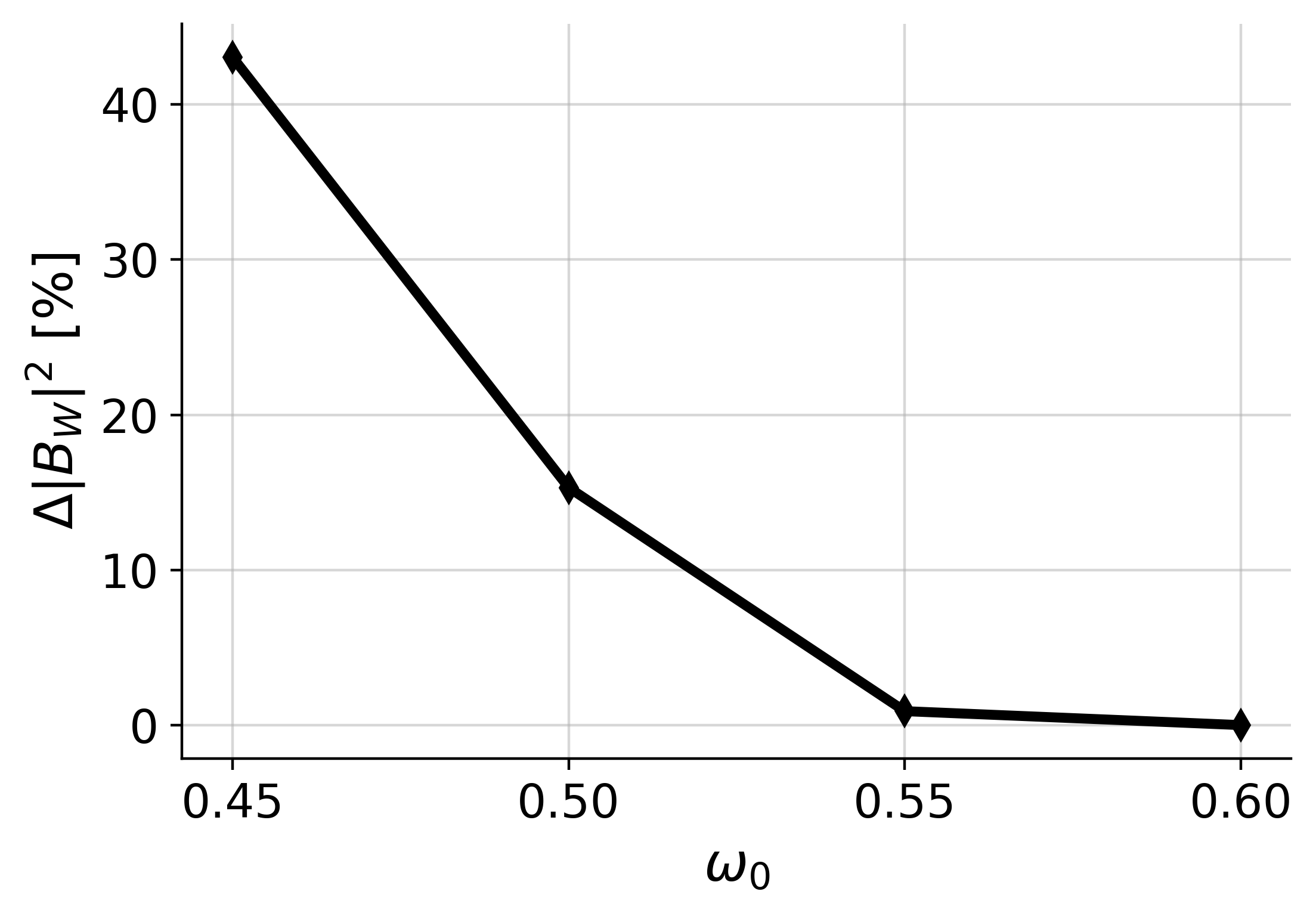}
    \end{subfigure}
    \caption{Cold electron heating, maximum initial growth rate, and primary wave damping driven by the perpendicular instability varying (a/b) cold electron density over total electron density $n_{c}/n_{e}$, (c/d) initial cold electron temperature $T_{c}(t=0)$, and (e/f) primary wave frequency $\omega_{0}$. Heating is proportional to the initial growth rate of the perpendicular instability for all parametric scans. The primary wave damping is $\Delta |B_{W}|^2 \sim \Delta K_{c} \sim  \Delta n_{c} T_{c}$, such that density increase overcome the reduction in heating and $\Delta |B_{W}|^2$ increases with increasing $n_{c}/n_{e}$. For $T_{c} > \SI{2}{\electronvolt}$, the perpendicular modes result in close to zero net heating of cold electrons and damping of the primary wave. Lastly, the perpendicular instability is more efficient for the lower band primary wave. }
    \label{fig:parametric_QLT_perp}
\end{figure}

In this section, we examine the parametric dependence of the secondary instability on the primary wave damping and cold electron heating at saturation ($t|\Omega_{ce}|=600$). Here, we let $t=0$ be the time of the primary wave reaching saturation (i.e, the onset of the secondary instability).  
Figure~\ref{fig:parametric_QLT_perp} presents the effects of varying cold electron density, temperature, and primary wave frequency on cold electron final temperature and the primary whistler wave magnetic energy density driven by the excitation of perpendicular electrostatic modes. 
The QLT results show that the cold electron heating rate is correlated to the initial maximum growth rate, which, for $n_{c}/n_{e} > 0.4$, reduces with increasing $n_{c}/n_{e}$. 
However, the primary wave damping rate increases with increasing $n_{c}/n_{e}$ since $\Delta |B_{W}|^2 \sim \Delta K_{c} \sim \Delta n_{c} T_{c}$, such that density increase overcomes the reduction in heating as seen in Figure~\ref{fig:reduction-in-heating} and $\Delta |B_{W}|^2$ increases with $n_{c}/n_{e}$.
The primary wave damping dependence on initial temperature is more straightforward as the cold electron density is now constant and set to $n_{c}/n_{e} = 0.8$. 
Note that increasing the cold electron temperature reduces the ratio of cold electron drift velocity to thermal velocity $|\vec{V}_{Dc}|/v_{tc}$, which is the key parameter/free energy source of the secondary instability. 
The results show that for $T_{c} > \SI{2}{\electronvolt}$, the perpendicular mode results in close to zero net heating and damping of the primary wave. 
Lastly, varying $\omega_{0}$ can significantly change the effectiveness of the perpendicular modes, as previously discussed in section~\ref{sec:parametric_linear}, such that setting $\omega_{0} = 0.45$ results in $40\%$ damping and $\omega_{0} = 0.55$ results in nearly zero net damping.  

\begin{figure}
    \centering
    \begin{subfigure}{0.4\textwidth}
        \centering
        \caption{Heating and initial maximum growth rate vs. $\frac{n_{c}}{n_{e}}$}
        \includegraphics[width=\textwidth]{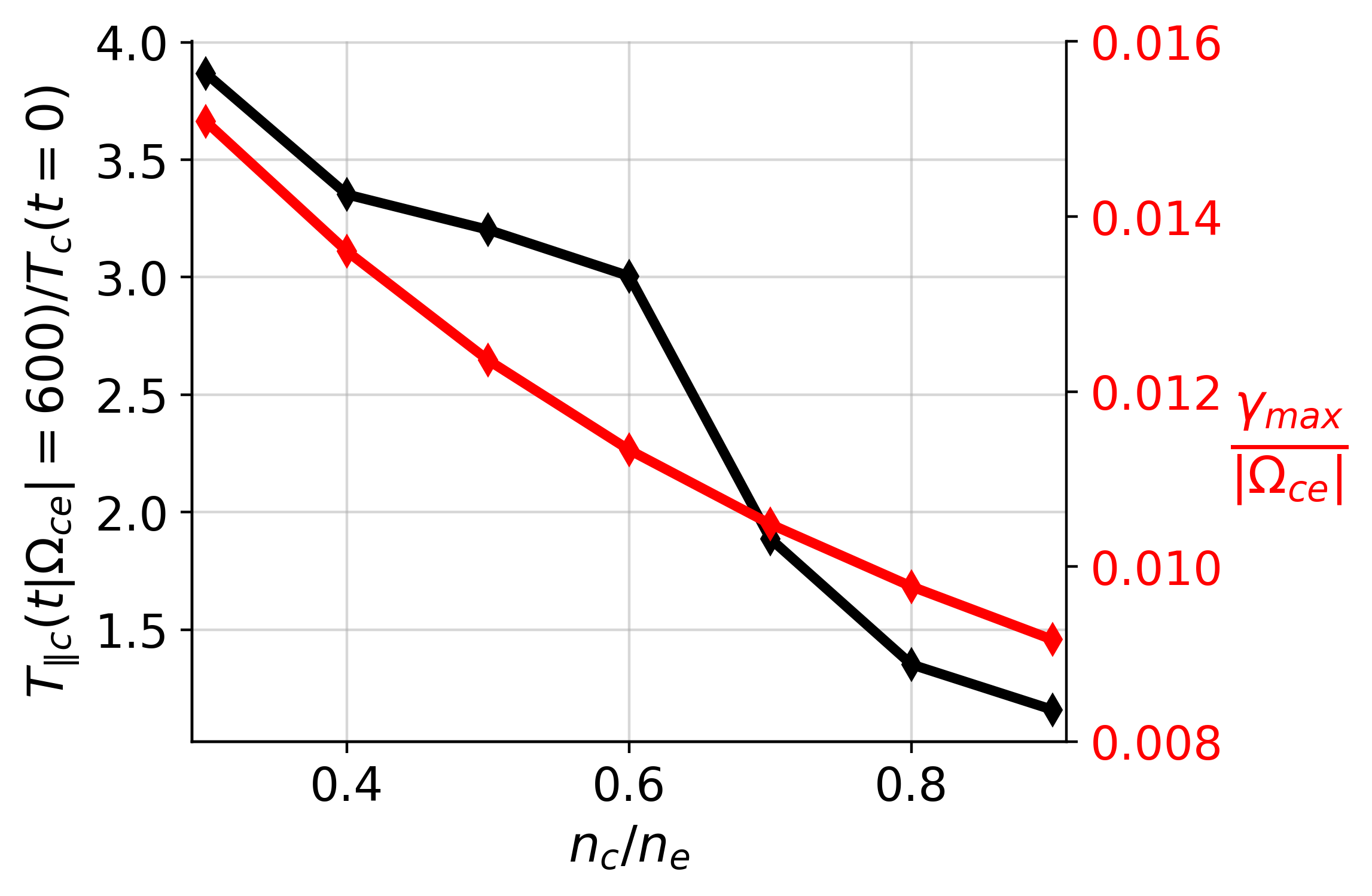}
    \end{subfigure}
    \begin{subfigure}{0.4\textwidth}
        \centering
        \caption{Primary whistler wave damping vs. $\frac{n_{c}}{n_{e}}$}
        \includegraphics[width=\textwidth]{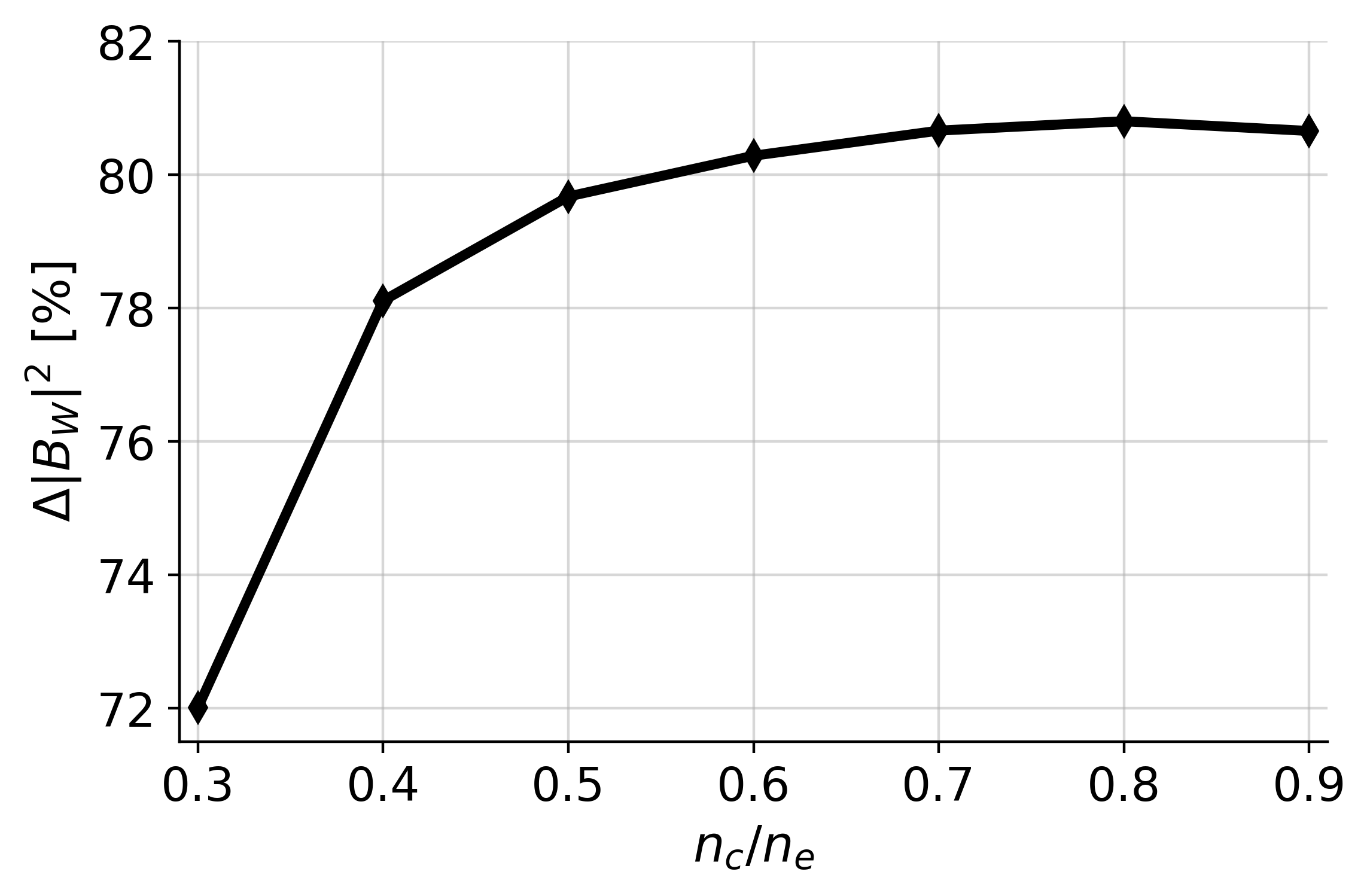}
    \end{subfigure}
    \begin{subfigure}{0.4\textwidth}
        \centering
        \caption{Heating and initial maximum growth rate vs. $T_{c}$}
        \includegraphics[width=\textwidth]{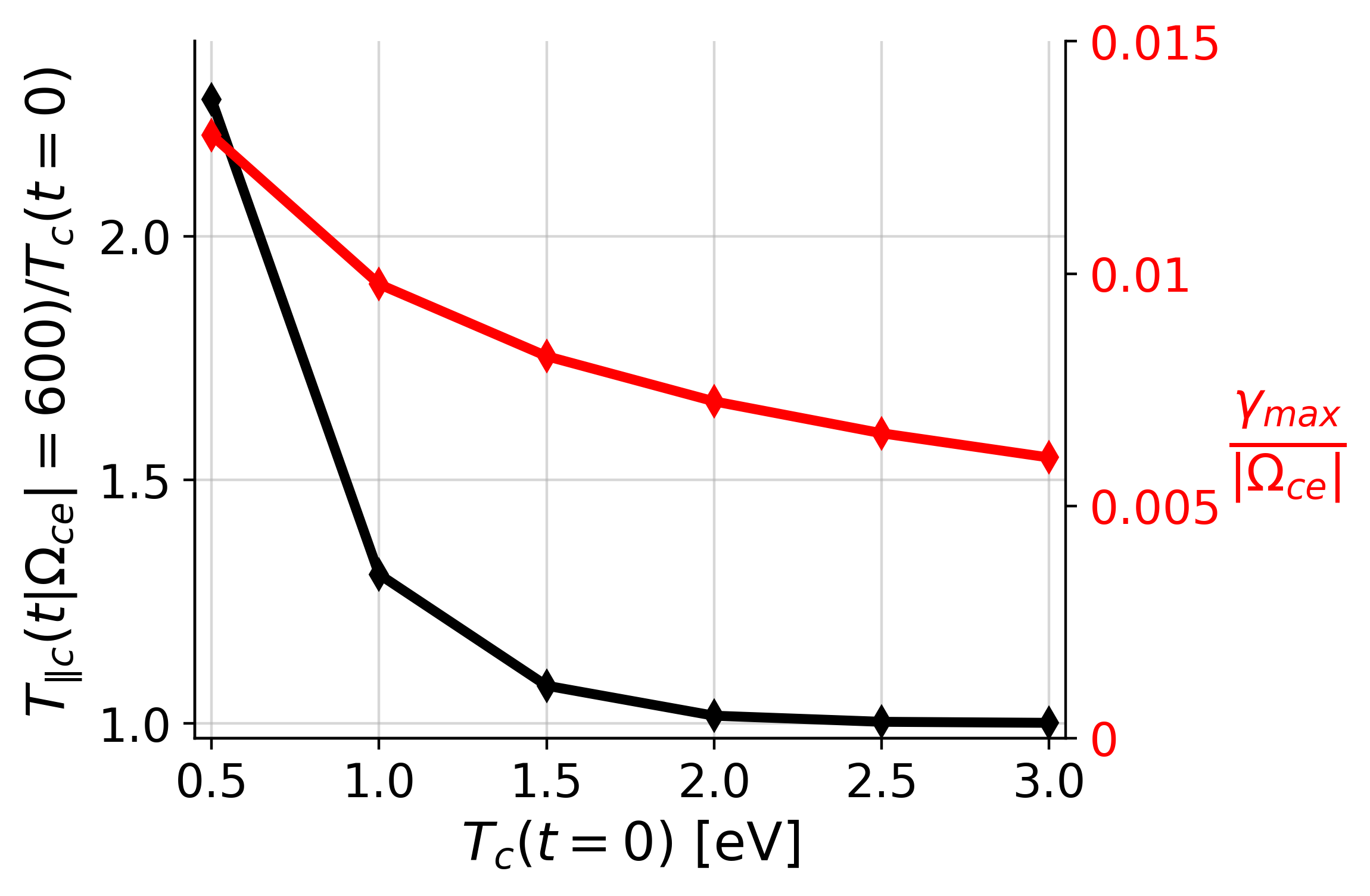}
    \end{subfigure}
    \begin{subfigure}{0.4\textwidth}
        \centering
        \caption{Primary whistler wave damping vs. $T_{c}$}
        \includegraphics[width=\textwidth]{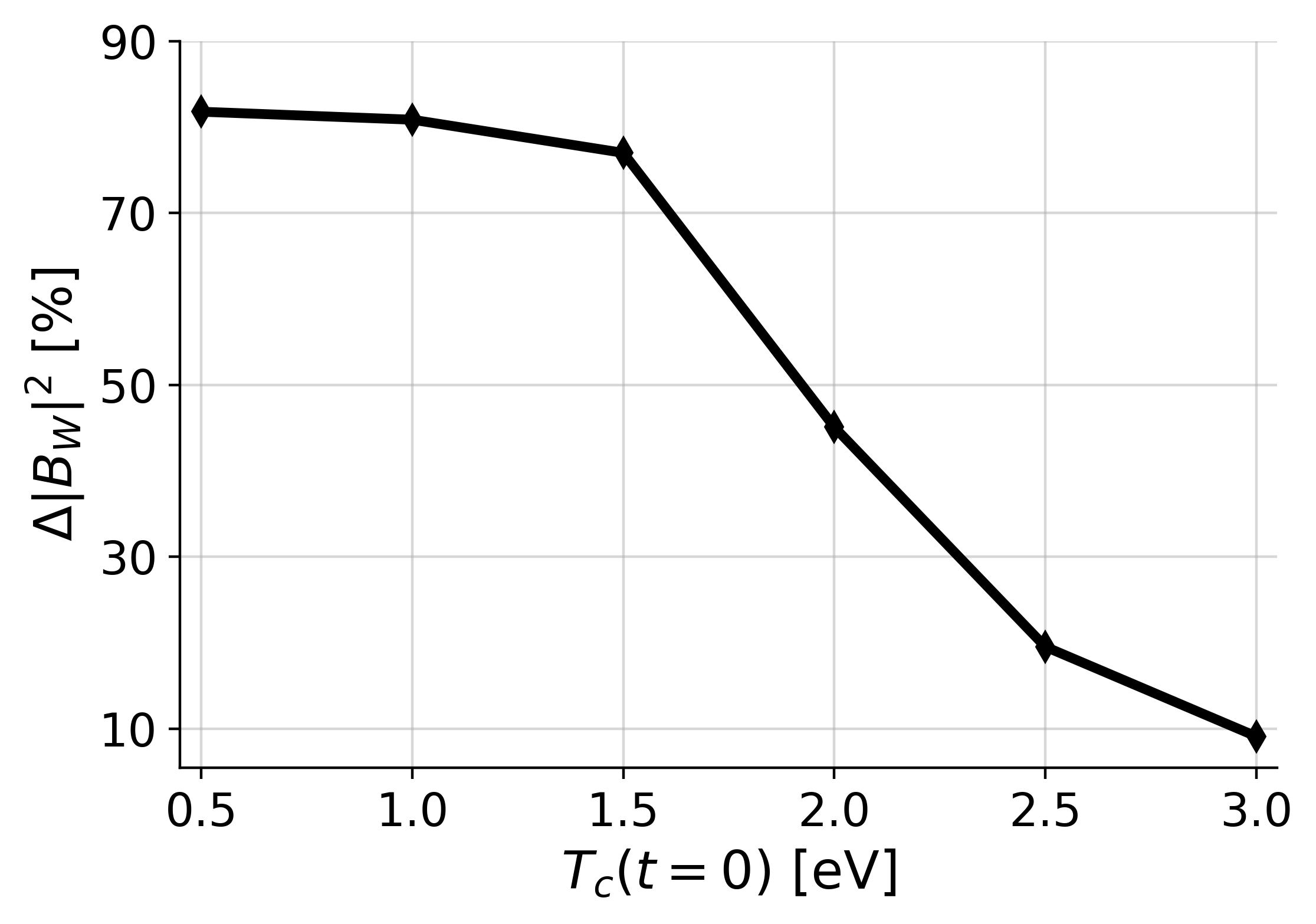}
    \end{subfigure}
    \begin{subfigure}{0.4\textwidth}
        \centering
        \caption{Heating and initial maximum growth rate vs. $\omega_{0}$}
        \includegraphics[width=\textwidth]{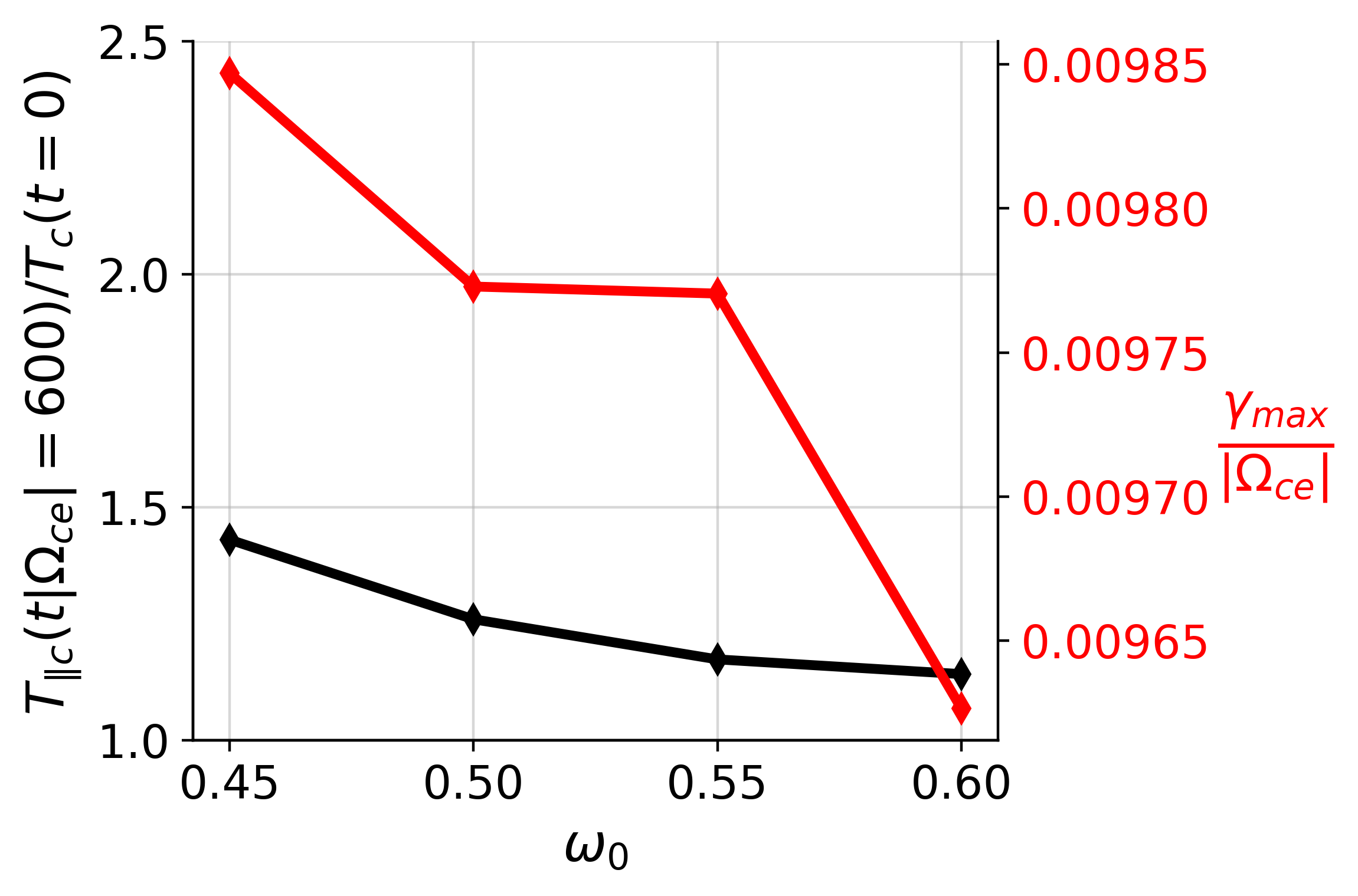}
    \end{subfigure}
    \begin{subfigure}{0.4\textwidth}
        \centering
        \caption{Primary whistler wave damping vs. $\omega_{0}$}
        \includegraphics[width=\textwidth]{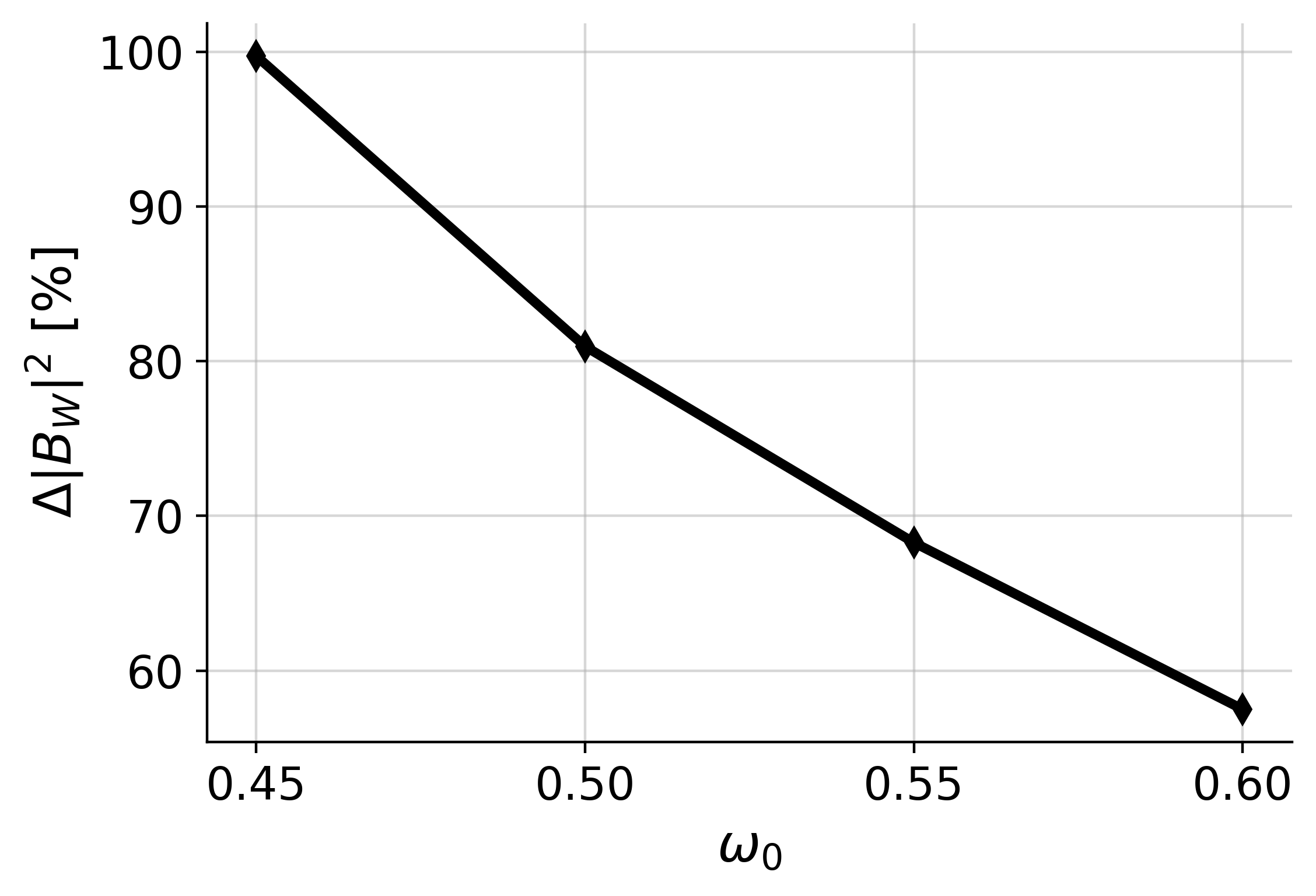}
    \end{subfigure}
    \caption{Same as Figure~\ref{fig:parametric_QLT_perp} for the oblique electrostatic whistler modes. The same relations hold. The primary wave damping is $\Delta |B_{W}|^2 \sim \Delta K_{c} \sim n_{c} \Delta T_{c}$, such that density increase overcome the reduction in heating and $\Delta |B_{W}|^2$ increases with increasing $n_{c}/n_{e}$. Moreover, the primary wave damps significantly in the presence of cold electrons in the sub-eV to \SI{2}{\electronvolt} range and lower-band primary wave. 
    }
    \label{fig:parametric_QLT_oblique}
\end{figure}

The cold electron density and temperature parametric scans for the oblique modes are presented in Figure~\ref{fig:parametric_QLT_oblique}. 
Similar to Figure~\ref{fig:parametric_QLT_perp}, the results in Figure~\ref{fig:parametric_QLT_oblique} compare the final primary wave magnetic energy density and cold electron temperature at $t|\Omega_{ce}|=600$ with the corresponding initial states, and all other parameters are fixed to those used in the PIC simulation described in section~\ref{sec:QLT_vs_PIC}.
The cold electron parallel heating from oblique modes decreases with increasing $n_{c}/n_{e}$, indicating an inverse proportionality. 
This is expected from linear theory, since the initial maximum growth rate is also inversely related to $n_{c}/n_{e}$ (given by the approximate dispersion relation~\eqref{approx-dispersion-relation}).
The primary wave damping rate increases slightly from $72\%$ to $81\%$ for $n_{c}/n_{e}$ ranging from $0.3$ to $0.9$, respectively.  
Similar to the perpendicular modes, the change in $n_{c}/n_{e}$ overcomes the reduction in heating and thus the primary wave damping rate increases with increasing $n_{c}/n_{e}$. 
Increasing the cold electron temperature can substantially change the primary wave damping rate and cold electron parallel heating. For instance, at $T_{c} = \SI{0.5}{\electronvolt}$ ($|V_{Dc}|/v_{tc} \approx 1.3$) the primary wave gets damped by 85\% and for $T_{c} = \SI{3}{\electronvolt}$ by $10\%$ ($|V_{Dc}|/v_{tc} \approx 0.53$). 
Additionally, the primary wave frequency $\omega_{0}$ highly impacts the efficiency of the oblique modes, where for $\omega_{0} = 0.45$ the primary wave gets damped by $100\%$ and for $\omega_{0} = 0.6$ by $60\%$. 
Additionally, (not shown here) as the primary frequency grows above $\omega_{0} > 0.5$, the secondary modes have a higher frequency and interact with the cold plasma mainly via cyclotron resonance (first harmonic), which heats the cold electrons in the perpendicular direction more than in the parallel direction. 
This is opposed to resonance via Landau damping, which is the main interaction for the cases presented before, that heats the plasma in the parallel direction. 
We highlight that although for $T_{c} = \SI{0.5}{\electronvolt}$, the maximum growth rate of the perpendicular modes ($\approx 0.015$) is larger than the oblique maximum growth rate ($\approx 0.0125$), the oblique modes are still the most dominant mechanism for damping the primary whistler wave and remain so for all the cases considered in this parameter regime.

\section{Conclusions}\label{sec:conclusions}
We investigate the role of cold electrons in the damping of parallel-propagating whistler waves through secondary drift-driven instabilities, using a moment-based QLT approach. 
We verified the developed QLT framework using PIC simulations and confirmed that our moment-based QLT captures accurately the most important aspects of the secondary instabilities saturation.
The developed QLT framework elucidates the fundamental physics of the secondary instabilities and provides a computationally efficient method for quantifying energy exchange from the primary whistler wave to cold electrons without requiring fully kinetic simulations that resolve the cold electron scales.
The main findings of this study are as follows. 
(1) Secondary oblique whistler wave excitation is persistent for a wide range of primary wave frequencies and cold electron densities and temperatures. 
(2) Oblique whistler waves (in comparison to perpendicular ECDI-like modes) are the main source of energy drainage from the primary parallel-propagating whistler wave.
(3) Both oblique and perpendicular ECDI-like modes are most effective in damping the primary wave when the cold electron drift velocity exceeds its thermal velocity, such that the amplitude of the primary whistler wave does not need to be large for the instability onset if the cold electrons are sufficiently cold. 
This highlights the need for future missions with adequate instrumentation to accurately measure the cold particle populations~\cite{delaznno_2021_cold_impact, maldonado_2023_cold_meassurements_Frontiers}. 
(4) Both oblique and perpendicular modes are most effective at damping parallel whistler waves when the primary wave frequency is in the lower-band range.

A recent study by~\citet{an_2017_whistler_pic} compares the saturated magnetic energy density from 2D3V PIC simulations with whistler-mode chorus wave observations in the inner magnetosphere measured by the Van Allen Probes~\cite{mauk_2013_van_allen}.
The 2D3V PIC simulations include hot electrons and immobile protons, neglecting cold electrons. 
\citet{an_2017_whistler_pic} results indicate that the simulated whistler wave magnetic energy density at saturation is approximately four orders of magnitude larger than Van Allen Probes observations (see Figure~5 in~\cite{an_2017_whistler_pic}).
This significant discrepancy can be potentially explained by the drift-driven secondary instabilities, which are not captured in the simulations presented in~\cite{an_2017_whistler_pic} because cold electrons are omitted. 
As discussed in section~\ref{sec:parametric_QLT}, the secondary oblique whistler modes strongly damp, by nearly 100\%, the primary wave when the primary wave is in the lower-band frequency range. 
Connecting these secondary instabilities to inner magnetosphere observations remains an important avenue for future investigation.

Moreover, a future goal of ours is to extend the framework to non-uniform background magnetic fields (e.g., a dipolar magnetic field) and non-uniform density, along with broadband whistler waves and wavepackets. 
It is also important to understand how the primary wave amplitude influences the regime in which the drift-driven instabilities are dominant over other nonlinear wave-wave interaction processes that result in whistler wave damping, e.g., parametric~\cite{ke_2017_whistler_parameteric, umeda_2017_whistler_parametric, karbashewski_2022_parametric, forslund_1972_whistler_parametric} and modulational instabilities~\cite{karpman_1977_whistler_modulation}.
In principle, the secondary instability discussed here can potentially be present for other primary low-frequency electromagnetic waves, such as magnetosonic waves~\cite{kitsenko_1973_magnetosonic,  mcbride_1973_magnetosonic}, electromagnetic ion-cyclotron waves, etc.~\cite{sizonenko_1967_emic, khazanov_1996_emic, bortnik_2010_emic_JGR, khazanov_2017_emic_JGR, bortnik_2011_emic_JGR}. 
Lastly, the discussed secondary instabilities lead to a predator-prey relationship between the primary and secondary waves, in which the secondary electrostatic modes (predator) damp the whistler modes (prey).
\citet{shao_2025_ech_whistler, gao_2022_oblique_JGR} also found a predator-prey relationship between electron Bernstein and whistler waves, but with the roles reversed. 
In their scenario, electron Bernstein modes are driven unstable by energetic electrons with a loss-cone distribution.
When such electrons also exhibit anisotropy, both electron Bernstein and whistler modes can be excited; in that case, whistler waves suppress electron Bernstein waves by filling the loss cone~\cite{shao_2025_ech_whistler}, so that electron Bernstein modes (prey) are damped through the excitation of whistler waves (predator). 
Further investigation into the long-term dynamics of such interactions, as well as their competition with the secondary instability discussed above, is needed.

\appendix

\section{Primary whistler anisotropy-driven instability} \label{sec:QLT_primary}
The whistler anisotropy instability, also referred to as the electromagnetic electron cyclotron instability, is driven by anisotropy in the hot electron population, characterized by $A_{h}>0$ (perpendicular temperature larger than parallel temperature).
Previous studies~\cite{gary_2012_linear, brice_1971_cold, cuperman_1974_cold, cuperman_1975_cold_JGR, delaznno_2021_cold_impact, schriver_1991_crres, frantsuzov_2022_whistler_cold} have shown that although hot electrons transfer energy to the whistler waves, the presence of cold electrons can significantly influence whistler wave properties, including the frequency, growth rate, wave normal angle, and saturation amplitude. 
The underlying mechanism is that, for a fixed whistler wave frequency $\omega_{r}$, an increase in the cold electron density $ n_c/n_e $ leads to a larger parallel wavenumber $ k_{\|}$. This increase in $k_{\|}$ reduces the resonant parallel velocity, $v_{\|}^{\mathrm{res}} = [\omega_r - |\Omega_{ce}|]/k_{\|}$, allowing a larger fraction of hot electrons to be resonant with the wave, thereby enhancing the growth rate and wave-particle energy transfer over specific ranges of cold electron density~\cite{cuperman_1974_cold}. 

The whistler dispersion relation for a plasma with multiple bi-Maxwellian species is 
\begin{equation}\label{dispersion_relation_whistler_general}
    \sum_{s} \omega_{ps}^2\left(\xi_{s}^{0} Z(\xi_{s}^{1})  + A_{s} \left( 1+ \xi_{s}^{1} Z(\xi_{s}^{1})\right)\right) = k_{\|}^2c^2,
\end{equation}
where $\xi_{s}^{n} \coloneqq [\omega + n\Omega_{cs}]/\sqrt{2} |k_{\|}| v_{ts}$; see~\cite[\S 7]{gary_1993_theory} for a detailed derivation. 
In section~\ref{sec:QLT_vs_PIC}, we consider isotropic cold electrons/ions and anisotropic hot electrons, in which the dispersion relation~\eqref{dispersion_relation_whistler_general} simplifies to  
\begin{equation}\label{dispersion_relation_whistler}
     \frac{m_{e}}{m_{i}} \xi_{i}^{0} Z(\xi_{i}^{1}) + \frac{n_{c}}{n_{e}} \xi_{c}^{0} Z(\xi_{c}^{1}) + \frac{n_{h}}{n_{e}}  \left[\xi_{h}^{0} Z(\xi_{h}^{1})  + A_{h} \left[ 1+ \xi_{h}^{1} Z(\xi_{h}^{1})\right]\right]= \frac{c^2 k_{\|}^2}{\omega_{pe}^2}
\end{equation}
Figure~\ref{fig:dispersion_relation_whistler} shows the dispersion relation results for the PIC simulation described in section~\ref{sec:QLT_vs_PIC}.
The dispersion relation results show that the most unstable whistler mode is around $\omega_{0} \approx 0.5|\Omega_{ce}|$ and $k_{\|0} d_{e} \approx 1$. 
This is consistent with the PIC results shown in section~\ref{sec:QLT_vs_PIC}.

Similar to the moment-based QLT approach described in section~\ref{sec:QLT_secondary}, we can derive moment-based QLT equations for the primary whistler wave anisotropy instability~\cite{safraz_2016_jgr_whistler, yoon_2017_qlt}, evolving the hot electron temperature and primary whistler magnetic power spectral density:
\begin{align*}
    \frac{\mathrm{d} T_{\perp h}(t)}{\mathrm{d} t} &= \frac{\sqrt{\pi} e^2}{2 m_{e}c^2}\int \mathrm{d} k_{\|} \frac{|\vec{\hat{B}}_{W}(k_{\|}, t)|^2 }{k_{\|}^2} \mathrm{Re}\{\xi_{h}^{0} - \xi_{h}^{1}\}\exp(-\mathrm{Re}\{\xi_{h}^{1}\}^2)\left[A_{h} [\mathrm{Re}\{\omega\} - |\Omega_{ce}|]  + \mathrm{Re}\{\omega\}\right],  \\
    \frac{\mathrm{d} T_{\| h}(t)}{\mathrm{d} t} &=-2\frac{\mathrm{d} T_{\perp h}}{\mathrm{d} t} + \frac{\sqrt{\pi} e^2}{m_{e}c^2}  \int \mathrm{d} k_{\|} \frac{|\vec{\hat{B}}_{W}(k_{\|}, t)|^2 }{k_{\|}^2} \mathrm{Re}\{\xi_{h}^{0}\}\exp(-\mathrm{Re}\{\xi_{h}^{1}\}^2)\left[A_{h} [\mathrm{Re}\{\omega\} - |\Omega_{ce}|] + \mathrm{Re}\{\omega\}\right], \\
    \partial_{t} |\vec{\hat{B}}_{W}(k_{\|}, t)|^2  &= 2 \gamma  |\vec{\hat{B}}_{W}(k_{\|}, t)|^2. 
\end{align*}
The QLT equations above are coupled to the whistler wave dispersion relation in Eq.~\eqref{dispersion_relation_whistler}. 
Figure~\ref{fig:QLT_vs_PIC_primary} presents a comparison between the moment-based QLT approach and those obtained from a 1D3V PIC simulation presented in~\cite{roytershteyn_2021_pop} with $n_{c} = 0.8n_{e}$, $\omega_{pe} = 4 |\Omega_{ce}|$, $T_{c} = \SI{10}{\electronvolt}$, $\alpha_{c} = 0.0249 d_{e} |\Omega_{ce}|$, and $T_{\| h}/T_{c} = 200$. 
The 1D3V PIC discretization parameters are: a periodic domain of size $L_{z} = 20 \pi d_{e}$, a grid resolution of $n_{z} = 15,200$ cells, $N_{ppc} = 10^{4}$ particles per cell per species, and a timestep of $\Delta t \omega_{pe} \approx 0.0029$. 
The two simulation approaches exhibit qualitative agreement in the evolution of the hot electron anisotropy and magnetic energy density.
Other studies by~\cite{kim_2017_jgr_qlt_whistler, tao_2017_grl_whistler, lee_2018_QLT_whistler} have shown similar agreement between moment-based QLT and PIC for describing the whistler anisotropy instability. 
Note that the 1D3V PIC simulation will not experience damping of the primary whistler wave since the secondary instability is driven by transverse drifts, which require at least a 2D3V simulation.

\renewcommand{\thefigure}{\arabic{figure}}
\begin{figure}
    \centering
    \begin{subfigure}{0.48\textwidth}
        \centering
        \caption{Whistler anisotropy instability frequency}
        \includegraphics[width=\textwidth]{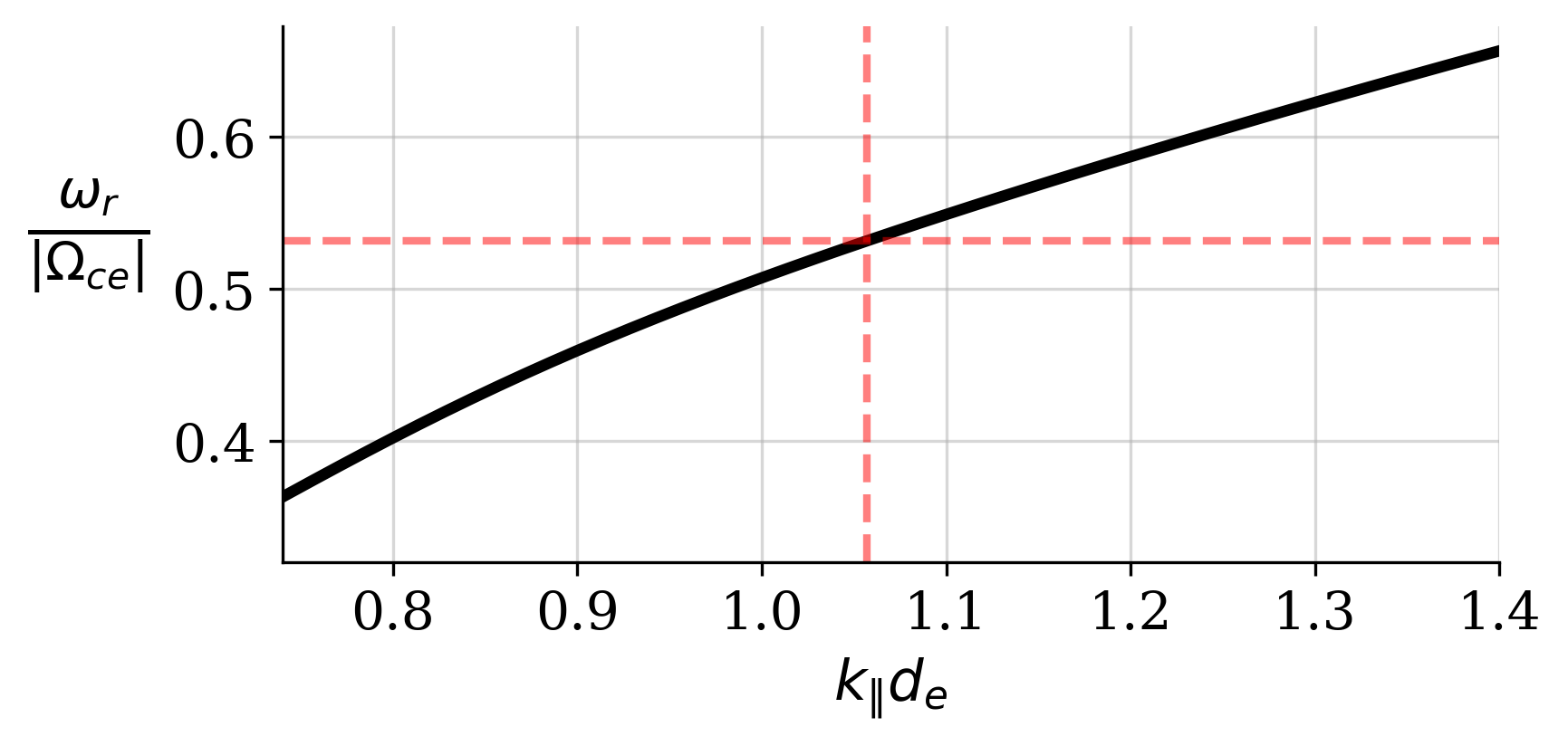}
    \end{subfigure}
    \begin{subfigure}{0.48\textwidth}
        \centering
        \caption{Whistler anisotropy instability growth rate}
        \includegraphics[width=\textwidth]{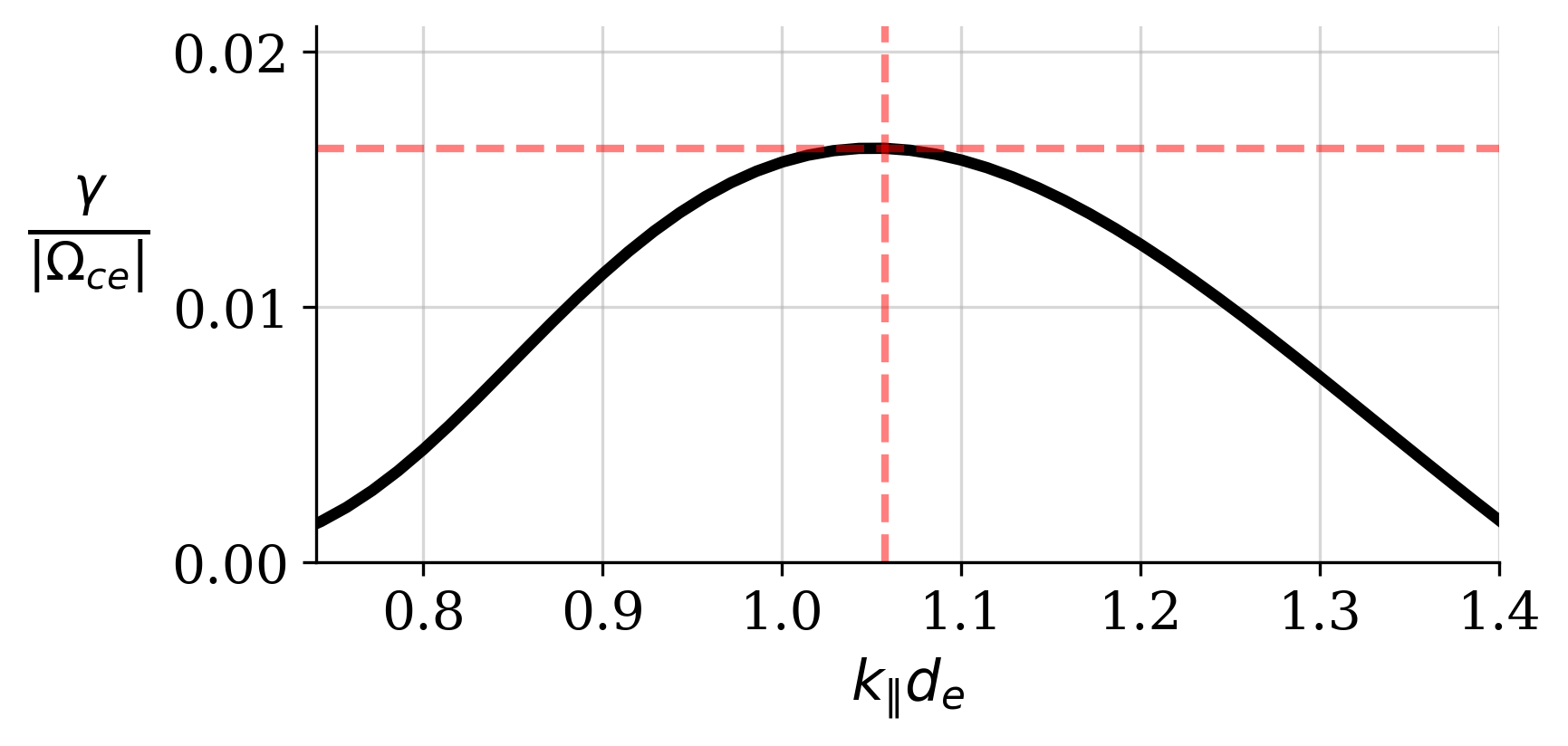}
    \end{subfigure}
    \caption{Frequency and growth rate of the whistler anisotropy instability for the PIC simulation parameters discussed in section~\ref{sec:QLT_vs_PIC}. The frequency and parallel wavenumber corresponding to the most unstable mode are indicated by red dashed vertical and horizontal lines. }
    \label{fig:dispersion_relation_whistler}
\end{figure}

\begin{figure}[h!]
    \centering
    \begin{subfigure}{0.48\textwidth}
        \centering
        \caption{Hot electron plasma beta evolution}
        \includegraphics[width=\textwidth]{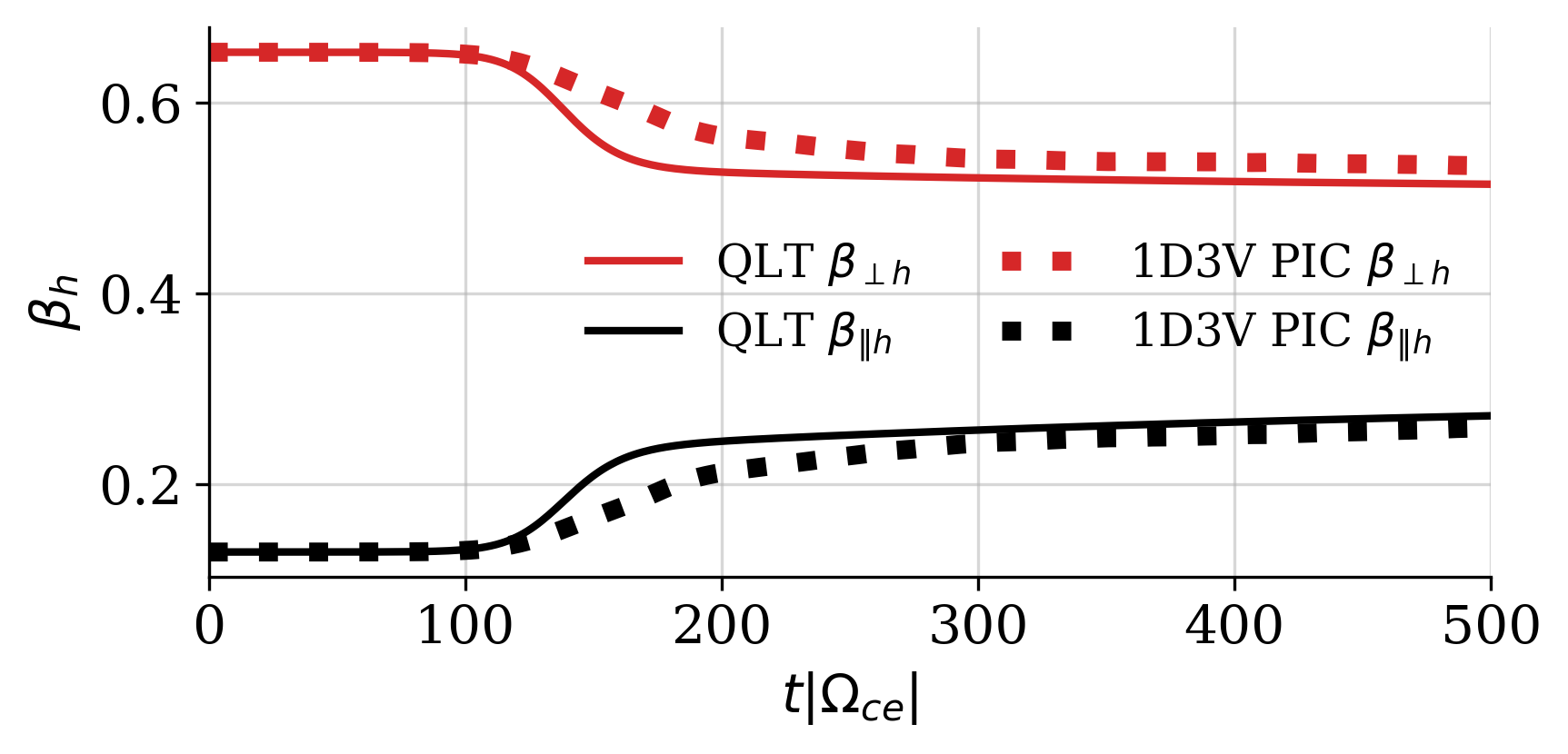}
    \end{subfigure}
        \begin{subfigure}{0.48\textwidth}
        \centering
        \caption{Magnetic energy density}
        \includegraphics[width=\textwidth]{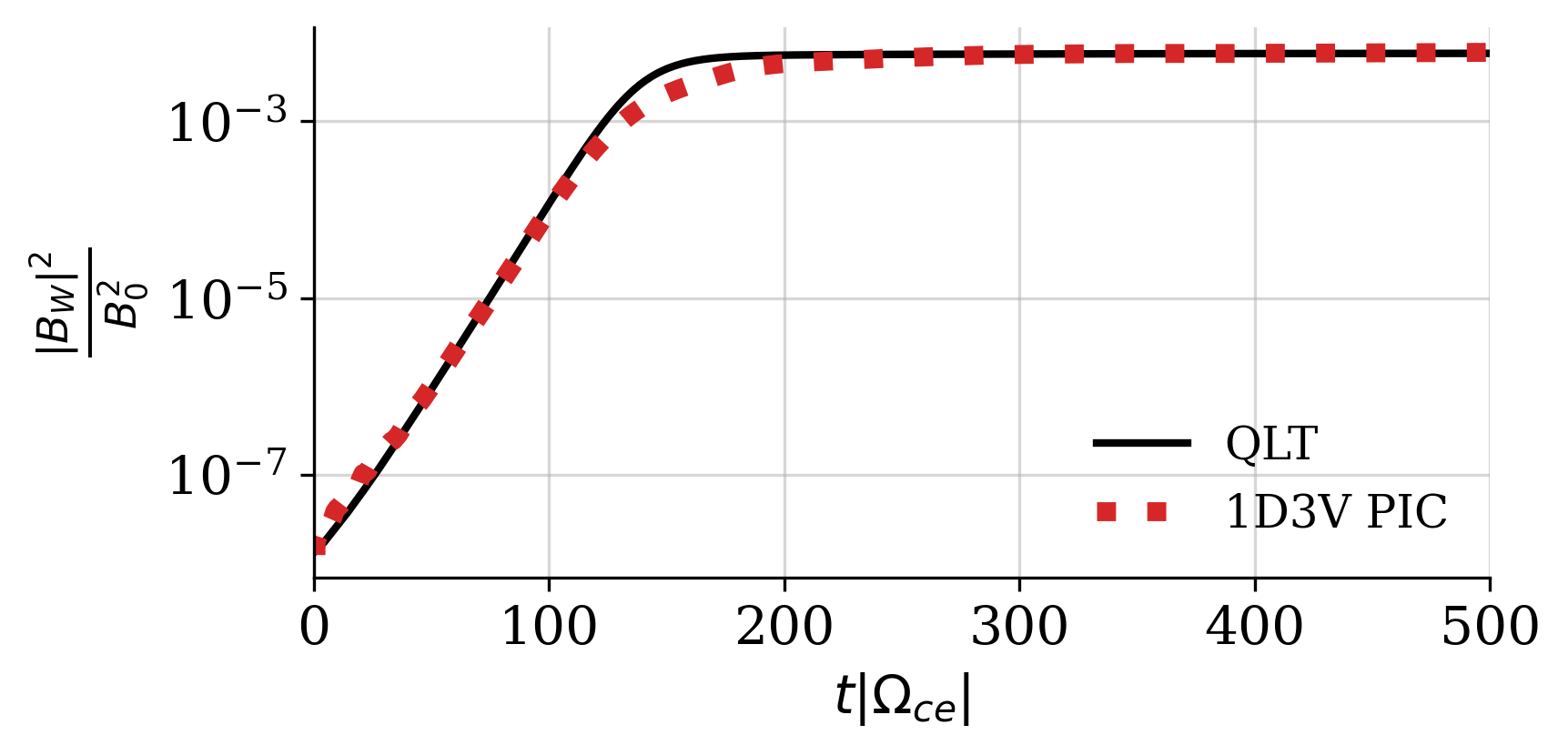}
    \end{subfigure}
    \caption{A comparison of 1D3V PIC vs. moment-based QLT for describing the whistler anisotropy instability evolution of the (a) hot electron plasma beta and (b) magnetic energy density. }
    \label{fig:QLT_vs_PIC_primary}
\end{figure}

\section{Energy conservation described via quantities in the co-drifting with cold electrons frame of reference} \label{sec:energy_conservation}
In the laboratory frame, total energy is a conserved quantity $\mathrm{d} \mathcal{E}^{\mathrm{lab}}_{\mathrm{tot}} / \mathrm{d} t = 0$, such that 
\begin{equation}\label{total_energy_definition}
    \mathcal{E}^{\mathrm{lab}}_{\mathrm{tot}}(t) \coloneqq \underbrace{\sum_{s} \int \mathrm{d}^{3} x \int \mathrm{d}^{3} v \left[\frac{1}{2}|\vec{v}|^2 f_{s}(\vec{x}, \vec{v}, t) \right]}_{\text{kinetic energy } K^{\mathrm{lab}}_{s}(t)} + \underbrace{\int \mathrm{d}^{3} x \frac{|\delta \vec{E}(\vec{x}, t)|^2}{8\pi}}_{\text{electrostatic energy}} +  \underbrace{ \left[\frac{\omega_{0}^2}{c^2 k_{\|0}^2} + 1\right ] \frac{| B_{W}(t)|^2}{8\pi}}_{\text{primary wave energy}}
\end{equation}
The total kinetic energy in the (non-inertial) frame co-drifting with cold electrons $K_{s}(t)$, see Eq.~\eqref{co-drifting_frame}, is  
\begin{align}
     K_{s}(t) &\coloneqq \int \mathrm{d}^{3} x' \int \mathrm{d}^{3} v' \left[\frac{1}{2}|\vec{v}'|^2 f_{s}(\vec{x}', \vec{v}', t) \right] \nonumber \\
    &= \int \mathrm{d}^{3} x' \int \mathrm{d}^{3} v' \left[\frac{1}{2}|\vec{v} - \vec{V}_{Dc}(t)|^2 f_{s}(\vec{x}', \vec{v}', t) \right] \nonumber \\
    &= K^{\mathrm{lab}}_{s}(t) -  \int \mathrm{d}^{3} x' \int \mathrm{d}^{3} v' \vec{V}_{Dc}(t) \cdot \vec{v} f_{s}(\vec{x}', \vec{v}', t) + \frac{1}{2} |\vec{V}_{Dc}(t)|^2 \int \mathrm{d}^{3} x' \int \mathrm{d}^{3} v' f_{s}(\vec{x}', \vec{v}', t) \nonumber\\
    &= K^{\mathrm{lab}}_{s}(t) - \vec{V}_{Dc}(t) \cdot \vec{M}^{\mathrm{lab}}_{s}(t) + \frac{1}{2} |\vec{V}_{Dc}(t)|^2 n_{s}. \label{kinetic_energy_lab}
\end{align}
Moreover, the electrostatic energy in the laboratory frame can be described as 
\begin{align}
    \int \mathrm{d}^3 x |\delta \vec{E}(\vec{x}, t)|^2 &= \int \mathrm{d}^{3} k  |\delta \vec{E}(\vec{k}, t)|^2 =  \int \mathrm{d}^{3} k \int \mathrm{d}^{3} x \exp(-i \vec{k} \cdot \vec{x}) |\delta \vec{E}(\vec{x}, t)|^2 \nonumber \\
    &= \int \mathrm{d}^{3} k \int \mathrm{d}^{3} x \exp\left(-i \vec{k} \cdot \vec{x}' - i \int_{0}^{t} \vec{V}_{Dc}(\tau) \cdot \vec{k} \mathrm{d} \tau \right) |\delta \vec{E}(\vec{x}, t)|^2 \nonumber \\
    &= \int \mathrm{d}^{3} k \exp\left(- i k_{\perp} \int_{0}^{t} \vec{V}_{x, Dc}(\tau) \mathrm{d} \tau \right) |\delta \vec{\tilde{E}}(\vec{k}, t)|^2 \nonumber \\
    &= \int \mathrm{d}^{3} k \sum_{n=-\infty}^{\infty} J_{n} \left( \frac{k_{\perp}|\vec{V}_{Dc}|}{\omega_{0}} \right) \exp(i n \omega_{0} t) |\delta \vec{\tilde{E}}(\vec{k}, t)|^2. \label{electrostatic_energy_lab}
\end{align}
Inserting Eq.~\eqref{kinetic_energy_lab} and Eq.~\eqref{electrostatic_energy_lab} in Eq.~\eqref{total_energy_definition} and considering that $\omega_{0}^2/c^2 k_{\|0}^2 \ll 1$ for whistler waves results in 
\begin{align}
    \mathcal{E}^{\mathrm{lab}}_{\mathrm{tot}}(t) &= \sum_{s} K_{s}(t) + \vec{V}_{Dc}(t) \cdot \sum_{s} \vec{M}^{\mathrm{lab}}_{s}(t) -\frac{1}{2} |\vec{V}_{Dc}(t)|^2  \label{kinetic_energy_lab_2} \\
    &+ \frac{1}{8\pi}  \int \mathrm{d}^{3} k \sum_{n=-\infty}^{\infty} J_{n}\left(\frac{k_{\perp}|\vec{V}_{Dc}|}{\omega_{0}} \right) \cos(n \omega_{0} t) |\delta \vec{\tilde{E}}(\vec{k}, t)|^2
    + \frac{| B_{W}(t)|^2}{8\pi}. \nonumber
\end{align}
We apply a low-pass filter on Eq.~\eqref{kinetic_energy_lab_2} to eliminate the terms that change on small time scales $\sim \omega_{0}^{-1}$
\begin{align*}
    \frac{\omega_{0}}{2\pi}\int_{t}^{t+\frac{2\pi}{\omega_{0}}} \mathcal{E}^{\mathrm{lab}}_{\mathrm{tot}}(\tau) \mathrm{d} \tau &= \sum_{s} K_{s}(t)  +\frac{1}{2} |\vec{V}_{Dc}(t)|^2  + \frac{1}{8\pi}  \int \mathrm{d}^{3} k J_{0}\left(\frac{k_{\perp}|\vec{V}_{Dc}|}{\omega_{0}} \right) |\delta \vec{\tilde{E}}(\vec{k}, t)|^2
    + \frac{| B_{W}(t)|^2}{8\pi}.
\end{align*}
For the parameters considered in the manuscript $k_{\perp} |\vec{V}_{Dc}|/\omega_{0} \lesssim 1$, so 
\begin{align*}
    \frac{\omega_{0}}{2\pi}\int_{t}^{t+\frac{2\pi}{\omega_{0}}} \mathcal{E}^{\mathrm{lab}}_{\mathrm{tot}}(\tau) \mathrm{d} \tau &= \sum_{s} K_{s}(t) +\frac{1}{2} |\vec{V}_{Dc}(t)|^2  + \frac{1}{8\pi}  \int \mathrm{d}^{3} k |\delta \vec{\tilde{E}}(\vec{k}, t)|^2
    + \frac{| B_{W}(t)|^2}{8\pi}\\
    &= \sum_{s} K_{s}(t)  + \frac{1}{8\pi}  \int \mathrm{d}^{3} k |\delta \vec{\tilde{E}}(\vec{k}, t)|^2
    + \frac{| B_{W}(t)|^2}{8\pi} \left[1 + 4\pi\left|\frac{\omega_{0}}{k_{\|0}} \frac{1}{B_{0}} \frac{\Omega_{ce}}{\omega_{0}  + \Omega_{ce}}\right|^2 \right],
\end{align*}
and by considering only the change in energy of cold electrons and the electric and magnetic fields, we arrive at Eq.~\eqref{dBdt}. 

Figure~\ref{fig:PIC_energy_conservation} shows the energy conservation in the laboratory frame of the 2D3V PIC simulation described in section~\ref{sec:simulation-setup-PIC-QLT}. 
The ions contribute negligibly to the total energy fluctuations as shown in Figure~\ref{fig:full_energy}.
Additionally, the changes in the electric field do not significantly impact the total energy fluctuations. 
The dominant contributions arise from the thermal energy of the hot and cold electron populations, the kinetic energy of the cold electrons, and the magnetic field energy. 
Since the QLT model, see Eq.~\eqref{dBdt}, retains only fluctuations in the cold electron thermal and kinetic energies, along with the electric and magnetic field energies, we use the PIC simulation to estimate the resulting energy conservation error. 
From Figure~\ref{fig:approx_energy}, this energy conservation error is approximately $10^{-2}\%$. 
Including the dynamics of the hot electron thermal energy would further improve energy conservation; however, incorporating its coupling to the monochromatic primary whistler wave is nontrivial and is therefore left for future work.

\renewcommand{\thefigure}{\arabic{figure}}
\begin{figure}[ht!]
    \centering
    \begin{subfigure}{0.5\textwidth}
        \centering
        \caption{All 2D3V PIC simulation energy fluctuation components}
        \includegraphics[width=\textwidth]{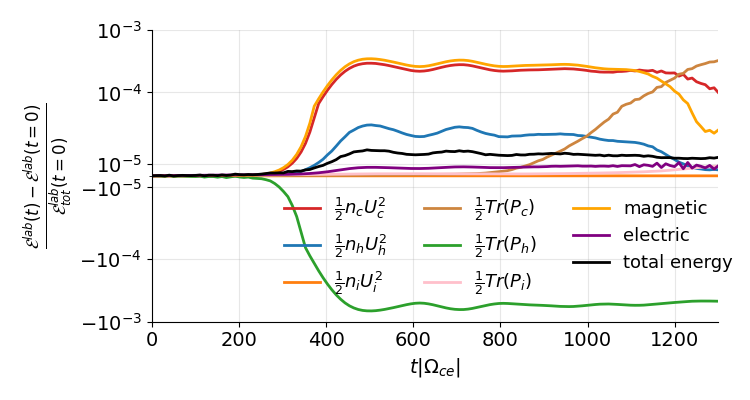}
        \label{fig:full_energy}
    \end{subfigure}
    \hspace{-10pt}
    \begin{subfigure}{0.5\textwidth}
        \centering
        \caption{Only cold electrons and electric and magnetic fields}
        \includegraphics[width=\textwidth]{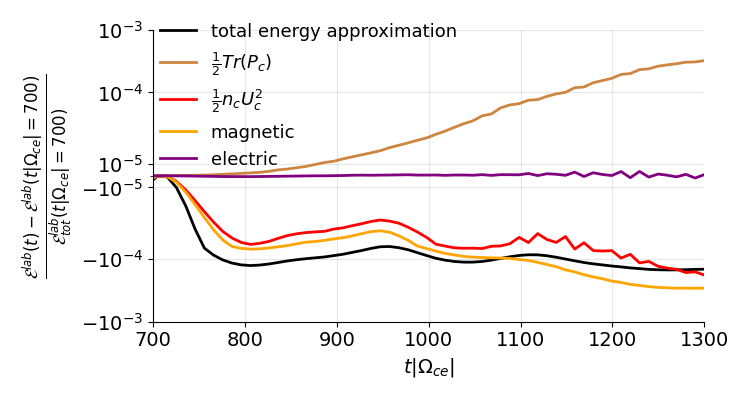}
        \label{fig:approx_energy}
    \end{subfigure}
    \caption{The 2D3V PIC simulation, see section~\ref{sec:simulation-setup-PIC-QLT}, (a) all energy components and (b) only the cold electrons and electric and magnetic field components. The 2D3V PIC simulation total energy conservation relative error is $ \approx 0.001\%$ due to numerical discretization as shown in subfigure~(a), and $\approx 0.01\%$ if only cold electrons and electric and magnetic field components are considered from $t|\Omega_{ce}| = 700$ as shown in subfigure~(b). The dominant contributions of the total energy fluctuations arise from the thermal energy of the hot and cold electron populations, the kinetic energy of the cold electrons, and the magnetic field energy.  }
    \label{fig:PIC_energy_conservation}
\end{figure}

\section*{Acknowledgment}
O.I. is grateful for discussions with Chris Holland.
O.I. was partially supported by the Los Alamos National Laboratory (LANL) Student Fellowship sponsored by the Center for Space and Earth Science (CSES). 
CSES is funded by LANL's Laboratory Directed Research and Development (LDRD) program under project number 20210528CR. O.I. was partially supported by the Strategic Enhancement of Excellence through Diversity Fellowship at the University of California, San
Diego in the Department of Mechanical and Aerospace Engineering.
The LANL LDRD Program supported G.L.D. under project number 20250577ER. LANL is operated by Triad National Security, LLC, for the National Nuclear Security Administration of the US Department of Energy (Contract No. 89233218CNA000001). 
V.R. was partially supported by NASA grant 80NSSC22K1014. 
S.J. was supported for this work by CSES. PIC simulations utilized in this work were performed using resources of the National Energy Research Scientific Computing Center (NERSC), a U.S. Department of Energy Office of Science User Facility located at Lawrence Berkeley National Laboratory, operated under Contract No. DE-AC02-05CH11231; resources provided by the NASA High-End Computing (HEC) Program through the NASA Advanced
Supercomputing (NAS) Division at Ames Research Center; and
resources at the Texas Advanced Computing Center (TACC) at The University of Texas at Austin.

\bibliography{references}

\end{document}